\RequirePackage{fix-cm}
\documentclass[twocolumn]{svjour3}          
\smartqed  
\usepackage{graphicx}
\usepackage{amsmath}
\usepackage[numbered,autolinebreaks,useliterate]{mcode}
\usepackage{subcaption}
\usepackage{bm}
\usepackage{xfrac}
\usepackage{natbib}
\journalname{Preprint}
\begin{document}

\title{A detailed introduction to density-based topology optimisation of fluid flow problems with implementation in MATLAB
\thanks{This paper is dedicated to my former supervisor and colleague Ole Sigmund, the pioneer of educational topology optimisation codes and so much more.}
}

\titlerunning{A detailed introduction to density-based topology optimisation of fluid flow problems}        

\author{Joe Alexandersen}

\institute{J. Alexandersen \at
              Department of Mechanical and Electrical Engineering \\
              University of Southern Denmark \\
              Campusvej 55, DK-5230 Odense M \\
              Tel.: +45 65507465\\
              \email{joal@sdu.dk}
}

\date{Received: date / Accepted: date}

\maketitle

\begin{abstract}
This article presents a detailed introduction to density-based topology optimisation of fluid flow problems. The goal is to allow new students and researchers to quickly get started in the research area and to skip many of the initial steps, often consuming unnecessarily long time from the scientific advancement of the field. This is achieved by providing a step-by-step guide to the components necessary to understand and implement the theory, as well as extending the supplied MATLAB code. The continuous design representation used and how it is connected to the Brinkman penalty approach, for simulating an immersed solid in a fluid domain, is illustrated. The different interpretations of the Brinkman penalty term and how to chose the penalty parameters are explained. The accuracy of the Brinkman penalty approach is analysed through parametric simulations of a reference geometry. The chosen finite element formulation and the solution method is explained. The minimum dissipated energy optimisation problem is defined and how to solve it using an optimality criteria solver and a continuation scheme is discussed. The included MATLAB implementation is documented, with details on the mesh, pre-processing, optimisation and post-processing. The code has two benchmark examples implemented and the application of the code to these is reviewed. Subsequently, several modifications to the code for more complicated examples are presented through provided code modifications and explanations. Lastly, the computational performance of the code is examined through studies of the computational time and memory usage, along with recommendations to decrease computational time through approximations.
\end{abstract}

\section{Introduction} \label{sec:intro}

\subsection{Topology optimisation}
The topology optimisation method emerged from sizing and shape optimisation at the end of the 1980s in the field of solid mechanics. The homogenisation method by \citet{Bendsoee1988} is cited as being the seminal paper on topology optimisation. Topology optimisation is posed as a material distribution technique that answers the question ``where should material be placed?'' or alternatively ``where should the holes be?''. This differentiates it from the classical disciplines of sizing and shape optimization, since it does not need an initial structure with the topology defined \textit{a priori}. Although a range of topology optimisation approaches exists, such as the level set and phase field methods, the most popular method remains the so-called ``density-based'' method. The review papers by \citet{Sigmund2013} and \citet{Deaton2014} give a general overview of topology optimisation methods and applications.

Due to the maturity of the method for solid mechanics, most finite element analysis (FEA) and computer-aided design (CAD) packages now have built-in topology optimisation capabilities. Further, many public codes have been made available, often together with educational articles explaining their use. The first published code was the famous ``99-line'' MATLAB code by \citet{Sigmund2001}, used broadly around the world in teaching and initial experiments with topology optimisation. Subsequently, this code was updated to a more efficent ``88-line'' version by \citet{Andreassen2011} and most recently a hyper-efficient ``neo99'' version by \citet{Ferrari2020}. Over the years, many educational codes have been published for topology optimisation using e.g. level set methods \citep{Challis2010,Wei2018}, unstructured polygonal meshes \citep{Talischi2012}, energy-based homogenisation \citep{Xia2015}, truss ground structures \citep{Zegard2014} and large-scale parallel computing \citep{Aage2015}. The recent review paper by \citet{Wang2021} provides a comprehensive overview of educational articles in the field of structural and multidisciplinary optimisation.

\subsection{Fluid flow}

The density-based topology optimisation approach was extended to the design of Stokes flow problems by \citet{Borrvall2003} and further to Navier-Stokes flow by \citet{GersborgHansen2005}. Since those seminal papers, topology optimisation has been applied to a large range of flow-based problems. The recent review paper by \citet{Alexandersen2020} details the many contributions to this still rapidly developing field. Topology optimisation of pure fluid flow has now become reasonably mature, with only minor recent contributions for steady laminar flow \citep{Alexandersen2020}. Therefore, research efforts should be focused on building further upon this technology, calling for an educational article and a simple code that allows for quickly getting started and skipping initial repetitive implementations.

\citet{Pereira2016} presented an extension of their polygonal mesh code \citep{Talischi2012} to Stokes flow topology optimisation. Whereas \citet{Pereira2016} focused on stable discretisations of polygonal elements for \textit{Stokes} flow, this educational article focuses on giving a \textbf{detailed and complete introduction} to density-based topology optimisation of fluid flow problems governed by the \textit{Navier-Stokes} equations. The aim is to serve as the first point of contact for new students and researchers to quickly get started in the research area and to skip many of the initial steps, often consuming unnecessarily long time from the scientific advancement of the field.
The goal is not to give an overarching overview of the literature on the subject, so references herein are rather sparse on purpose and the reader is referred to the recent review paper instead \citep{Alexandersen2020}.

This article provides a step-by-step introduction to all the needed steps, along with a relatively simple and efficient implementation in MATLAB - where compatibility of the main files with the open-source alternative GNU Octave has been ensured.
The code presented in this paper builds on the ``88-line'' code by \citet{Andreassen2011} and uses the same basic code structure and variable names.

\subsection{Paper layout}

The article is structured as follows:
\begin{itemize}
    \item Section \ref{sec:designrep} introduces the design representation used for density-based topology optimisation of fluid flow problems;
    \item Section \ref{sec:physics} presents the physical formulation and an in-depth investigation of the error convergence for important problem parameters;
    \item Section \ref{sec:discrandsol} details the finite element discretisation and the state solution procedure used;
    \item Section \ref{sec:optimisation} discusses the optimisation formulation, sensitivity analysis and design solution procedure;
    \item Section \ref{sec:incex} introduces the two benchmark examples included in the code;
    \item Section \ref{sec:code} presents a description of the accompanying MATLAB implementation;
    \item Section \ref{sec:examples} shows results obtained using the code for the benchmark examples;
    \item Section \ref{sec:modifications} discusses modifications to the code for solving more advanced problems;
    \item Section \ref{sec:compperf} discusses the computational time and memory usage;
    \item Section \ref{sec:conclusion} presents concluding remarks.
\end{itemize}

\section{Design representation} \label{sec:designrep}

This section describes the design representation introduced to model solid domains immersed in a fluid, allowing for density-based topology optimisation. First, the true separated discrete domains that should be represented are introduced, and secondly, a continuous design representation is described.

\begin{figure}
    \begin{subfigure}{\columnwidth}
    \centering
    \includegraphics[height=0.6\linewidth]{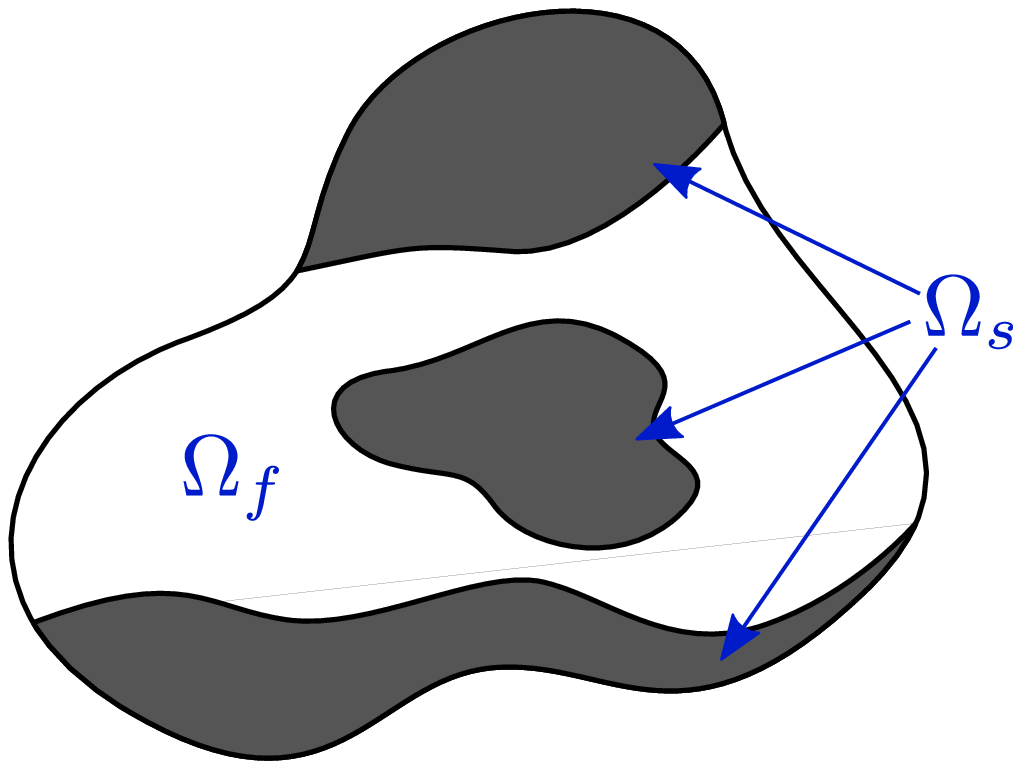}
    \caption{Discrete separation of domains}
    \label{fig:arbitrary_disc}
    \end{subfigure}
    \\
    \begin{subfigure}{\columnwidth}
    \centering
    \includegraphics[height=0.6\linewidth]{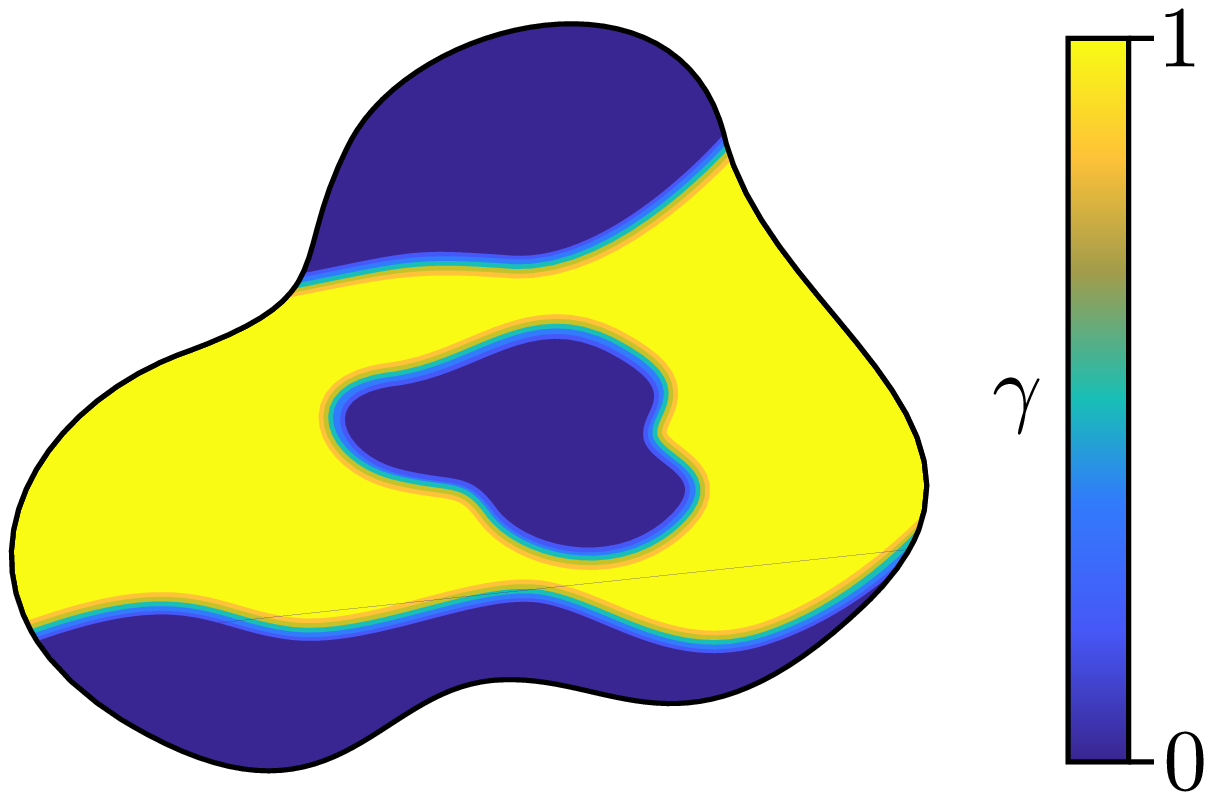}
    \caption{Continuous design representation}
    \label{fig:arbitrary_dens}
    \end{subfigure}
    \caption{Arbitrary computational domain consisting of fluid and solid domains.}
    \label{fig:arbitrary}
\end{figure}

\subsection{Discrete separate domains}
Figure \ref{fig:arbitrary_disc} shows an arbitrary example of a truly discrete problem with separated fluid and solid domains, denoted as $\Omega_f$ and $\Omega_s$, respectively. Both $\Omega_f$ and $\Omega_s$ can be comprised as the union of non-overlapping subdomains, as illustrated for $\Omega_s$ in Figure \ref{fig:arbitrary_disc}. The interface between the solid and fluid domains is clearly defined due to the separation of the two domains.

\subsection{Continuous relaxed representation}
In order to perform density-based topology optimisation using gradient-based methods, the discrete representation must be relaxed.
Firstly, a spatially-varying field, $\gamma(\mathbf{x})$, is introduced, which will be termed the design field. It is equivalent to the characteristic function of the fluid domain $\Omega_f$:
\begin{equation}
    \gamma(\mathbf{x}) = 
    \left\{
\begin{array}{rl}
    1 & \; \mathrm{ if } \; \mathbf{x}\in\Omega_f \\
    0 & \; \mathrm{ if } \; \mathbf{x}\in\Omega_s
\end{array}
\right.
\end{equation}
Secondly, the design field is relaxed from binary values to a continuous field, allowing intermediate values: $\gamma(\mathbf{x}) \in \left[ 0;1 \right]$.
Figure \ref{fig:arbitrary_dens} shows a continuous relaxed representation of the discrete separated domains of Figure \ref{fig:arbitrary_disc}. The continuous representation has an implicit representation of the solid-fluid interface, represented by a transition region of intermediate design field values as shown in Figure \ref{fig:arbitrary_dens}. Generally, during the optimisation process there will be many regions of intermediate values, whereas a converged design will have significantly less, if any -- depending on various details as will be discussed later.
By coupling the continuous design representation to coefficients of the governing equations, it is possible to model an immersed geometry which is implicitly defined by the design field.

\section{Physical formulation} \label{sec:physics}

This section presents the physical formulation of the fluid flow problems treated in this article. The Navier-Stokes equations are the governing equations and a so-called Brinkman penalty term is introduced to facilitate density-based topology optimisation.

\subsection{Governing equations}
This article restricts itself to steady-state laminar and incompressible flow, governed by the Navier-Stokes equations:
\begin{subequations}
\begin{align}
    \rho u_{j}\dfrac{\partial u_i}{\partial x_j} - \mu \dfrac{\partial}{\partial x_j}\left( \dfrac{\partial u_i}{\partial x_j} + \dfrac{\partial u_j}{\partial x_i} \right) + \dfrac{\partial p}{\partial x_i} - f_{i} &= 0  \label{eq:navierstokes-a} \\
    \dfrac{\partial u_i}{\partial x_i} &= 0  \label{eq:navierstokes-b}
\end{align} \label{eq:navierstokes}
\end{subequations}
\hspace{-1ex}where $u_{i}$ is the $i$-th component of the velocity vector $\textbf{u}$, $p$ is the pressure, $\rho$ is the density, $\mu$ is the dynamic viscosity, and $f_{i}$ is the $i$-th component of the body force vector $\textbf{f}$.
The Reynolds number is commonly used to describe a flow and is defined as:
\begin{equation} \label{eq:reynoldsnumber}
    Re = \frac{\rho U L}{\mu}
\end{equation}
where $U$ and $L$ are the reference velocity and length-scale, respectively.
The Reynolds number describes the ratio of inertial to viscous forces in the fluid: if $Re < 1$, the flow is dominated by viscous diffusion; if $Re > 1$, the flow is dominated by inertia.

\subsection{Brinkman penalty term}

In order to facilitate topology optimisation, an artificial body force is introduced, termed the Brinkman penalty term:
\begin{equation}
f_{i} = -\alpha u_{i} \label{eq:brinkman}
\end{equation}
which represents a resistance term and a momentum sink, drawing energy from the flow. The Brinkman penalty factor, $\alpha$, is a spatially-varying parameter defined as follows for the separated domains introduced in Figure \ref{fig:arbitrary_disc}:
\begin{equation} \label{eq:alpha}
    \alpha(\mathbf{x})=\left\{
    \begin{array}{ll}
        0 & {\rm if}\;\mathbf{x} \in \Omega_{f} \\
        \infty & {\rm if}\;\mathbf{x} \in \Omega_{s}
    \end{array}
    \right.
\end{equation}
\hspace{-1ex}where the 0 in the fluid domain recovers the original equations and the infinite value inside the solid domain theoretically ensures identically zero velocities. This allows for extending the Navier-Stokes equations to the entire computational domain:
\begin{equation}
\left.
\begin{array}{rll}
    \rho u_{j}\dfrac{\partial u_i}{\partial x_j} - \mu \dfrac{\partial}{\partial x_j} & \left( \dfrac{\partial u_i}{\partial x_j} + \dfrac{\partial u_j}{\partial x_i} \right) \\
    & + \dfrac{\partial p}{\partial x_i} + \alpha\left( \gamma(\mathbf{x}) \right) u_{i}& = 0 \\
    \\
    & \textcolor{white}{\left( \dfrac{\partial u_i}{\partial x_j} + \dfrac{\partial u_j}{\partial x_i} \right)} \dfrac{\partial u_i}{\partial x_i}& = 0
\end{array}
\right\} \rm{for}\, \mathbf{x} \in \Omega
\label{eq:navierstokes_brink}
\end{equation}
where solid domains are modelled as immersed geometries through the Brinkman penalty term.

The Brinkman penalty term can be interpreted in three different ways:
\begin{enumerate}
    \item Volumetric penalty term to ensure zero velocities in the solid regions
    \item Friction term from out-of-plane viscous resistance
    \item Friction term from an idealised porous media
\end{enumerate}
The first interpretation is a purely algorithmic approach, where the only purpose is to simulate an immersed fully solid geometry inside a fluid domain. The second and third are both physical interpretations, but require different lower and upper bounds of Equation \ref{eq:alpha} formulated based on physical parameters. The second interpretation is limited to only two-dimensional problems, where the computational domain has a finite and relatively small out-of-plane thickness. The first and third interpretations can be applied to three-dimensional problems.

The seminal work by \citet{Borrvall2003} for Stokes flow and \citet{GersborgHansen2005} for Navier-Stokes flow, both introduced a design parametrisation based on the out-of-plane channel height for two-dimensional problems following the second interpretation above. Both works also note the comparison to the third interpretation and \citet{GersborgHansen2005} argues for the first interpretation as the ultimate goal of topology optimisation for fluid flow. The recent review paper by \citet{Alexandersen2020} shows that the dominant interpretation is the first, a purely algorithmic approach to ensure zero velocities. This is because it easily extends to three-dimensions and no physical reflections are required. However, as it will be discussed in Sections \ref{sec:minpenalty} and \ref{sec:maxpenalty} and shown in Section \ref{sec:varparstudy}, it is highly relevant to couple the parameters to the physical interpretations to ensure correct scaling of the equations.

\subsubsection{Numerical relaxation}

In practise, the bounds of the Brinkman penalty factor are defined as follows, based on the design field, $\gamma$, introduced in Figure \ref{fig:arbitrary_dens}:
\begin{equation}
    \alpha(\gamma(\mathbf{x}))=\left\{
    \begin{array}{ll}
        \alpha_{\rm{min}} & {\rm if}\; \gamma(\mathbf{x}) = 1 \\
        \alpha_{\rm{max}} & {\rm if}\; \gamma(\mathbf{x}) = 0
    \end{array}
    \right. \label{eq:alpha_dens}
\end{equation}
For numerical reasons, a finite value must be used in the solid domain, $\alpha_\textrm{max}$, being large enough to ensure negligible velocities but small enough to ensure stability. The choice of this maximum value will be discussed in Sections \ref{sec:maxpenalty} and \ref{sec:varparstudy}. Furthermore, if two-dimensional problems with a finite out-of-plane thickness are treated, a minimum value, $\alpha_\textrm{min}$, should remain to account for the out-of-plane viscous resistance as discussed in Section \ref{sec:minpenalty}.

\begin{figure}
    \centering
    \includegraphics[width=0.9\linewidth]{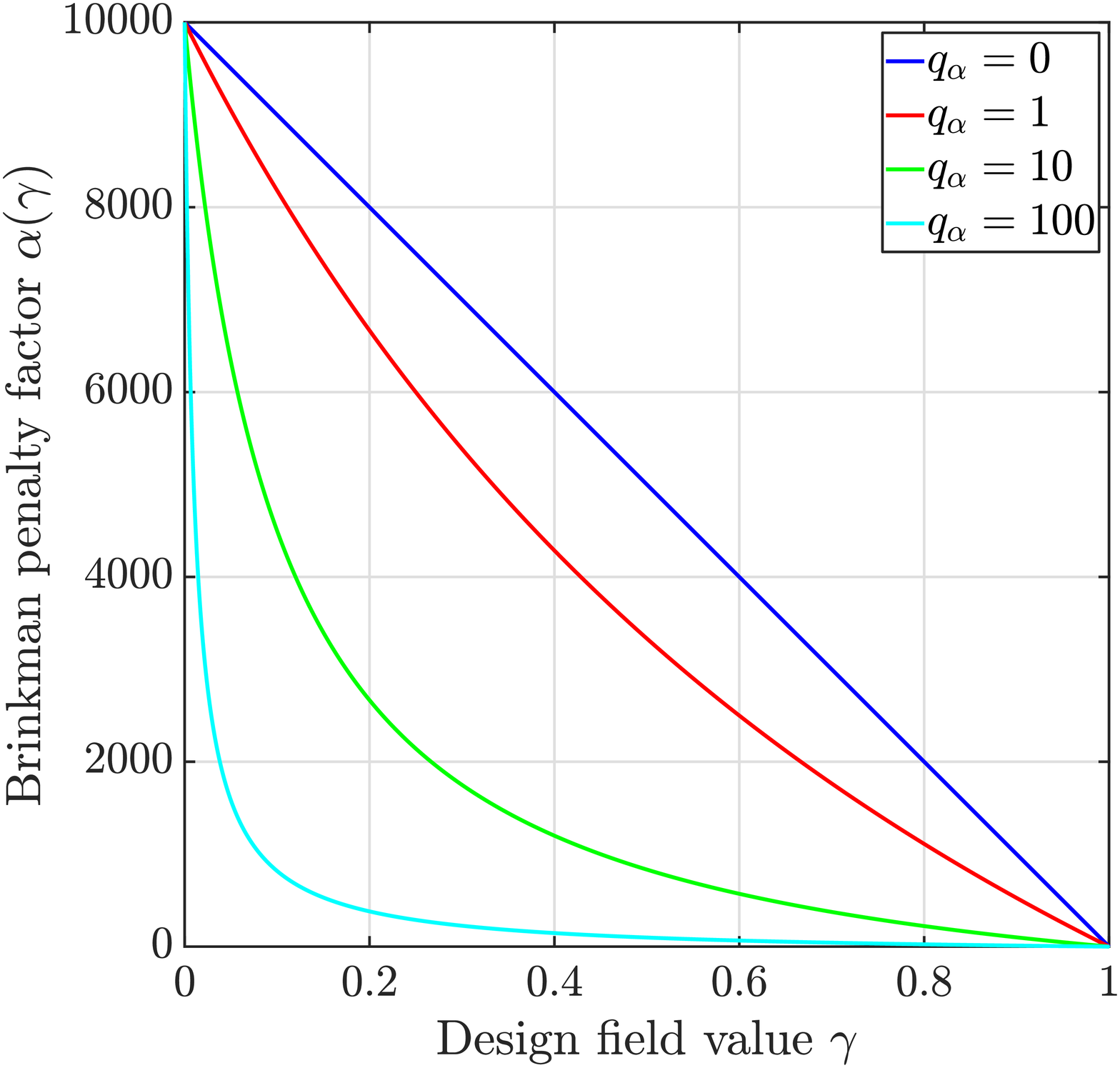}
    \caption{Interpolation of $\alpha$ for various $q_{\alpha}$.}
    \label{fig:interpol_alpha}
\end{figure}
In order to interpolate between solid and fluid domains, the Brinkman penalty factor is interpolated using the following function:
\begin{equation} \label{eq:alpha_interp}
    \alpha(\gamma) = \alpha_{\rm{min}} + (\alpha_{\rm max}-\alpha_{\rm{min}}) \frac{1-\gamma}{1+q_{\alpha}\gamma}
\end{equation}
where $q_{\alpha}$ is a parameter determining the shape of the interpolation, as illustrated in Figure \ref{fig:interpol_alpha} for $\alpha_\textrm{min} = 0$ and $\alpha_\textrm{max} = 10^4$. This function is not the same as introduced by \citet{Borrvall2003}, but rather the Rational Approximation of Material Properties (RAMP) function \citep{Stolpe2001}. This ensures a linear interpolation is directly achieved by setting $q_\alpha$ to 0, whereas the function used by \citet{Borrvall2003} only has the linear case as an asymptotic limit. Effectively, $q_\alpha$ herein is the inverse of the $q$ used by \citet{Borrvall2003}.

\subsubsection{Minimum penalty factor} \label{sec:minpenalty}
The seminal work by \citet{Borrvall2003} for Stokes flow and \citet{GersborgHansen2005} for Navier-Stokes flow, introduced topology optimisation of fluid flow problems using a design parametrisation based on the out-of-plane channel height for two-\allowbreak dimensional problems. By assuming parallel plates distanced $2h$ apart and a fully-developed pressure-driven flow (Poiseuille flow), an out-of-plane parabolic velocity profile is assumed between the two plates. This allows for analytical through-thickness integration, yielding an additional term to the momentum equations from the out-of-plane viscous resistance in the form of Equation \ref{eq:brinkman}. When treating two-dimensional problems with a finite out-of-plane thickness and lateral no-slip walls, the minimum Brinkman penalty must be set to:
\begin{equation} \label{eq:alphah}
    \alpha_\textrm{min} = \frac{5\mu}{2h^2}
\end{equation}
where $h$ is half of the domain thickness. However, when treating three-dimensional problems, the minimum Brinkman penalty factor should be set to 0 to recover the original unencumbered Navier-Stokes momentum equations.

\subsubsection{Maximum penalty factor} \label{sec:maxpenalty}
Finding the correct maximum penalty factor is not trivial. It must be large enough to ensure negligible flow through solid regions, but small enough to ensure numerical stability. While the Brinkman term is often seen purely as a volumetric penalty approach for imposing the no-flow conditions inside the solid, it is beneficial to ensure consistent dimensional scaling of the penalty factor. This can be done by relating it to the friction factor from a fictitious porous media:
\begin{equation} \label{eq:alphakappa}
    \alpha_{\rm{max}} = \frac{\mu}{\kappa}
\end{equation}
where $\kappa$ is the permeability of the fictitious porous media. This permeability must then be small enough to ensure negligible velocities inside the solid regions, but as will be shown in Section \ref{sec:twopipe_amax}, the minimum dissipated energy optimisation is pretty forgiving. For a given constant permeability, Equation \ref{eq:alphakappa} ensures an almost constant ratio between the order of magnitude for the velocities in the flow and solid regions for a range of Reynolds numbers, as demonstrated in Section \ref{sec:varparstudy}.

\subsubsection{Note on physical interpretations}

It is easily seen from Equations \ref{eq:alphah} and \ref{eq:alphakappa}, that the two physical interpretations are analogous by connecting the permeability to the out-of-plane thickness:
\begin{equation}
    \kappa = \frac{2h^2}{5}
\end{equation}
as noted by both \citet{Borrvall2003} and \citet{GersborgHansen2005}.

\subsubsection{Dissipated energy}

This article focuses on minimising the dissipated energy by a flow through a channel structure.
The dissipated energy due to viscous resistance
is defined as:
\begin{equation} \label{eq:disenergy_visc}
    \phi = \frac{1}{2} \int_{\Omega_{f}} \mu \frac{\partial u_i}{\partial x_j} \left(  \frac{\partial u_i}{\partial x_j} +  \frac{\partial u_j}{\partial x_i} \right) \,dV 
\end{equation}
The dissipated energy is often used as the objective functional of topology optimisation and it is, thus, relevant to investigate the accuracy of this measure when using the Brinkman penalty method.

\subsection{Effect of varying parameters} \label{sec:varparstudy}

This section presents the effect of varying the Reynolds number $Re$ and Brinkman penalty factor $\alpha$ on the accuracy of the Brinkman-penalised Navier-Stokes equations. 

\subsubsection{Example setup}

\begin{figure}
\begin{subfigure}{\columnwidth}
  \centering
  \includegraphics[width=\linewidth]{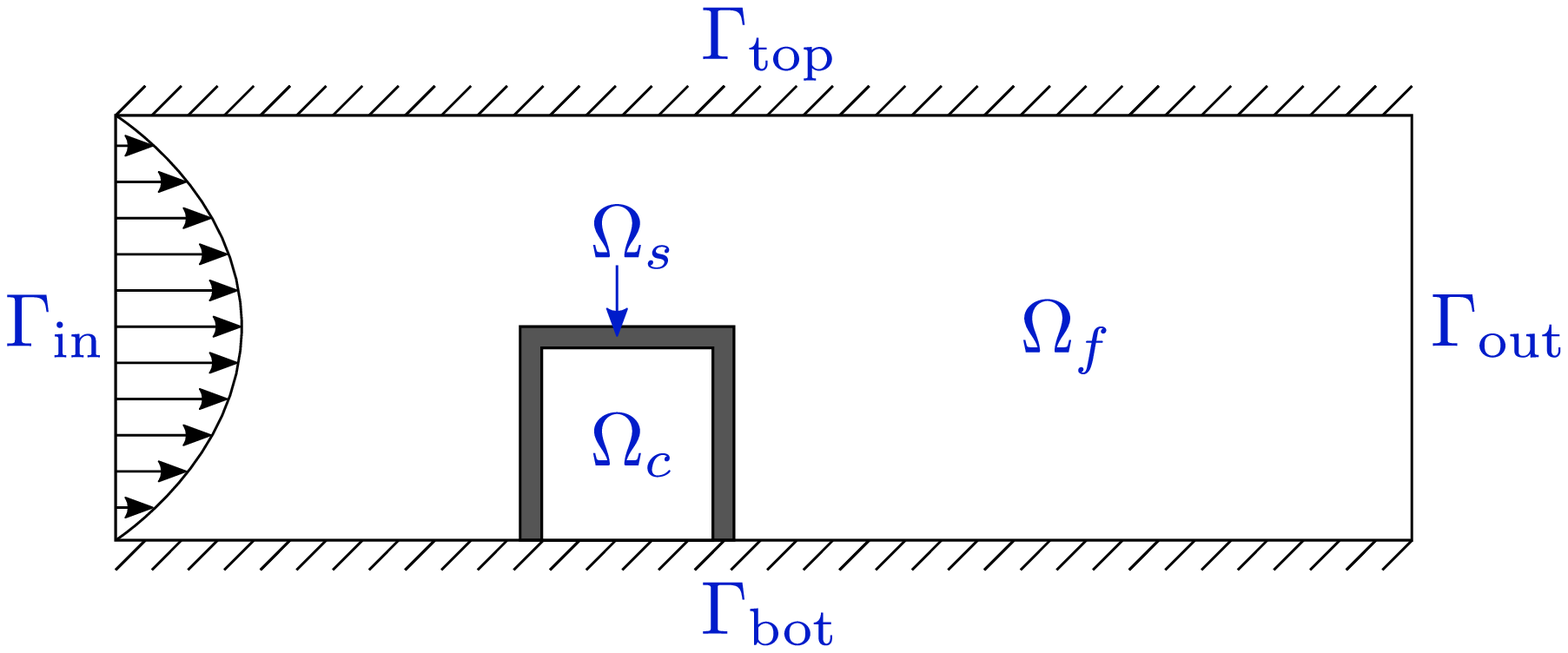}
  \caption{Domain and boundary definitions}
  \label{fig:headlesshorse-bcs}
\end{subfigure}
\\
\begin{subfigure}{\columnwidth}
  \centering
  \includegraphics[width=\linewidth]{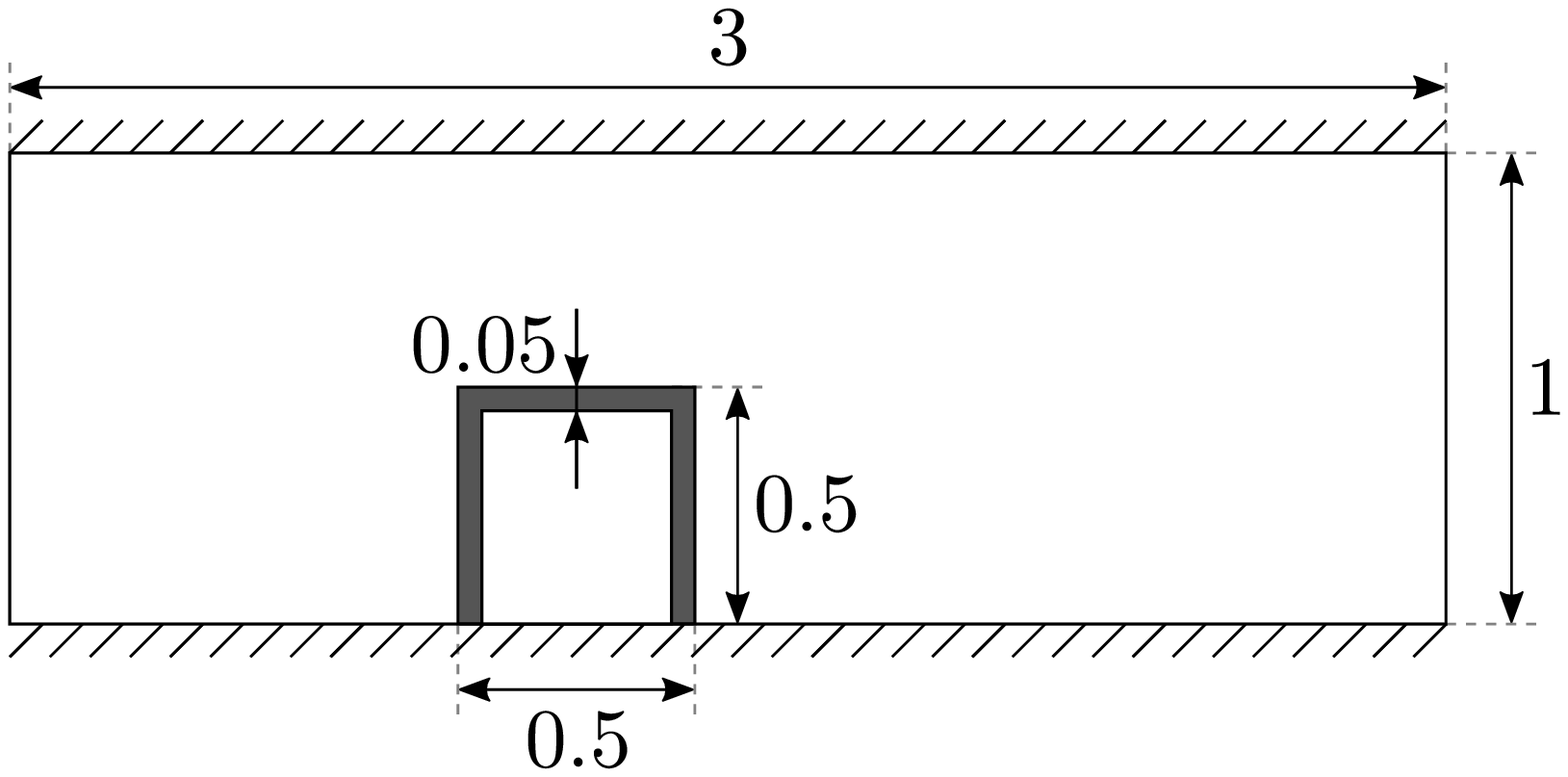}
  \caption{Dimensionless lengths}
  \label{fig:headlesshorse-dims}
\end{subfigure}
\caption{Problem setup for the example used in Section \ref{sec:varparstudy}.}
\label{fig:headlesshorse}
\end{figure}
Figure \ref{fig:headlesshorse} presents the problem setup used to explore the accuracy of the Brinkman-penalised Navier-Stokes equations. A hollow box obstacle is placed inside a channel, where a fully-developed parabolic flow with a maximum of $U_\textrm{in}$ enters at the left-hand side inlet and exits at the right-hand side zero normal-stress outlet. This model will be solved using both a discrete design representation, modelling only the fluid domain, and a continuous design representation, with the Brinkman-penalised Navier-Stokes equations in the entire computational domain. 
The same locally-refined mesh is used for both models to ensure that the discretisation error does not affect the comparison in a significant sense.

The porous media definition in Equation \ref{eq:alphakappa} is chosen for the Brinkman penalty factor and the problem is non-dimensionalised by setting the inlet velocity to $U_\textrm{in} = 1$, the density to $\rho = 1$, the dynamic viscosity to $\mu = \sfrac{1}{Re}$, and the permeability to $\kappa = Da$. The Darcy number, $Da = \frac{\kappa}{L^2}$, is the non-dimensional permeability. The Reynolds and Darcy numbers will be varied.

\subsubsection{Fluid solution}
\begin{figure*}
\begin{subfigure}{\columnwidth}
  \centering
  \includegraphics[width=\linewidth]{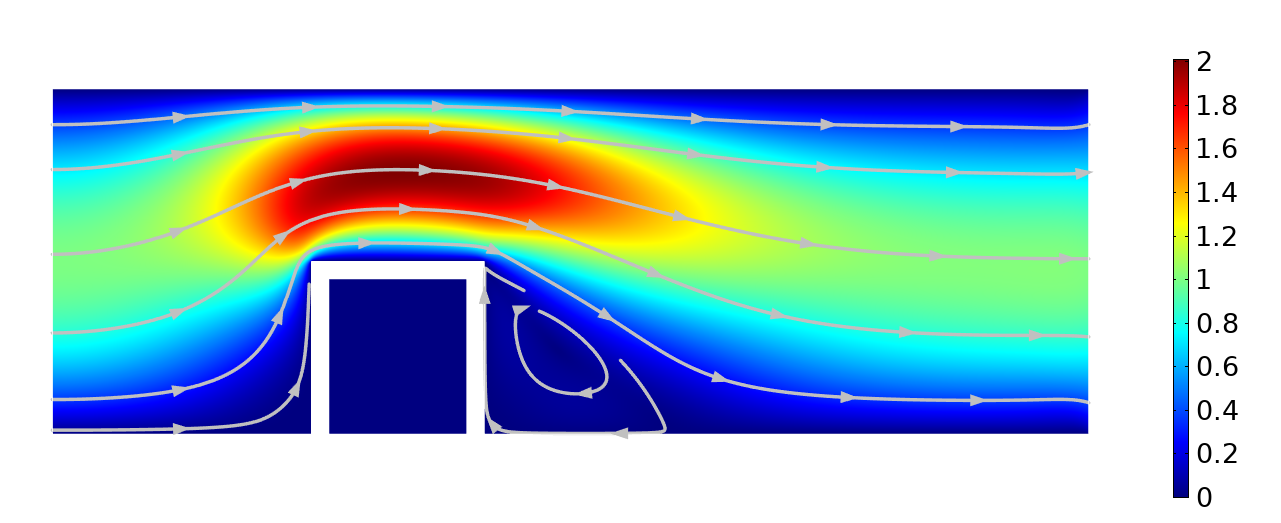}
  \caption{Velocity magnitude}
  \label{fig:headless_sepdom-vel}
\end{subfigure}
\hfill
\begin{subfigure}{\columnwidth}
  \centering
  \includegraphics[width=\linewidth]{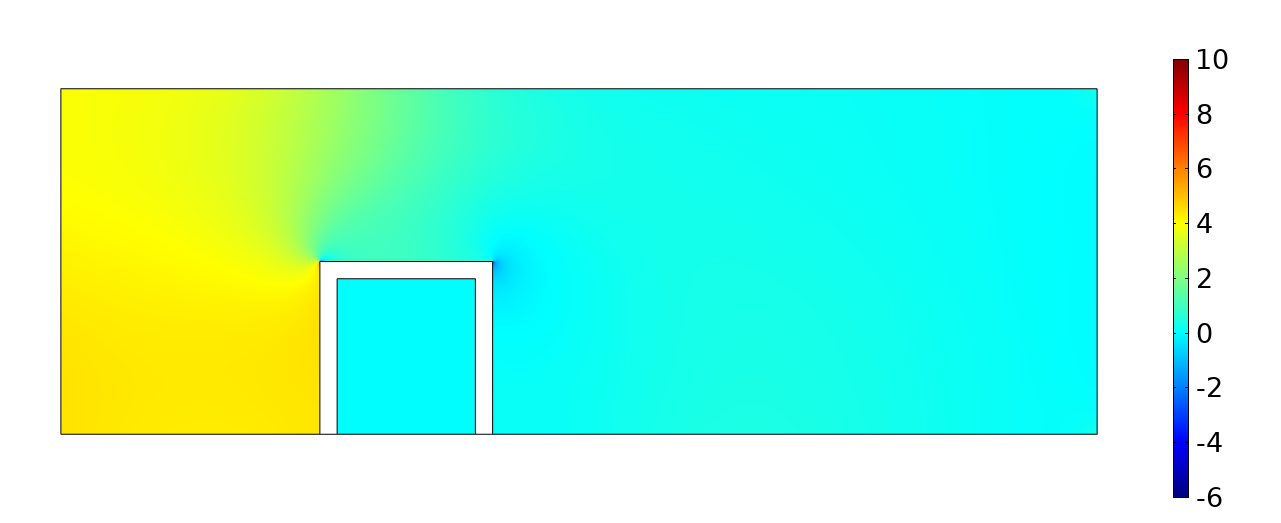}
  \caption{Pressure field}
  \label{fig:headless_sepdom-pres}
\end{subfigure}
\caption{Velocity magnitude and pressure fields using the discrete design representation for $Re=20$.}
\label{fig:headless_sepdom}
\end{figure*}
Figure \ref{fig:headless_sepdom} shows the velocity magnitude and pressure fields using the discrete design representation for $Re=20$.
Since the fluid cavity inside the solid structure is disconnected, the pressure is here set equal to a reference pressure of 0. It can be seen that the flow moves above the obstacle and leaves a re-circulation zone behind it.

\begin{figure}
\begin{subfigure}{\columnwidth}
  \centering
  \includegraphics[width=\linewidth]{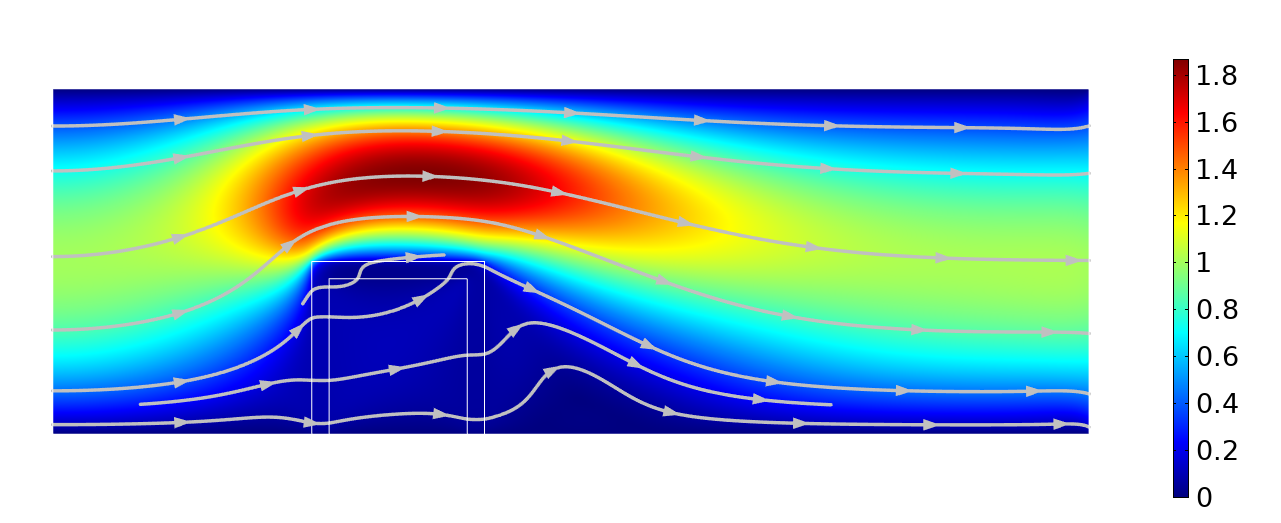}
  \caption{$Da=10^{-4}$}
  \label{fig:headless_brinkdom_vel-Da1em4}
\end{subfigure}
\\
\begin{subfigure}{\columnwidth}
  \centering
  \includegraphics[width=\linewidth]{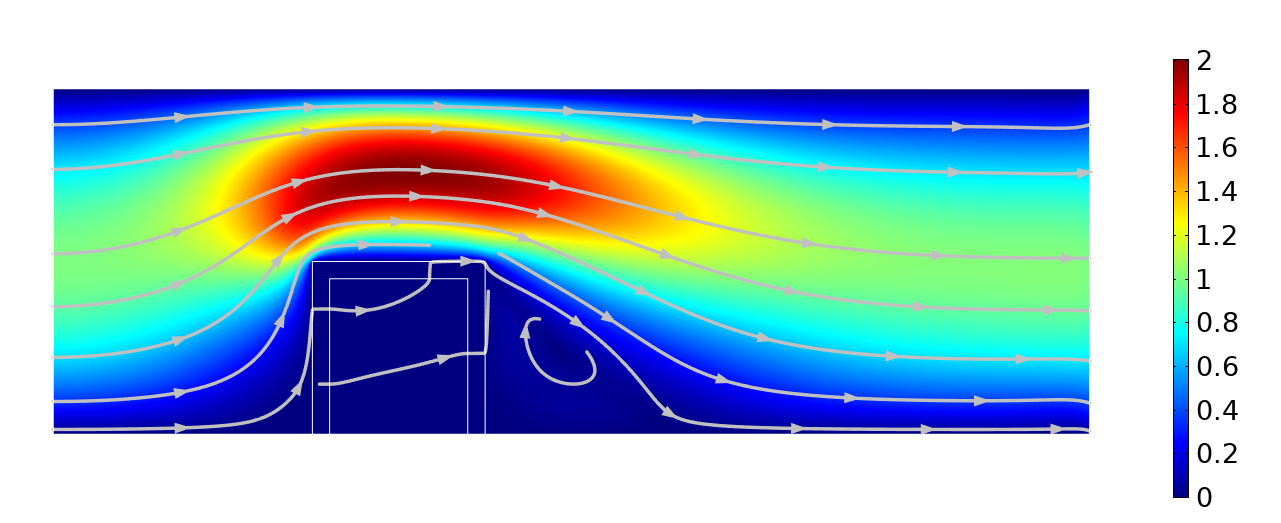}
  \caption{$Da=10^{-6}$}
  \label{fig:headless_brinkdom_vel-Da1em6}
\end{subfigure}
\\
\begin{subfigure}{\columnwidth}
  \centering
  \includegraphics[width=\linewidth]{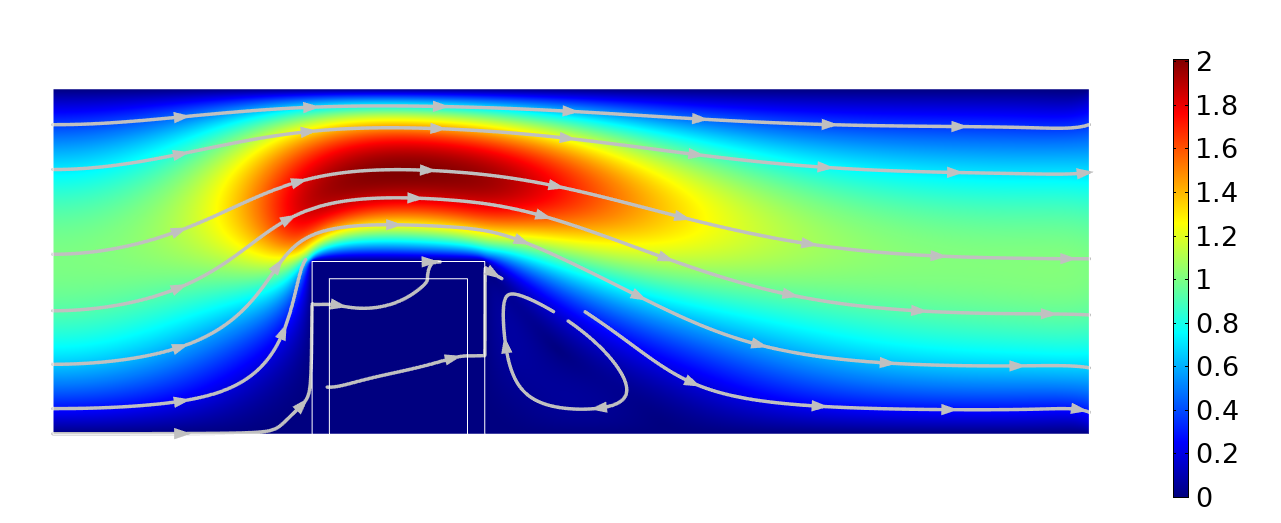}
  \caption{$Da=10^{-8}$}
  \label{fig:headless_brinkdom_vel-Da1em8}
\end{subfigure}
\caption{Velocity magnitude fields using the Brinkman approach for $Re=20$ and $Da\in\{10^{-4},10^{-6},10^{-8}\}$.}
\label{fig:headless_brinkdom_vel}
\end{figure}
Figure \ref{fig:headless_brinkdom_vel} shows the velocity magnitude field using the Brinkman approach for $Re=20$ and $Da\in\{10^{-4},\allowbreak 10^{-6},\allowbreak 10^{-8}\}$.
For the highest permeability case, $Da = 10^{-4}$, it is seen that a significant amount of fluid passes through the solid structure and clearly non-zero velocities exist in the cavity. When this is increased to $Da = 10^{-6}$, the flow solution becomes more accurate, but the recirculation zone behind the obstacle is not fully captured. This is improved for the last case, $Da = 10^{-8}$, but even here flow is passing through the obstacle. This will always be the case for the Brinkman penalty approach, but the velocities can be driven towards zero by decreasing the permeability sufficiently. 

\begin{figure}
\begin{subfigure}{\columnwidth}
  \centering
  \includegraphics[width=\linewidth]{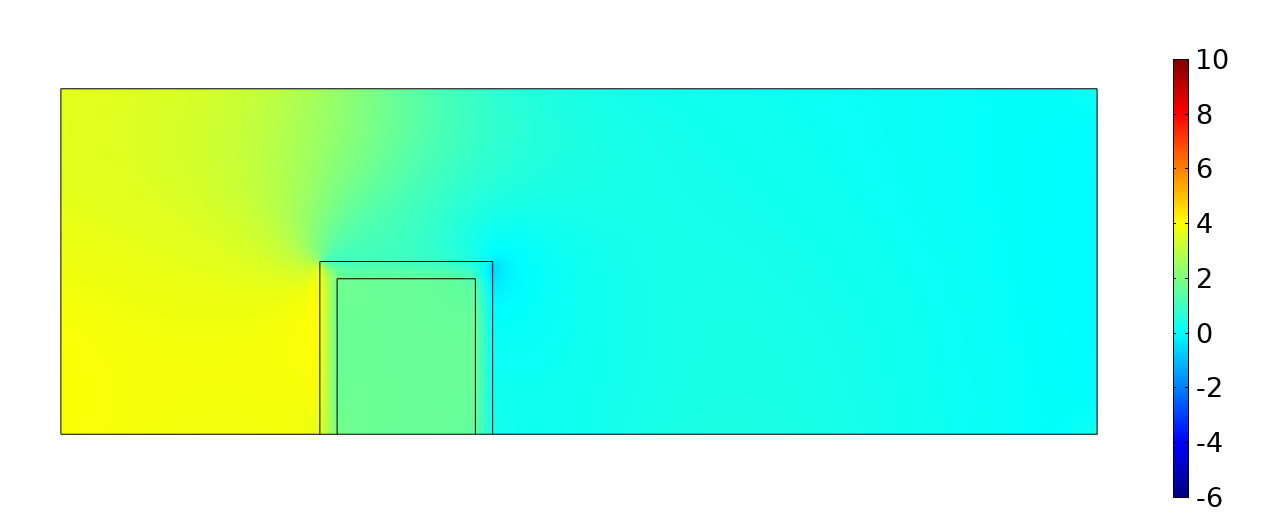}
  \caption{$Da=10^{-4}$}
  \label{fig:headless_brinkdom_pres-Da1em4}
\end{subfigure}
\\
\begin{subfigure}{\columnwidth}
  \centering
  \includegraphics[width=\linewidth]{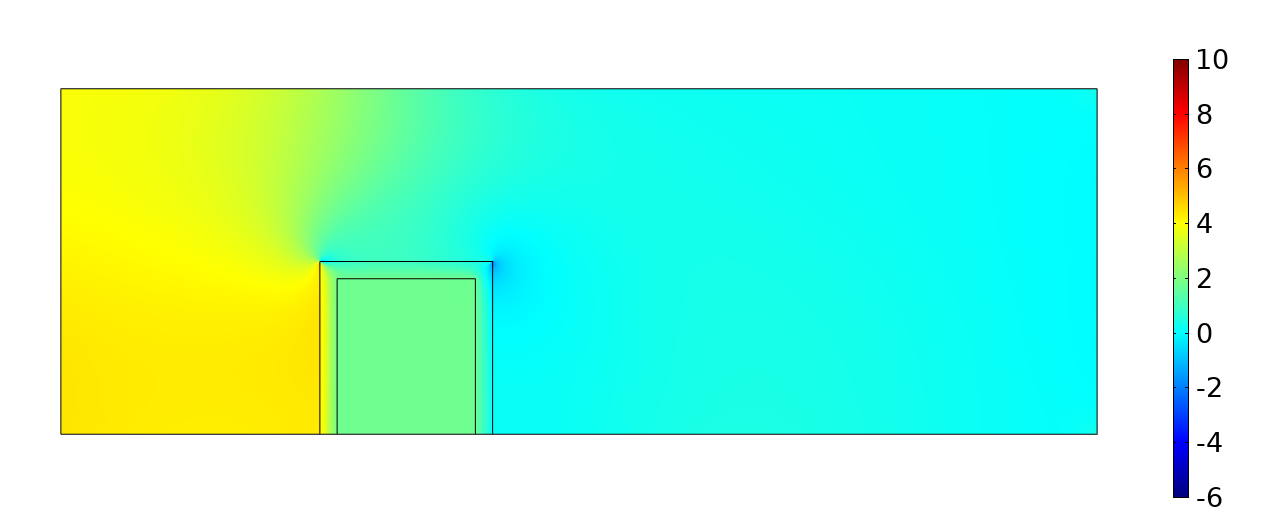}
  \caption{$Da=10^{-6}$}
  \label{fig:headless_brinkdom_pres-Da1em6}
\end{subfigure}
\\
\begin{subfigure}{\columnwidth}
  \centering
  \includegraphics[width=\linewidth]{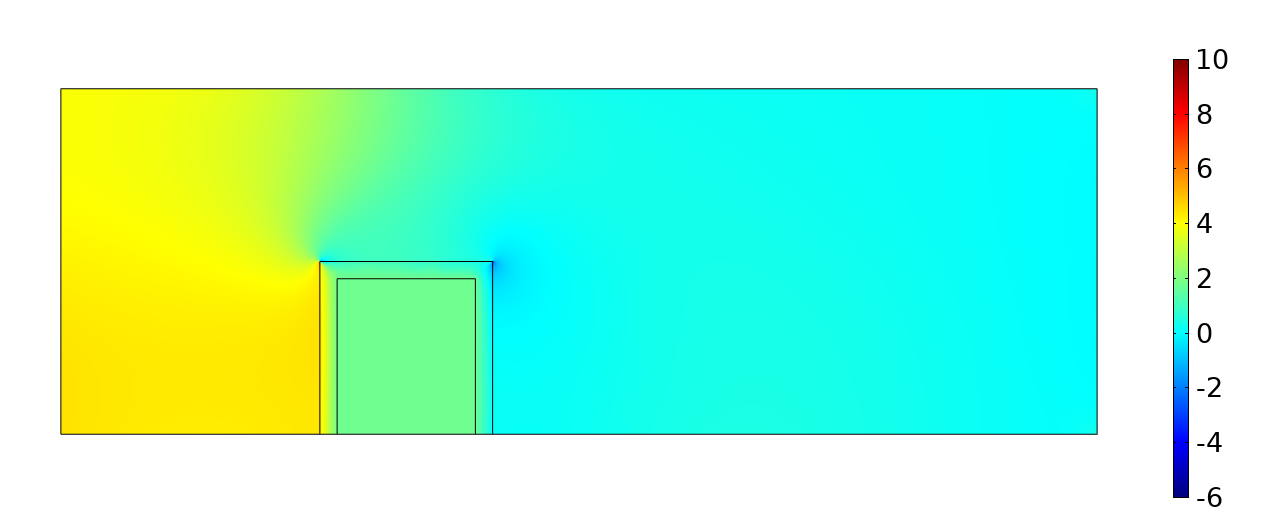}
  \caption{$Da=10^{-8}$}
  \label{fig:headless_brinkdom_pres-Da1em8}
\end{subfigure}
\caption{Pressure fields using the Brinkman approach for $Re=20$ and $Da\in\{10^{-4},10^{-6},10^{-8}\}$.}
\label{fig:headless_brinkdom_pres}
\end{figure}
Figure \ref{fig:headless_brinkdom_pres} shows the pressure field using the Brinkman approach for $Re = 20$ and $Da\in\{10^{-4},\allowbreak 10^{-6},\allowbreak 10^{-8}\}$. Due to the continuous nature of the design representation, pressure gradients exist inside the solid structure and a non-zero pressure exists in the cavity. The cavity pressure approximately becomes the mean of the high and low surface pressures - for this case $\approx (4 + 0)/2 = 2$.
For fluid dissipation problems, this is not really an issue because this pressure is not important for the overall dissipation and since disconnected fluid regions never occur in the optimised designs. But for fluid-structure-interaction, where the fluid exerts a pressure on the solid, the internal non-zero pressure can be an issue \citep{Lundgaard2018}.

\subsubsection{Error convergence study}

To clearly show the convergence of the approximate solution from the continuous representation towards the true solution from the discrete representation, a number of error measures are computed. The spatial average of the velocity magnitude inside the solid structure and inside the cavity are computed as:
\begin{subequations}
\begin{equation}
    \varepsilon_{u,\mathrm{s}} = \frac{1}{| \Omega_s |} \int_{\Omega_s} \| \mathbf{u} \|_{2} \,dV
\end{equation}
\begin{equation}
    \varepsilon_{u,\mathrm{c}} = \frac{1}{| \Omega_c |} \int_{\Omega_c} \| \mathbf{u} \|_{2} \,dV
\end{equation}
\end{subequations}
These are seen as directly equivalent to a relative error measure, since the true solution for these quantities are zero and the reference (inlet) value is 1.
Furthermore, the relative error is found for the dissipated energy and the pressure drop:
\begin{subequations}
\begin{equation}
    \varepsilon_{\phi} = \frac{| \phi_{\mathrm{disc}} - \phi_{\mathrm{cont}} |}{\phi_{\mathrm{disc}}}
\end{equation}
\begin{equation}
    \varepsilon_{\Delta p} = \frac{| {\Delta p}_{\mathrm{disc}} - {\Delta p}_{\mathrm{cont}} |}{{\Delta p}_{\mathrm{disc}}}
\end{equation}
\end{subequations}
where subscripts `disc' and `cont' denote discrete and continuous design representations, respectively, $\phi$ is the dissipated energy of the fluid subdomain defined in Equation \ref{eq:disenergy_visc} and $\Delta p$ is the pressure drop from inlet to outlet as computed from:
\begin{equation}
    \Delta p = \frac{1}{| \Gamma_\mathrm{in} |} \int_{\Gamma_\mathrm{in}} p \,dS - \frac{1}{| \Gamma_\mathrm{out} |} \int_{\Gamma_\mathrm{out}} p \,dS
\end{equation}

\begin{figure}
    \centering
    \includegraphics[width=0.95\linewidth]{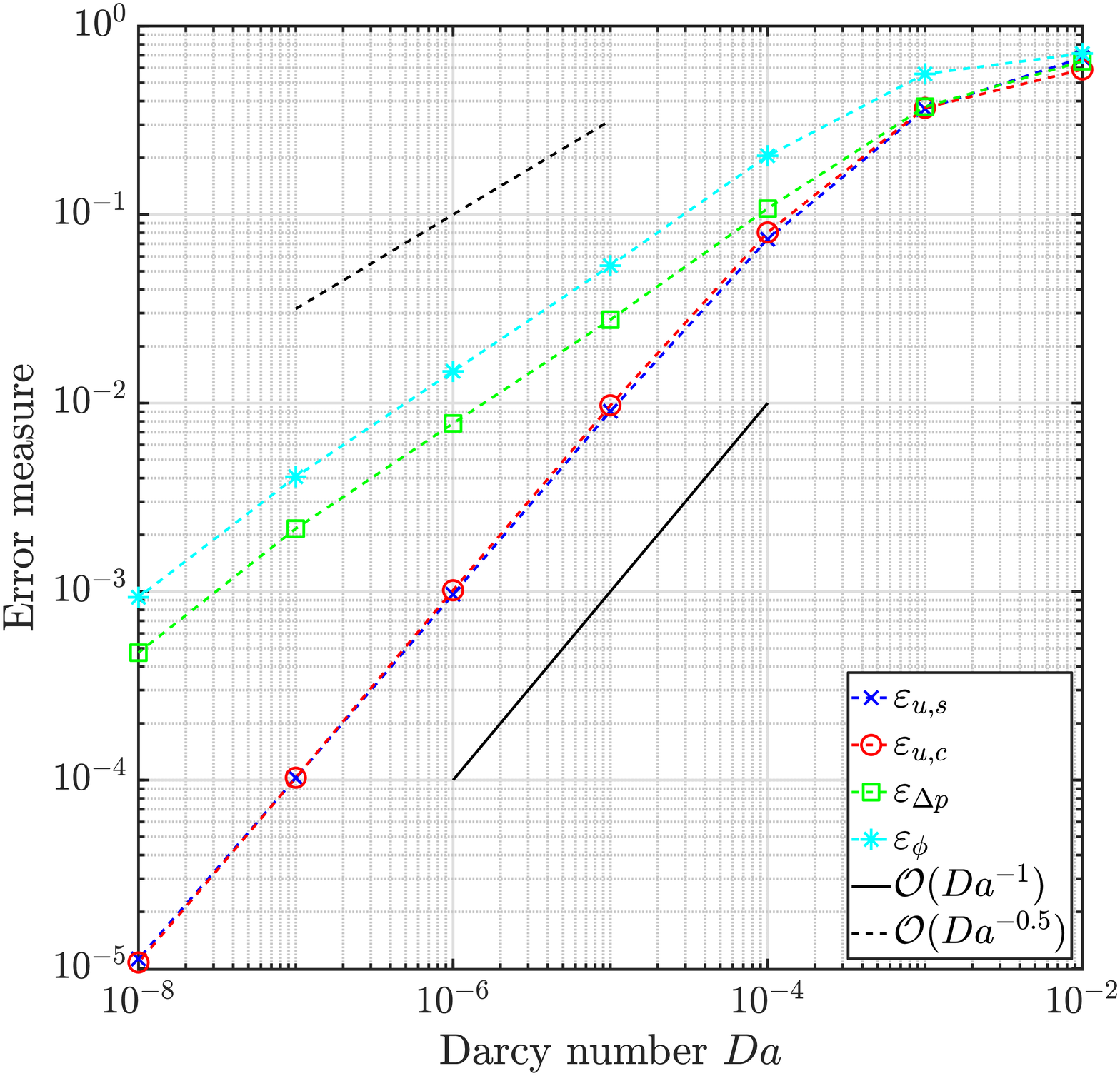}
    \caption{Error measures as a function of permeability for a Reynolds number of $Re=20$ and
    $Da \in \{ 10^{-2},10^{-3},10^{-4},10^{-5},10^{-6},10^{-7},10^{-8} \}$
    .}
    \label{fig:headless_errorVkappa}
\end{figure}
Figure \ref{fig:headless_errorVkappa} shows the error measures as a function of the Darcy number for $Re=20$. It is clearly seen that the velocity error measures converge at an order of 1 and the pressure and dissipation measures at an order of \sfrac{1}{2}. This shows that compared to the velocity field, the pressure field needs a significantly higher Brinkman penalty factor to reach a certain accuracy. The dissipated energy also requires a larger Brinkman penalty factor to achieve a certain accuracy. However, this is not necessarily important for minimum dissipated energy optimisation as will be demonstrated in Section \ref{sec:twopipe_amax}. Additional accuracy becomes important if these functionals are used as constraints where the limit represents a physical value. The constraint looses its physicality, if the functional is not resolved to a large enough accuracy\footnote{This is very similar to stress constraints in structural topology optimisation, where a maximum stress limit becomes meaningless, unless the stresses and aggregated maximum is accurate \cite{Le2010,Silva2019}.}. However, if treated in relative terms as in Section \ref{sec:flowreversal}, then this is less of an issue. The accuracy of the pressure field is also very important if it is used to drive additional physics, such as fluid-structure interaction (FSI) which was clearly shown by \citet{Lundgaard2018}.

\begin{figure}
    \centering
    \includegraphics[width=0.95\linewidth]{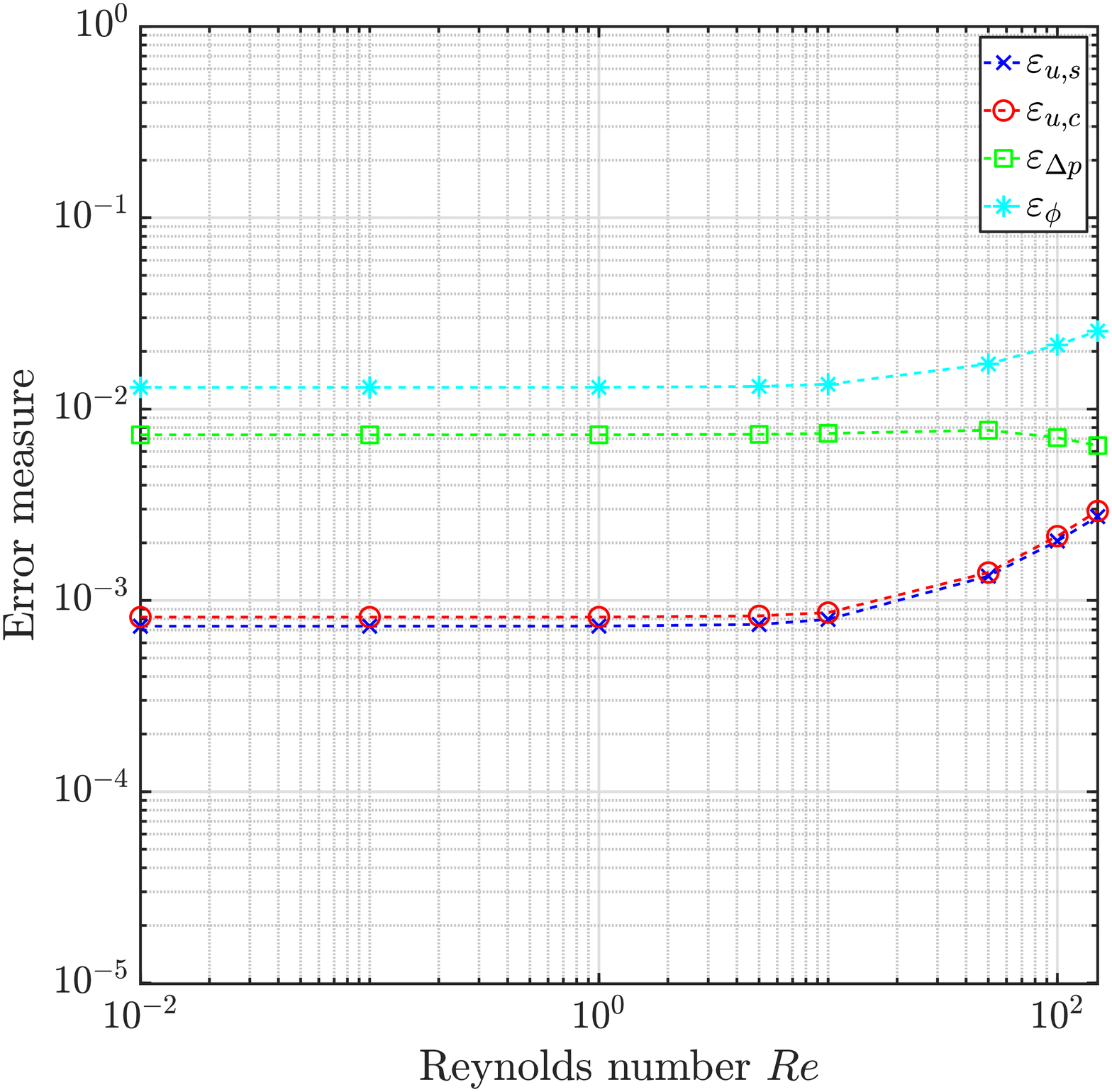}
    \caption{Error measures as a function of Reynolds number for a Darcy number of $Da=10^{-6}$ and $Re\in \{ 0.01,0.1,1,5,10,50,100,150 \}$.}
    \label{fig:headless_errorVrenum}
\end{figure}
Figure \ref{fig:headless_errorVrenum} shows the error measures as a function of the Reynolds number for $Da=10^{-6}$. It can be seen that defining the Brinkman penalty factor using Equation \ref{eq:alphakappa} ensures a constant error for a range of Reynolds numbers. However, as the Reynolds number becomes larger than 10, the error begins to slowly grow. This makes sense, since the porous media analogy is based on scaling with the viscous diffusion term and a Brinkman porous media model. As the Reynolds number increases, the convective term begins to dominate in Equation \ref{eq:navierstokes_brink} and, thus, it makes sense that the penalty term should scale differently. However, the slight increase in the errors are not seen as significant for the moderate Reynolds numbers treatable with the present steady-state implementation. An alternative scaling law proposed by \citet{Kondoh2012} is discussed in Appendix \ref{app:kondoh}.

\section{Discretisation and solution} \label{sec:discrandsol}

\subsection{Finite element formulation} \label{sec:fea}
The governing equations in Equation \ref{eq:navierstokes} are discretised using a stabilised continuous Galerkin finite element formulation.
\begin{figure}
    \centering
    \includegraphics[width=0.75\columnwidth]{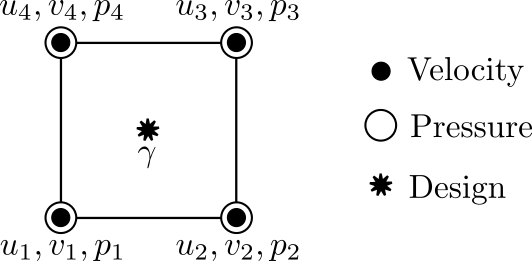}
    \caption{Element degrees-of-freedom (DOFs) and nodal numbering.}
    \label{fig:element}
\end{figure}
Both the velocity and pressure fields are approximated using piecewise bi-linear interpolation functions and the design field is approximated using piecewise constant values, rendering a \textit{u1p1g0} element as shown in Figure \ref{fig:element}. This yields 8 velocity degrees-of-freedom (DOFs), 4 pressure DOFs, and 1 design DOF per element. Due to the equal-order interpolation for velocity and pressure, Pressure-Stabilising Petrov-Galerkin (PSPG) stabilisation is applied to get a stable finite element formulation. Furthermore, to avoid spurious oscillations due to convection, Streamline Upwind Petrov-Galerkin (SUPG) stabilisation is also applied. The detailed formulation is given in Appendix \ref{app:FEM}.

The discretisation gives rise to a non-linear system of equations, which is posed in residual form:
\begin{equation} \label{eq:residual}
\mathbf{r}\hspace*{-1pt}\left( \mathbf{s}, \bm{\gamma} \right) = \mathbf{A}\hspace*{-2pt}\left( \mathbf{s}, \bm{\gamma} \right) \mathbf{s} = \mathbf{0}
\end{equation}
where $\mathbf{A}$ is the system coefficient matrix, $\mathbf{s} = \left[ \mathbf{u}^{T} \mathbf{p}^{T} \right]^{T}$ is the vector of global state DOFs, $\mathbf{u}$ is the vector of velocity DOFs, $\mathbf{p}$ is the vector of pressure DOFs, and $\bm{\gamma}$ is the vector of design field values. There are many non-linear and design-dependent terms in the system: the convection term is dependent on the velocity field; the Brinkman term is dependent on the design field; and the stabilisation terms are dependent on both the velocity and design field (either directly through convection and Brinkman terms or indirectly through the stabilisation parameter). Thus, a robust non-linear solver is needed and the Newton solver is detailed in the subsequent section. Furthermore, derivation of accurate partial derivatives, such as the Jacobian matrix and those necessary for optimisation, can be cumbersome. Therefore, these derivatives will be automatically computed using the Symbolic Toolbox in MATLAB.

\subsection{Newton solver} \label{sec:newton}

To solve the non-linear system of equations, a damped Newton method is used. At each non-linear iteration, $k$, the following linearised problem is solved for the Newton step, $\bm{\Delta}\mathbf{s}_{k}$:
\begin{equation}
    \left. \frac{\partial \mathbf{r}}{\partial \mathbf{s}}\right|_{k-1} \bm{\Delta}\mathbf{s}_{k} = - \mathbf{r}(\mathbf{s}_{k-1})
\end{equation}
where the solution is then updated according to:
\begin{equation}
    \mathbf{s}_{k} = \mathbf{s}_{k-1} + \delta_{k} \bm{\Delta}\mathbf{s}_{k}
\end{equation}
with $\delta_{k} \in \left[ 0.01; 1 \right]$ being the damping factor\footnote{The lower bound is chosen to avoid stagnation.}. In order to gain robust convergence for moderate Reynolds numbers, the damping factor is updated according to the minimum of a quadratic fit of the residual norm dependence on $\delta$. The residual and its 2-norm is calculated for $\delta \in \left\{0, 0.5, 1 \right\}$, with the current value ($\delta=0$) already known. A quadratic equation is then fitted to the residual norms with the unique minimising damping factor expressed analytically. The damping factor is then set to:
\begin{equation}
    \delta_{k} = \textrm{max}\left( 0.01, \textrm{min}\left( 1, \frac{3 ||\mathbf{r}_{0}||_{2} - 4||\mathbf{r}_{0.5}||_{2} + ||\mathbf{r}_{1}||_{2}}{4||\mathbf{r}_{0}||_{2} - 8||\mathbf{r}_{0.5}||_{2} + 4||{\mathbf{r}}_{1}||_{2}} \right) \right)
\end{equation}
where $\mathbf{r}_{\delta}$ is the residual evaluated for the solution updated with a given $\delta$.

The Newton solver is run until the current residual norm has been reduced relative to the initial below a predefined threshold. To speed up convergence, the state solution from the previous design iteration is used as the initial solution for the Newton solver. If the solver does not converge before a maximum iteration number is reached, it tries again from a zero (except for boundary conditions) initial solution. If that also fails, the Reynolds number may be too large to have a steady-state solution for the problem. It is also possible that the problem is so non-linear, that it might need ramping of the inlet velocity or Reynolds number.

\section{Optimisation formulation} \label{sec:optimisation}

\subsection{Objective functional} \label{sec:objective}

The dissipated energy was originally introduced as the objective functional by \citet{Borrvall2003}. Equation \ref{eq:disenergy_visc} is the dissipated energy in the fluid domain due to viscous resistance only. The addition of the Brinkman penalty term introduces a body force, which must be taken into account in the dissipated energy:
\begin{equation} \label{eq:disenergy_strong}
    \phi =  \frac{1}{2} \int_{\Omega} \left( \mu \frac{\partial u_i}{\partial x_j} \left(  \frac{\partial u_i}{\partial x_j} +  \frac{\partial u_j}{\partial x_i} \right) +\alpha(\mathbf{x})u_{i}u_{i} \right) \,dV
\end{equation}
which is now integrated over the entire computational domain.

\subsection{Volume constraint} \label{sec:volume}

In order to avoid trivial solutions, a maximum allowable volume of fluid is imposed as a constraint. The relative volume of fluid is in continuous form defined as:
\begin{equation}
    V =  \frac{1}{| \Omega |}\int_{\Omega} \gamma(\mathbf{x}) \,dV
\end{equation}
which represents the volumetric average of the design field.
The relative fluid volume should be restricted to be below $V_{f} \in ]0;1[$, which is the fraction of the total domain allowed to be fluid.
After discretisation as described in Section \ref{sec:discrandsol}, the relative volume is computed by:
\begin{equation}
    V = \frac{1}{n_{el}} \sum_{e=1}^{n_{el}} \gamma_{e}
\end{equation}
where $n_{el}$ is the number of elements and it has been exploited that all elements have the same volume in the current implementation.

\subsection{Optimisation problem}

The final discretised optimisation problem is:
\begin{equation}\label{eq:optprob}
\begin{split}
 \underset{\bm{\gamma}}{\text{minimise:}} &\quad f_{\phi} \left(\textbf{s}(\bm{\gamma}),\bm{\gamma}\right) = \phi \\
\text{subject to:} & \quad V(\bm{\gamma}) \leq V_{f}\\
\text{with:}& \quad \textbf{r}(\textbf{s}(\bm{\gamma}),\bm{\gamma})= \mathbf{0}\\
& \quad 0 \leq \gamma_{i} \leq 1, \; i=1,\ldots,n_{el}
\end{split}
\end{equation}

\subsection{Adjoint sensitivity analysis} \label{sec:adjoint}

The gradients of any functionals dependent on the state field are found using adjoint sensitivity analysis. A general description is given in Appendix \ref{app:adjoint}. The sensitivities for a generic functional $f$ are found from:
\begin{equation} \label{eq:sensitivities}
\frac{d f}{d\gamma_{e}} = \frac{\partial f}{\partial\gamma_{e}} - \bm{\lambda}^T \frac{\partial\mathbf{r}}{\partial\gamma_{e}}
\end{equation}
where $\bm{\lambda}$ is the vector of adjoint variables for the state residual constraint found from the adjoint problem:
\begin{equation} \label{eq:adjointproblem}
\frac{\partial \mathbf{r}}{\partial\mathbf{s}}^{T} \bm{\lambda} = \frac{\partial f}{\partial \mathbf{s}}^{T} 
\end{equation}
The above is valid for any equation system formulated as a residual, $\mathbf{r}$, and any state-dependent functional, $f$, be it objectives or constraints. For a given system, two partial derivatives of the residual are necessary, and for a given functional, two partial derivatives of the functional are necessary: with respect to the design field and the state field. These partial derivatives can be found analytically by hand, but in the presented implementation they are found automatically using symbolic differentiation in MATLAB.

\subsection{Optimality criteria optimiser} \label{sec:optcrit}
An optimality criteria (OC) based optimisation algorithm is used in the presented code. The implemented OC method is the modified version suggested by \citet{Bendsoee2004} for compliant mechanism problems, where the design gradients can be both positive and negative:
\begin{equation} \label{eq:OC}
    \gamma^{new}_{e} = \left\{
    \begin{array}{lll}
    \gamma^{L}_{e} & & \textrm{if}\;\; \gamma_{e}{B_{e}}^{\eta} \leq \gamma^{L}_{e} \\
    \gamma^{U}_{e} & & \textrm{if}\;\; \gamma_{e}{B_{e}}^{\eta} \geq \gamma^{U}_{e} \\
    \gamma_{e}{B_{e}}^{\eta} & & \rm{otherwise}
    \end{array}
    \right.
\end{equation}
where $\eta = \frac{1}{3}$ is a damping factor and the upper bounds at each iteration are given as:
\begin{subequations}
\begin{equation}
    \gamma^{L}_{e} = \textrm{max}(0,\gamma_{e}-m)
\end{equation}
\begin{equation}
    \gamma^{U}_{e} = \textrm{min}(1,\gamma_{e}+m)
\end{equation}
\end{subequations}
with $m$ being the movelimit.
The optimality criteria ratio is given as: 
\begin{equation} \label{eq:OCratio}
    B_{e} = \textrm{max}\left( \varepsilon,\frac{-\frac{\partial \phi}{\partial \gamma_{e}}}{l \frac{\partial \mathcal{V}}{\partial \gamma_{e}}} \right)
\end{equation}
where $\varepsilon$ is a small positive number (e.g. $10^{-10}$) and $l$ is the Lagrange multiplier for the volume constraint. This means that positive gradients (less fluid is better) are ignored and the design process is driven only by the negative gradients (more fluid is better). However, this has worked extremely well for the problems at hand, because mostly the dissipated energy gradients are negative.

The Lagrange multiplier is found using bisection. To speed up the process, the upper bound estimate suggested by \citet{Ferrari2020} is modified for $\eta = \frac{1}{3}$:
\begin{equation} \label{eq:OCupper}
    \bar{l} = \left( \frac{1}{n_{el}V_{f}} \sum_{e=1}^{n_e} \gamma_{e} \left( -\frac{\frac{\partial \phi}{\partial \gamma_{e}}}{\frac{\partial V}{\partial \gamma_{e}}} \right)^{\frac{1}{3}} \right)^{3}
\end{equation}

\subsection{Continuation scheme} \label{sec:contin}
As shown by \citet{Borrvall2003}, the minimisation of dissipated energy is self-penalising for the Brinkman approach with linear interpolation of $\alpha(\gamma)$. This means that intermediate design values are not beneficial and the optimal solutions are discrete 0-1. However, it is often beneficial to relax the problem using the interpolation factor, $q_{\alpha}$, in the initial stages of the optimisation process in order to avoid convergence to particularly poor local minima. Therefore, it is common to solve the minimisation problem for a sequence of $q_{\alpha}$, known as a continuation scheme.

In the presented code, a heuristic approach is implemented to automatically choose the interpolation factor. The sequence for the continuation scheme is given by:
\begin{equation}
    q_{\alpha} \in \left\{ q^{0}_{\alpha}, \frac{q^{0}_{\alpha}}{2}, \frac{q^{0}_{\alpha}}{10}, \frac{q^{0}_{\alpha}}{20} \right\}
\end{equation}
where:
\begin{equation} \label{eq:qinit}
    q^{0}_{\alpha} = \frac{ \left(\alpha_\textrm{max}-\alpha_{0}\right) - x_{0}\left(\alpha_\textrm{max}-\alpha_\textrm{min}\right) }{x_{0}\left(\alpha_{0}-\alpha_\textrm{min}\right) }
\end{equation}
is the initial interpolation factor necessary to ensure an initial Brinkman penalty of $\alpha_{0}$ for an initial design field value of $x_{0}$. The interpolation factor is changed when the subproblem is considered converged or after a predefined number of design iterations.
For the problems implemented in the presented code,
an initial Brinkman penalty of $\alpha_{0}=\frac{2.5\mu}{0.1^2}$ works very well.
However, do bear in mind that this is a heuristic value and will not work for all problems and all settings.

Due to the non-convexity, the initial Brinkman penalty of the domain plays an important role in which direction the design will progress. The initial flow field determines the exploration of the design space, because if the flow is not moving through some parts of the design domain, sensitivities will often be near zero in those areas. Therefore, it is often beneficial to start with an initial Brinkman penalty that allows for the fluid to move into most areas of the domain. This was for instance recently observed in the topology optimisation of heat exchanger designs by \citet{Hoeghoej2020}.

\section{Benchmark examples} \label{sec:incex}

\subsection{Double pipe problem} \label{sec:incex1}

\begin{figure}
    \centering
    \includegraphics[width=\columnwidth]{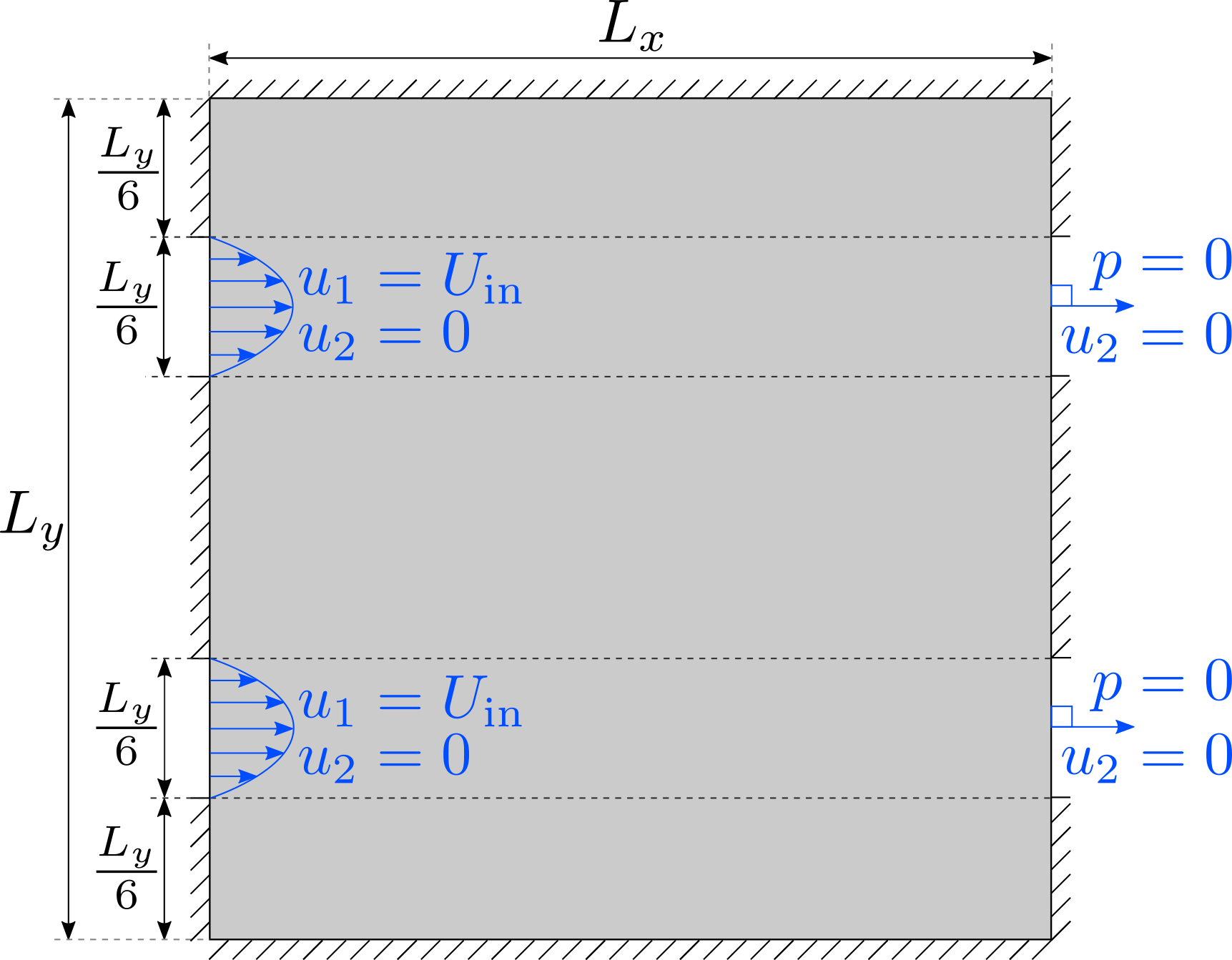}
    \caption{Problem setup for the double pipe problem.}
    \label{fig:example1_bcs}
\end{figure}
This problem was first introduced by \citet{Borrvall2003} and is illustrated in Figure \ref{fig:example1_bcs}. On the left-hand side, there are two inlets at which parabolic normal velocity profiles are prescribed with a maximum velocity of $U_{in}$. On the right-hand side, there are two zero-pressure outlets at which the flow is specified to exit in the normal direction. This is a more realistic boundary condition traditionally used in the computational fluid dynamics community, compared to prescribed parabolic flow profiles at the outlets \citep{Borrvall2003}. However, this does introduce new and better minima as was shown by \cite{Papadopoulos2021}, which makes the problem more difficult.

\subsection{Pipe bend problem} \label{sec:incex2}

\begin{figure}
    \centering
    \includegraphics[width=0.9\columnwidth]{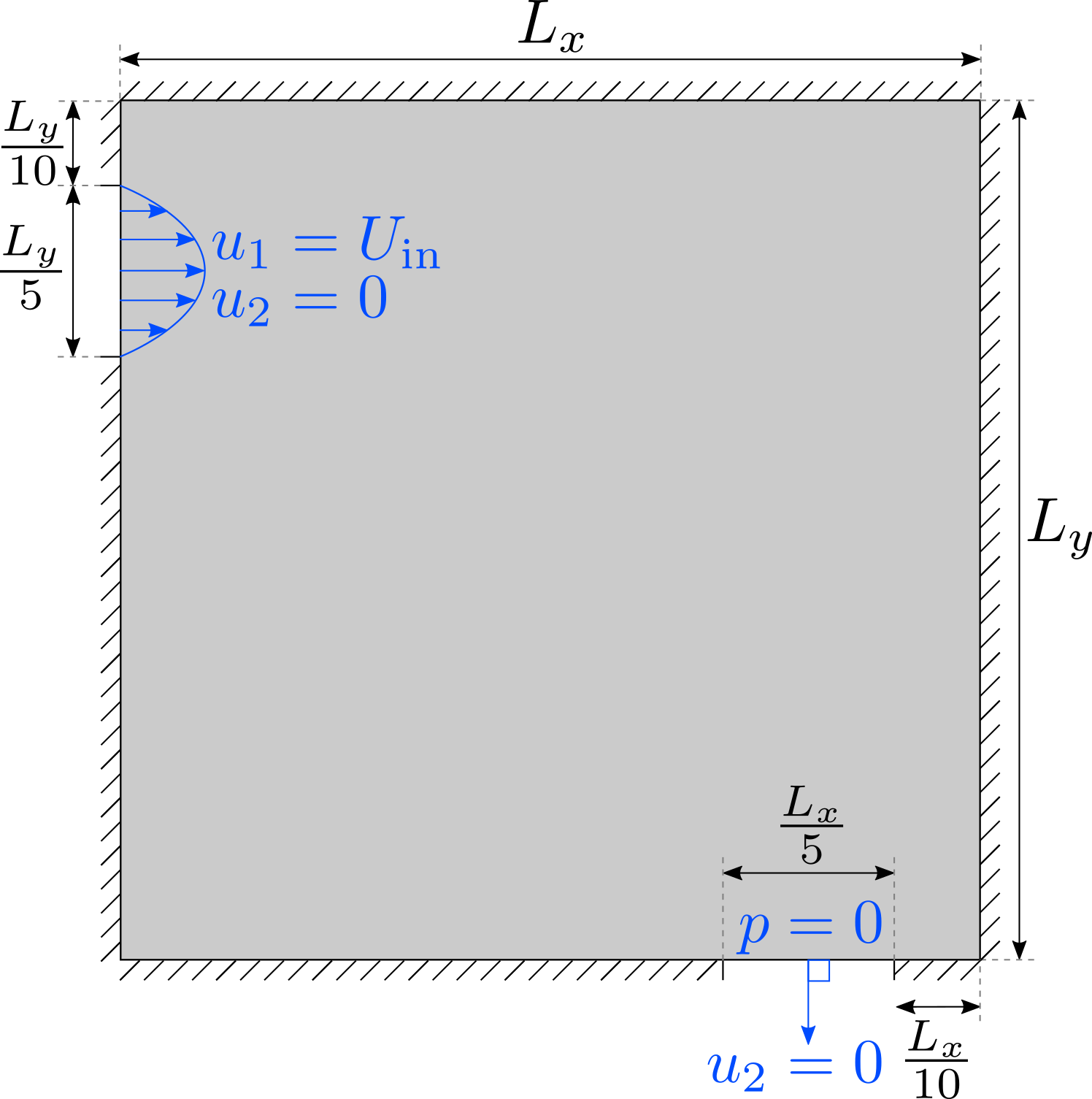}
    \caption{Problem setup for the pipe bend problem.}
    \label{fig:example2_bcs}
\end{figure}
This problem was also introduced by \citet{Borrvall2003} and is illustrated in Figure \ref{fig:example2_bcs}. On the left-hand side, there is an inlet at which a parabolic normal velocity profile is prescribed with a maximum velocity of $U_{in}$. On the bottom, there is a zero-pressure outlet at which the flow is specified to exit in the normal direction - a minor change from \citet{Borrvall2003} as for the double pipe problem.

\section{MATLAB implementation} \label{sec:code}

\subsection{Mesh and numbering} \label{sec:numbering}
\begin{figure}
    \centering
    \begin{subfigure}{\columnwidth}
    \centering
    \includegraphics[height=2.5cm]{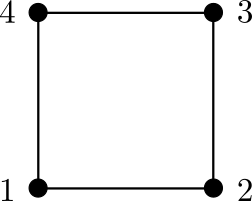}
    \caption{Local nodal numbering}
    \label{fig:numbering_element}
    \end{subfigure}
    \\
    \vspace{3ex}
    \begin{subfigure}{\columnwidth}
    \centering
    \includegraphics[height=2.5cm]{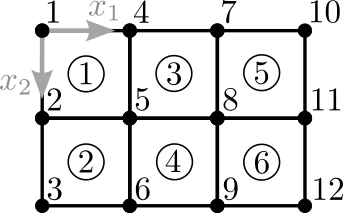}
    \caption{Global nodal and element numbering}
    \label{fig:numbering_global}
    \end{subfigure}
    \caption{Local and global numbering. Numbers without circles indicate nodal numbering and number with circles indicate element numbering.}
    \label{fig:numbering}
\end{figure}
Figure \ref{fig:numbering} shows the local and global numbering used in the presented implementation. As shown in Figure \ref{fig:numbering_global}, for the global numbering of both the nodes and elements begins in the upper-left corner, increasing from top to bottom and then from left to right. The origin of the global coordinate system is placed in the upper-left corner, with the first spatial direction, $x_1$, being positive from left to right and the second spatial direction, $x_2$, being positive from top to bottom. This orientation has been chosen to simplify easy plotting of the various spatial fields in MATLAB. The velocity and pressure DOFs follow the ordering of the nodes and the design field follows the ordering of the elements.

\subsection{Brief description of code}

The full code consists of multiple files, the main ones of which are available in Appendices \ref{app:topFlow}-\ref{app:analytical}. The full code base is available on GitHub \citep{GitHubLink2022}.

The main file \texttt{topFlow.m} (Appendix \ref{app:topFlow}) can be divided into three parts: pre-processing, optimisation loop, and post-processing. Each part will be very briefly described herein, except the definition of problem parameters which is treated in detail below since this will be of importance to the subsequent definition of examples and results. A detailed description of the code is available in the Supplementary Material.

\subsubsection{Definition of input parameter (lines 6-29)}

The pre-processing part starts with the setting of input parameters (lines 6-29) to define the problem and optimisation parameters.
In the presented code, two different problems are already implemented and the \texttt{probtype} variable determines which one is to be optimised:
\vspace{-12pt}
\begin{lstlisting}[firstnumber=7,firstline=7,lastline=8]
% PROBLEM TO SOLVE (1 = DOUBLE PIPE; 2 = PIPE BEND)
probtype = 1;
\end{lstlisting}
Next the dimensions of the domain and the discretisation are given:
\vspace{-12pt}
\begin{lstlisting}[firstnumber=9,firstline=9,lastline=12]
% DOMAIN SIZE
Lx = 1.0; Ly = 1.0;
% DOMAIN DISCRETISATION
nely = 30; nelx = nely*Lx/Ly;
\end{lstlisting}
where \texttt{Lx} and \texttt{Ly} are the dimensions in the \textit{x}- and \textit{y}-directions, respectively. The number of elements in the x-direction, \texttt{nelx}, is automatically determined from the supplied number of elements in the y-direction, \texttt{nely}, to ensure square elements. The implementation can handle rectangular elements, so \texttt{nelx} can be decoupled from \texttt{nely} if necessary.

The allowable volume fraction $V_{f}$ is stored in the \texttt{volfrac} variable and the initial design field value is set to be the same to fulfill the volume constraint from the start:
\vspace{-12pt}
\begin{lstlisting}[firstnumber=13,firstline=13, lastline=14,]
% ALLOWABLE FLUID VOLUME FRACTION
volfrac = 1/3; xinit = volfrac;
\end{lstlisting}
The inlet velocity $U_{in}$ is stored in the variable \texttt{Uin} and the fluid density, $\rho$ and  dynamic viscosity, $\mu$, are stored in the variables \texttt{rho} and \texttt{mu}, respectively:
\vspace{-12pt}
\begin{lstlisting}[firstnumber=15,firstline=15, lastline=16,numbers=right]
% PHYSICAL PARAMETERS
Uin = 1e0; rho = 1e0; mu = 1e0;
\end{lstlisting} 

The maximum and minimum Brinkman penalty factors are per default defined with respect to an out-of-plane thickness according to Equation \ref{eq:alphah} \citep{Borrvall2003} and the heuristic continuation scheme is automatically computed and stored in \texttt{qavec}:
\vspace{-12pt}
\begin{lstlisting}[firstnumber=17,firstline=17, lastline=22,numbers=right]
% BRINKMAN PENALISATION
alphamax = 2.5*mu/(0.01^2); alphamin = 2.5*mu/(100^2);
% CONTINUATION STRATEGY
ainit = 2.5*mu/(0.1^2);
qinit = (-xinit*(alphamax-alphamin) - ainit + alphamax)/(xinit*(ainit-alphamin));
qavec = qinit./[1 2 10 20]; qanum = length(qavec); conit = 50;
\end{lstlisting}
where \texttt{ainit} is the initial Brinkman penalty, \texttt{qinit} is the initial interpolation factor computed using Equation \ref{eq:qinit}, \texttt{qanum} is the number of continuation steps, and \texttt{conit} is the maximum number of iterations per continuation step.

Various optimisation parameters are then defined:
\vspace{-12pt}
\begin{lstlisting}[firstnumber=23,firstline=23, lastline=25,numbers=right]
% OPTIMISATION PARAMETERS
maxiter = qanum*conit; mvlim = 0.2; plotdes = 0;
chlim = 1e-3; chnum = 5;
\end{lstlisting}
where \texttt{maxiter} is the maximum number of total design iterations, \texttt{mvlim} is the movelimit $m$ for the OC update, \texttt{chlim} is the stopping criteria, \texttt{chnum} is the number of subsequent iterations the convergence criteria must be met, and \texttt{plotdes} is a Boolean determining whether to plot the design during the optimisation process (0 = no, 1 = yes).

The parameters for the Newton solver are defined:
\vspace{-12pt}
\begin{lstlisting}[firstnumber=26,firstline=26, lastline=27,numbers=right]
% NEWTON SOLVER PARAMETERS
nltol = 1e-6; nlmax = 25; plotres = 0;
\end{lstlisting}
where \texttt{nltol} is the required residual tolerance for convergence, \texttt{nlmax} is the maximum number of non-linear iterations, and \texttt{plotres} is a Boolean determining whether to plot the convergence of the residual norm over the optimisation process (0 = no, 1 = yes).

Finally, the option to export the contour of the final design as a DXF (Data Exchange Format) file can be activated:
\vspace{-12pt}
\begin{lstlisting}[firstnumber=28,firstline=28, lastline=29,numbers=right]
% EXPORT FILE
filename='output'; exportdxf = 0;
\end{lstlisting}
where \texttt{filename} is a specified filename and \texttt{exportdxf} is a Boolean determining whether to export the design (0 = no, 1 = yes).

\subsubsection{Other pre-processing (lines 30-66)}

Subsequently, various arrays are defined (lines 30-44) for the quick assembly of the matrices and vectors from the finite element analysis based on the \texttt{sparse} function in MATLAB. The boundary conditions for the available problems are defined in the separate MATLAB file \texttt{problems.m} (Appendix \ref{app:problems}) and matrices for quick enforcement of Dirichlet conditions are built (lines 45-51). Finally, various arrays, counters and constants are initialised (lines 52-66).

\subsubsection{Optimisation loop (lines 67-167)}

Prior to the optimisation loop start, some information is output to the screen and timers are initiated (lines 67-73). The optimisation starts using a \texttt{while} loop (line 74) and computation of the greyscale indicator \citep{Sigmund2007} and material interpolation per Equation \ref{eq:alpha_interp} (lines 76-80). The non-linear solver loop iteratively updates the solution (lines 81-112) using Newton's method according to Section \ref{sec:newton}. The finite element analysis is carried out according to Section \ref{sec:fea} (lines 85-93) using function \texttt{RES} and \texttt{JAC} to build to residual vector and Jacobian matrix, respectively. For the converged solution, the objective functional is evaluated (line 114) using the function \texttt{PHI} according to Section \ref{sec:objective} and the volume is evaluated for the current design (line 117) according to Section \ref{sec:volume}. The current solution is evaluated (lines 124-126) and if it is not considered converged, then the adjoint problem is formed and solved (lines 131-136) and the sensitivities are computed (lines 137-144) according to Section \ref{sec:adjoint}. Subsequently, the design field is updated using the optimality criteria optimiser (lines 145-154) according to Section \ref{sec:optcrit}. Before repeating the process for the new design, the interpolation parameter is updated if necessary according to the continuation scheme (lines 155-159) in Section \ref{sec:contin}.

The functions for evaluating the residual vector, Jacobian matrix, objective functional, and their partial derivatives, are all automatically built using the file \texttt{analyticalElement.m}. This script uses the MATLAB Symbolic Toolbox to symbolically derive and differentiate the required vectors and matrices.

\subsubsection{Post-processing (lines 168-170)}

The results are post-processed and plotted in an external file \texttt{postproc.m} (line 169), where 5 standard plots are made:
\begin{enumerate}
    \item Design field $\gamma(\mathbf{x})$
    \item Brinkman penalty field, $\alpha(\gamma(\mathbf{x}))$
    \item Velocity magnitude field, $U = \sqrt{u_i u_i}$
    \item Pressure field, $p$
    \item Velocity and design field along a cut-line
\end{enumerate}
Finally, if requested by setting \texttt{exportdxf = 1}, the contour of the final optimised design is exported to DXF format for verification analysis in external software.

\section{Examples} \label{sec:examples}

\subsection{Double pipe problem} \label{sec:twopipe}

This problem was introduced in Section \ref{sec:incex1} and is the default for the provided code when \texttt{probtype = 1}. It was originally introduced by \citet{Borrvall2003} for Stokes flow, but will herein also be optimised for Navier-Stokes flow.

\subsubsection{Stokes flow} \label{sec:twopipe_stokes}

\begin{table}
    \centering
    \begin{tabular}{l|l|l}
        Parameter & Value & Code \\ \hline
        $L_{x}$ & $1$ & \texttt{Lx = 1.0} \\
        $L_{y}$ & $1$ & \texttt{Ly = 1.0} \\
        $\rho$ & $10^{-3}$ & \texttt{rho = 1e-3} \\
        $\mu$ & $1$ & \texttt{mu = 1.0} \\
        $\alpha_\textrm{max}$ & $\sfrac{2.5\mu}{(0.01^2)}$ & \texttt{alphamax = 2.5*mu/(0.01\textasciicircum 2)} \\
        $\alpha_\textrm{min}$ & $\sfrac{2.5\mu}{(100^2)}$ & \texttt{alphamin = 2.5*mu/(100\textasciicircum 2)} \\
        $V_{f}$ & $\sfrac{1}{3}$ & \texttt{volfrac = 1/3} \\
        $n^{e}_{x}$ & $102$ & \texttt{nelx = 102} \\
        $n^{e}_{y}$ & $102$ & \texttt{nely = 102}
    \end{tabular}
    \caption{Parameter values used for double pipe solutions in Figures \ref{fig:doublepipe_Lx1-0} and \ref{fig:doublepipe_Lx1-5}.}
    \label{tab:example1values_BP}
\end{table}
In order to make the results directly comparable to those of \citet{Borrvall2003}, the parameter values are set as in Table \ref{tab:example1values_BP}. To simulate Stokes flow ideally the density would be set to $0$, but due to the current implementation not being able to handle that, the density is set to a small number, $\rho = 10^{-3}$ instead. This yields an inlet Reynolds number of $Re_\textrm{in} = 1.67\times10^{-4}$, which is sufficiently small to ensure Stokes flow.

\begin{figure}
    \centering
    \begin{subfigure}{\columnwidth}
    \hspace{14ex}
    \includegraphics[height=0.54\columnwidth]{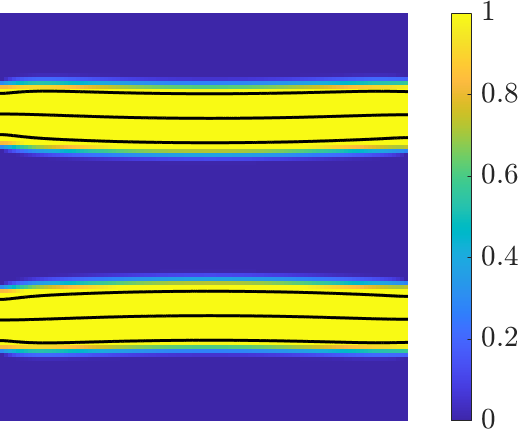}
    \caption{Design field}
    \label{fig:doublepipe_Lx1-0-a}
    \end{subfigure}
    \\ \vspace{1ex}
    \begin{subfigure}{\columnwidth}
    \hspace{14ex}
    \includegraphics[height=0.54\columnwidth]{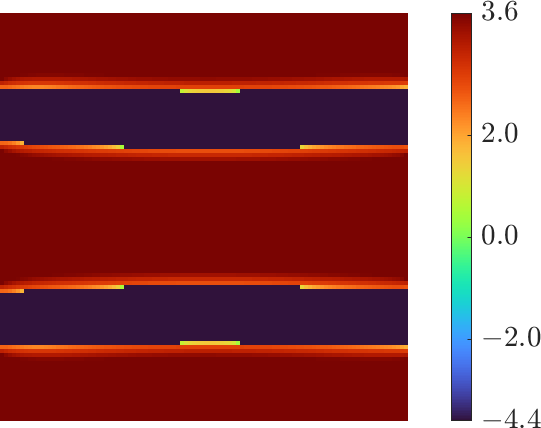}
    \caption{Brinkman penalty field}
    \label{fig:doublepipe_Lx1-0-b}
    \end{subfigure}
    \\ \vspace{2ex}
    \begin{subfigure}{\columnwidth}
    \hspace{14ex}
    \includegraphics[height=0.53\columnwidth]{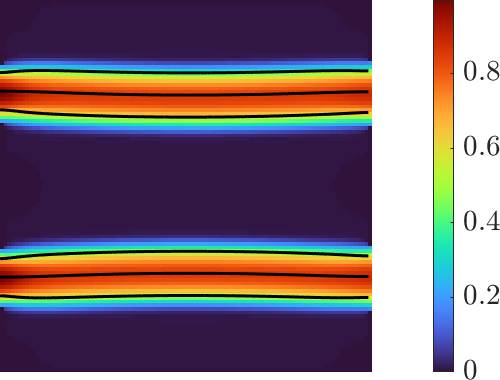}
    \caption{Velocity magnitude field}
    \label{fig:doublepipe_Lx1-0-c}
    \end{subfigure}
    \\ \vspace{2ex}
    \begin{subfigure}{\columnwidth}
    \hspace{14ex}
    \includegraphics[height=0.53\columnwidth]{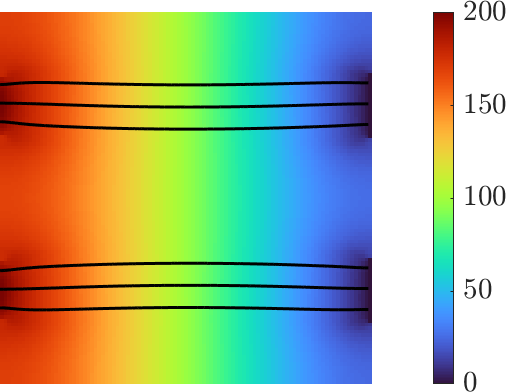}
    \caption{Pressure field}
    \label{fig:doublepipe_Lx1-0-d}
    \end{subfigure}
    \caption{Optimised design and corresponding physical fields for the double pipe problem with $L_{x} = 1$ and $Re_\textrm{in}=10^{-3}$. Final values: $\phi = 22.0956$ after $63$ iterations.}
    \label{fig:doublepipe_Lx1-0}
\end{figure}
Figure \ref{fig:doublepipe_Lx1-0} shows the optimised design and corresponding physical fields for a square domain with $Re_\textrm{in}=1$. From Figure \ref{fig:doublepipe_Lx1-0-b} it can be seen that the converged design in Figure \ref{fig:doublepipe_Lx1-0-a} ensures the lowest and highest Brinkman penalty factor in the fluid region and solid region, respectively. This results in negligible velocities in the solid region with the flow passing easily through the channels, as seen in Figure \ref{fig:doublepipe_Lx1-0-c}. Figure \ref{fig:doublepipe_Lx1-0-d} shows the pressure field, where it can be seen that a continuous pressure field exists even in the solid region. The final objective value is $\phi = 22.0956$, which is very close to the value of $25.67$ reported by \citet{Borrvall2003}. Differences can be attributed to different implementations, most likely the fact that the present uses stabilised linear finite elements in contrast to mixed-order quadratic-linear elements \citep{Borrvall2003}. The level of intermediate design variables since the final interpolation factors are almost the same, at 9.85 for the present code and an equivalent of 10 in \citet{Borrvall2003}, and the level of greyscale is therefore similar.

\begin{figure}
    \centering
    \begin{subfigure}{\columnwidth}
    \hspace{2ex}
    \includegraphics[height=0.54\columnwidth]{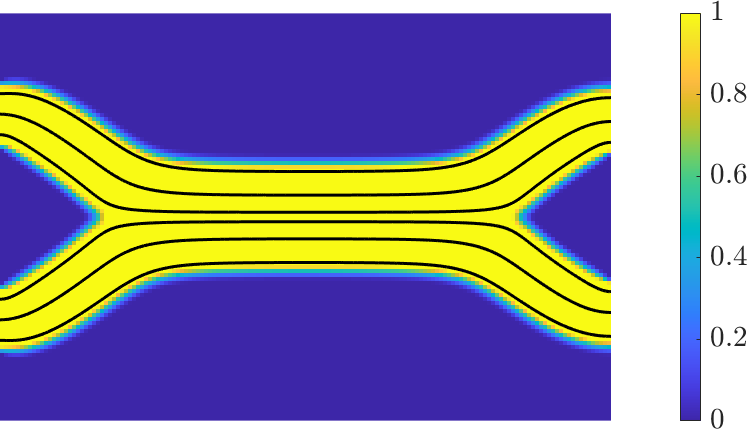}
    \caption{Design field}
    \label{fig:doublepipe_Lx1-5-a}
    \end{subfigure}
    \\ \vspace{1ex}
    \begin{subfigure}{\columnwidth}
    \hspace{2ex}
    \includegraphics[height=0.54\columnwidth]{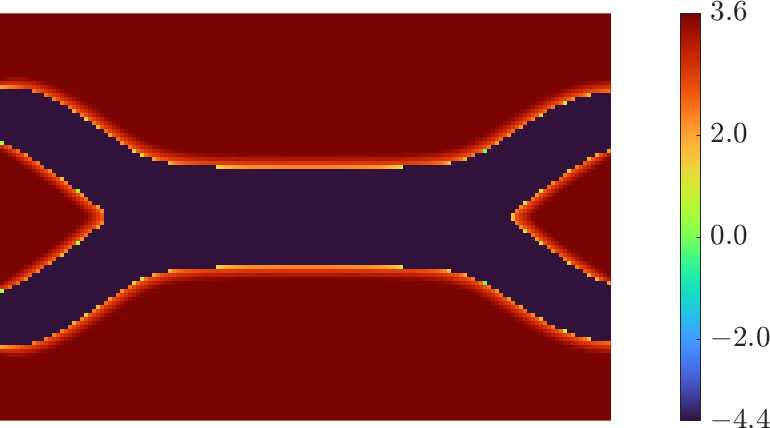}
    \caption{Brinkman penalty field}
    \label{fig:doublepipe_Lx1-5-b}
    \end{subfigure}
    \\ \vspace{2ex}
    \begin{subfigure}{\columnwidth}
    \hspace{2ex}
    \includegraphics[height=0.53\columnwidth]{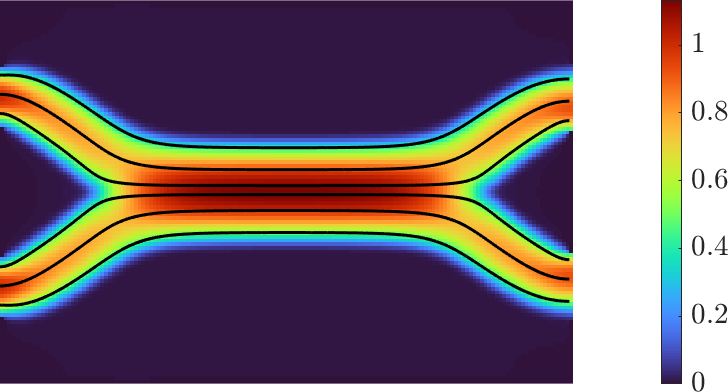}
    \caption{Velocity magnitude field}
    \label{fig:doublepipe_Lx1-5-c}
    \end{subfigure}
    \\ \vspace{2ex}
    \begin{subfigure}{\columnwidth}
    \hspace{2ex}
    \includegraphics[height=0.53\columnwidth]{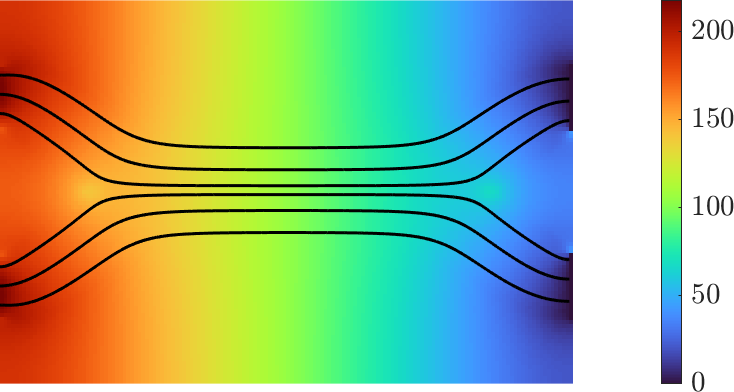}
    \caption{Pressure field}
    \label{fig:doublepipe_Lx1-5-d}
    \end{subfigure}
    \caption{Optimised design and corresponding physical fields for the double pipe problem with $L_{x} = 1.5$ and $Re_\textrm{in}=10^{-3}$. Final values: $\phi = 23.5732$ after $87$ iterations.}
    \label{fig:doublepipe_Lx1-5}
\end{figure}
An elongated domain is investigated by setting $L_{x} = 1.5$ similar to \citet{Borrvall2003}, with the number of elements increased proportionally in the same direction. Figure \ref{fig:doublepipe_Lx1-5} shows the optimised design and corresponding physical fields. The optimised design is now a ``double wrench'' design, where the channels merge at the centre to minimise viscous losses along the channel walls. The final objective value is $\phi = 23.5732$, which is very close to the value of $27.64$ reported by \citet{Borrvall2003}.

\subsubsection{Navier-Stokes}  \label{sec:twopipe_ns}

\begin{table}
    \centering
    \begin{tabular}{l|l|l}
        Parameter & Value & Code \\ \hline
        $\rho$ & $1$ & \texttt{rho = 1.0} \\
        $\mu$ & $\sfrac{1}{6 Re_\textrm{in}}$ & \texttt{mu = 1/(6*20)}
    \end{tabular}
    \caption{Updated values used for double pipe solutions in Figure \ref{fig:doublepipe_reinc}. Other values are given in Table \ref{tab:example1values_BP}.}
    \label{tab:example1values_RE}
\end{table}
In order to non-dimensionalise the problem and control it using the Reynolds number, $Re_\textrm{in}$, the density is set to $\rho = 1$ (\texttt{rho = 1.0}) and the viscosity is set to $\mu = \sfrac{1}{6 Re_\textrm{in}}$ (\texttt{mu = 1/(6*20)}). The 6 in the denominator ensures that the Reynolds number, $Re_\textrm{in}$, is based on the inlet height which is $\frac{L_{y}}{6}$ as shown in Figure \ref{fig:example1_bcs}. Other parameter values are as previously.

\begin{figure}
    \centering
    \begin{subfigure}{\columnwidth}
    \hspace{0.1\columnwidth}
    \includegraphics[height=0.55\columnwidth]{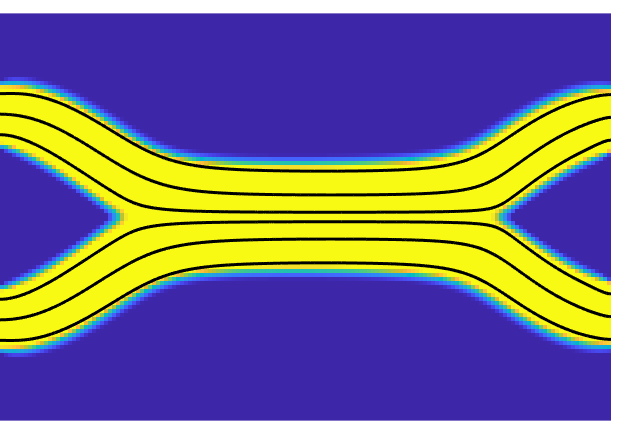}
    \caption{$Re_\textrm{in} = 20$}
    \label{fig:doublepipe_reinc-a}
    \end{subfigure}
    \\
    \begin{subfigure}{\columnwidth}
    \hspace{0.1\columnwidth}
    \includegraphics[height=0.55\columnwidth]{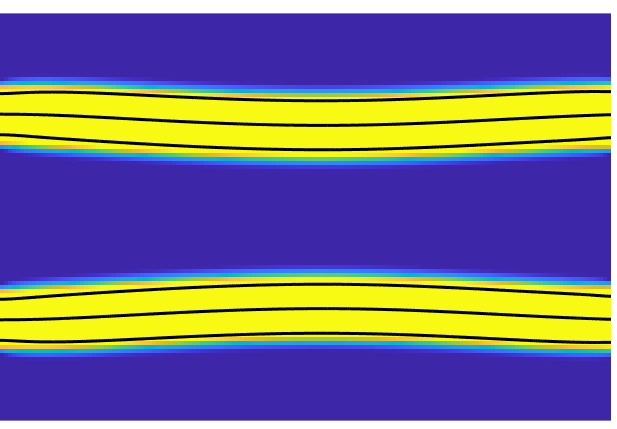}
    \caption{$Re_\textrm{in} = 40$}
    \label{fig:doublepipe_reinc-b}
    \end{subfigure}
    \caption{Optimised designs for the double pipe problem with increasing Reynolds number and $L_{x} = 1.5$. Final values: (a) $\phi = 0.2126$ after $102$ iterations; (b) $\phi = 0.1380$ after $59$ iterations.}
    \label{fig:doublepipe_reinc}
\end{figure}
Figure \ref{fig:doublepipe_reinc} shows the optimised designs obtained for increasing Reynolds number. It is observed that for $Re_\textrm{in}=40$ and above, the OC solver converges to the local minimum of two ``straight'' channels. It is observed using COMSOL simulations, that this local minimum is worse than the double wrench channel for Reynolds numbers up to 200. The MMA solver introduced in Section \ref{sec:MMA} is actually able to converge to a significantly better design for $Re_\textrm{in}=40$.

\subsubsection{Variation of $\alpha_\textrm{max}$} \label{sec:twopipe_amax}

To demonstrate the robustness of the proposed heuristic initial Brinkman penalty and continuation scheme, as well as the insensitivity of minimum dissipation optimisation to the maximum Brinkman penalty, the double pipe problem will be optimised for $Re_\textrm{in}=20$ using 5 additional $\alpha_\textrm{max}$. Getting to the double wrench channel for $Re_\textrm{in}=20$ seems pretty difficult and the poorer local minima double pipe design is obtained very easily. Having a constant fixed initial Brinkman penalty, $\alpha_\textrm{init}$, of the entire design domain ensures that the same initial flow field is obtained independent of maximum Brinkman penalty, $\alpha_\textrm{max}$.

The values predicted by the presented MATLAB code are compared to verification analyses using COMSOL of post-processed designs. The \texttt{export.m} file provides export capabilities of the design field contour to the DXF file format. The DXF file can then be imported into COMSOL and used to trim the computational domain. An example COMSOL file is included in the Supplementary Material.

\begin{figure}
    \centering
    \includegraphics[width=0.75\linewidth]{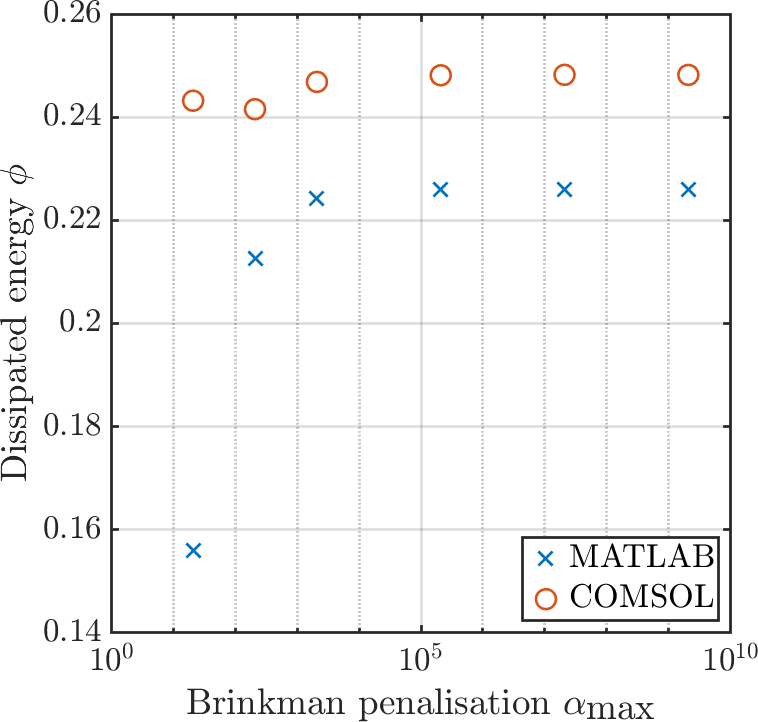}
    \caption{$Re_\textrm{in} = 20$}
    \label{fig:doublepipe_amax}
\end{figure}
Figure \ref{fig:doublepipe_amax} shows the dissipated energy of the final optimised designs. The final designs look basically identical to Figure \ref{fig:doublepipe_reinc-a} and are, thus, not shown here. It can be seen that the post-processed performance of the designs are more or less constant independent of $\alpha_\textrm{max}$. The values predicted by the continuous Brinkman model are seen to be quite far off for lower maximum penalties. However, this shows that even a low maximum Brinkman penalty, which in theory yields a large error in the prediction of the dissipated energy (Figure \ref{fig:headless_errorVkappa}), ends up giving the same or close to the same final design and performance. This insensitivity was already noted originally by \citet{Borrvall2003}, but not explicitly shown.

This conclusion is somewhat disconnected from the investigations of the flow accuracy in Section \ref{sec:varparstudy} and Figure \ref{fig:headless_brinkdom_vel}. However, in contrast to the reference geometry in Section \ref{sec:varparstudy}, the final optimised design for minimum dissipated energy do not have re-circulation zones, since this is detrimental to the energy functional. Therefore, for the optimised designs, a lower maximum Brinkman penalty appears to be forgiving.

\subsection{Pipe bend problem} \label{sec:pipebend}

This problem was introduced in Section \ref{sec:incex2} and is selected in the code by setting \texttt{probtype = 2}. It was originally introduced by \citet{Borrvall2003} for Stokes flow and also treated by \citet{GersborgHansen2005} for Navier-Stokes flow.

\subsubsection{Stokes flow}

\begin{table}
    \centering
    \begin{tabular}{l|l|l}
        Parameter & Value & Code \\ \hline
        $L_{x}$ & $1$ & \texttt{Lx = 1.0} \\
        $L_{y}$ & $1$ & \texttt{Ly = 1.0} \\
        $\rho$ & $10^{-3}$ & \texttt{rho = 1e-3} \\
        $\mu$ & $1$ & \texttt{mu = 1.0} \\
        $\alpha_\textrm{max}$ & $\sfrac{2.5\mu}{(1/0.01^2)}$ & \texttt{alphamax = 2.5*mu/(1/0.01\textasciicircum 2)} \\
        $\alpha_\textrm{min}$ & $\sfrac{2.5\mu}{(100^2)}$ & \texttt{alphamin = 2.5*mu/(100\textasciicircum 2)} \\
        $V_{f}$ & $0.25$ & \texttt{volfrac = 0.25} \\
        $n^{e}_{x}$ & $100$ & \texttt{nelx = 100} \\
        $n^{e}_{y}$ & $100$ & \texttt{nely = 100}
    \end{tabular}
    \caption{Parameter values used for pipe bend solution in Figure \ref{fig:example2_BP}.}
    \label{tab:example2values_BP}
\end{table}
In order to make the results comparable to those of \citet{Borrvall2003}, the parameter values are set as in Table \ref{tab:example2values_BP}. As previously, the density is set to a small number, $\rho = 10^{-3}$, yielding an inlet Reynolds number of $Re_\textrm{in} = 2\times10^{-4}$ which is small enough to approximate Stokes flow.

\begin{figure}
    \centering
    \includegraphics[height=0.55\columnwidth]{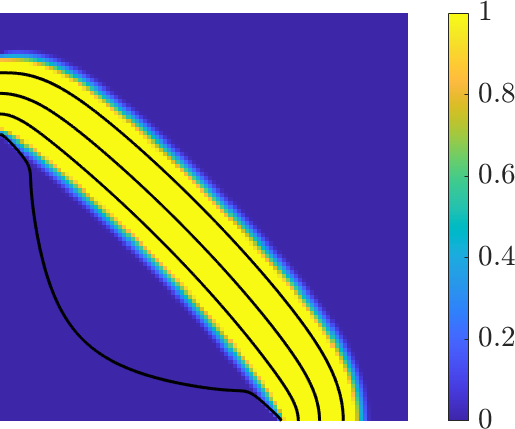}
    \caption{Optimised design for the pipe bend problem using values from \citet{Borrvall2003}. Final values: $\phi = 9.1862$ after $66$ iterations.}
    \label{fig:example2_BP}
\end{figure}
Figure \ref{fig:example2_BP} shows the optimised design obtained using the values from \citet{Borrvall2003} listed in Table \ref{tab:example2values_BP}. The design is a straight channel from the inlet to the outlet with a final objective value of $\phi = 22.0956$, which is very close to the value of $25.67$ reported by \citet{Borrvall2003}. The streamlines show that the maximum Brinkman penalty is not large enough to provide enough resistance to flow through the solid regions. However, as previously shown, while increasing the $\alpha_\textrm{max}$ may increase the accuracy of the flow modelling, almost identical optimised designs are obtained for minimum dissipated energy.

\subsubsection{Navier-Stokes flow}

\begin{table}
    \centering
    \begin{tabular}{l|l|l}
        Parameter & Value & Code \\ \hline
        $L_{x}$ & $1$ & \texttt{Lx = 1.0} \\
        $L_{y}$ & $1$ & \texttt{Ly = 1.0} \\
        $\rho$ & $10^{-3}$ & \texttt{rho = 1.0} \\
        $\mu$ & $\sfrac{1}{(5\,Re_\textrm{in})}$ & \texttt{mu = 1/(5*10)} \\
        $\alpha_\textrm{min}$ & $\sfrac{2.5\mu}{(1/5^2)}$ & \texttt{alphamin = 2.5*mu/(1/5\textasciicircum 2)} \\
        $\alpha_\textrm{max}$ & $10^{4}\alpha_\textrm{min}$ & \texttt{alphamax = 1e4*alphamin} \\
        $V_{f}$ & $0.25$ & \texttt{volfrac = 0.25} \\
        $n^{e}_{x}$ & $100$ & \texttt{nelx = 100} \\
        $n^{e}_{y}$ & $100$ & \texttt{nely = 100}
    \end{tabular}
    \caption{Parameter values used for pipe bend solution in Figure \ref{fig:example2_reinc}.}
    \label{tab:example2values_reinc}
\end{table}
In order to investigate the effect of inertia, the out-of-plane channel height is set to the width of the inlet. This ensures a hydraulic diameter of the inlet width/height and a well-defined Reynolds number. The problem is non-dimensionalised and controlled using the inlet Reynolds number, $Re_\textrm{in}$, leading to the parameter values listed in Table \ref{tab:example2values_reinc}. The 5 in the denominator of the viscosity ensures that the Reynolds number is based on the inlet width, which is $\frac{L_{y}}{5}$ as shown in Figure \ref{fig:example2_bcs}.

\begin{figure}
    \centering
    \includegraphics[height=0.65\columnwidth]{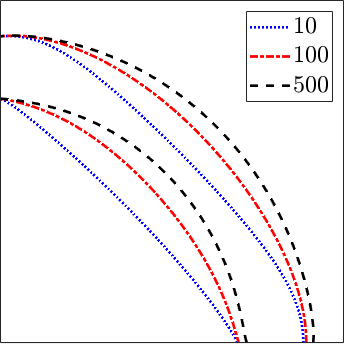}
    \caption{Design contours for the pipe bend problem with increasing Reynolds number. Final values: (a) $\phi = 2.484\times10^{-1}$ after $58$ iterations; (b) $\phi = 2.830\times10^{-2}$ after $63$ iterations; (c) $\phi = 6.502\times10^{-3}$ after $40$ iterations.}
    \label{fig:example2_reinc}
\end{figure}
Figure \ref{fig:example2_reinc} shows the contours of the optimised designs for increasing Reynolds number. For a low Reynolds number, $Re_\textrm{in} = 10$, a more or less straight pipe is formed similar to that for Stokes flow. When the Reynolds number increases, inertia begins to play a role and the curvature of the pipe changes to accommodate this.
\begin{table}
    \centering
    \begin{tabular}{r|c|c|c}
                     & \multicolumn{3}{c}{Optimised for:} \\
        Analysed at: & $Re=10$ & $Re=100$ & $Re=500$ \\ \hline
             $Re=10$ [$\times10^{-1}$] & $\mathbf{2.484}$ & $2.731$ & $3.071$ \\
            $Re=100$ [$\times10^{-2}$] & $2.889$ & $\mathbf{2.830}$ & $3.105$ \\
            $Re=500$ [$\times10^{-3}$] & $8.459$ & $6.553$ & $\mathbf{6.502}$
    \end{tabular}
    \caption{Cross-check table of the optimised designs for the pipe bend problem shown in Figure \ref{fig:example2_reinc}. Values should be multiplied by. The best performing design for each analysis Reynolds number is highlighted in bold.}
    \label{tab:pipebend_cross}
\end{table}
This is confirmed by the cross-check of the three designs with the objective values in Table \ref{tab:pipebend_cross}. A cross-check is strictly necessary prior to drawing conclusions from designs optimised for different conditions. A cross-check is where the optimised designs are tested for the other flow conditions and compared. A successful cross-check is where the design optimised for certain conditions is also the best for these conditions. This means it has the lowest value in its own row, such that the diagonal has the best values as highlighted in bold in Table \ref{tab:pipebend_cross}.

\section{Modifications} \label{sec:modifications}

This Section presents a series of extensions to the provided code, showing how it is possible to extend it to solve a range of more complicated examples including fixed regions and other objective functionals.

\subsection{Fixed regions of solid and fluid} \label{sec:passive}

It is common that the design domain is only a subset of the computational domain, where the rest is defined as fixed solid or fluid domains. In the discretised problem this can be handled by restricting optimisation to the \textit{active} elements in the defined design domain, with \textit{passive} elements in the fixed domains remaining constant at their prescribed value.

\begin{figure}
    \centering
    \includegraphics[width=\columnwidth]{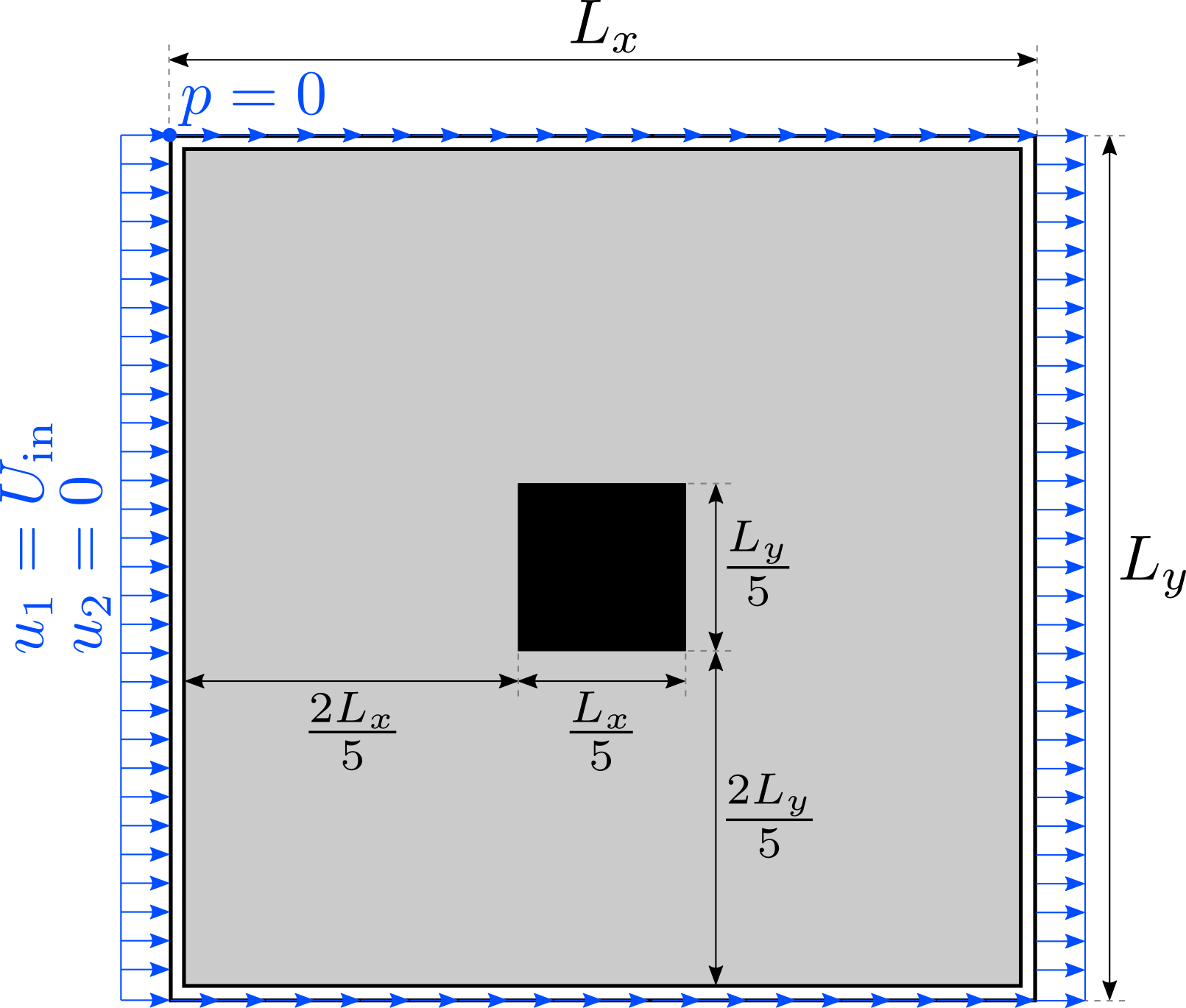}
    \caption{Problem setup for the modified rugby ball problem.}
    \label{fig:rugby_bc}
\end{figure}
The implementation of passive and active elements will be illustrated using the so-called rugby ball problem, introduced by \citet{Borrvall2003} in the context of topology optimisation, but it has its roots in the work by \cite{Pironneau1973}. Figure \ref{fig:rugby_bc} shows the external flow problem with a free flow velocity of $U_\textrm{in}$ applied around the entire outer boundary and a single zero-pressure point constraint in the upper-left corner. A thin region next to the boundary all the way around the outside will be prescribed as fluid regions and a solid square will be prescribed in the centre of the domain. The solid square is not imposed originally, but is done so here to demonstrate the definition of both types of passive domains. 
\begin{table}
    \centering
    \begin{tabular}{l|l|l}
        Parameter & Value & Code \\ \hline
        $L_{x}$ & $1$ & \texttt{Lx = 1.0} \\
        $L_{y}$ & $1$ & \texttt{Ly = 1.0} \\
        $\rho$ & $10^{-3}$ & \texttt{rho = 1e-3} \\
        $\mu$ & $1$ & \texttt{mu = 1.0} \\
        $\alpha_\textrm{max}$ & $\sfrac{2.5\mu}{(0.01^2)}$ & \texttt{alphamax = 2.5*mu/(0.01\textasciicircum 2)} \\
        $\alpha_\textrm{min}$ & $\sfrac{2.5\mu}{(100^2)}$ & \texttt{alphamin = 2.5*mu/(100\textasciicircum 2)} \\
        $V_{f}$ & $0.94$ & \texttt{volfrac = 0.94} \\
        $n^{e}_{x}$ & $100$ & \texttt{nelx = 100} \\
        $n^{e}_{y}$ & $100$ & \texttt{nely = 100}
    \end{tabular}
    \caption{Parameter values used for rugby ball solutions in Figure \ref{fig:rugbyball_result}.}
    \label{tab:rugbyvalues}
\end{table}
Table \ref{tab:rugbyvalues} lists the parameter values used herein.

First of all, the new problem should be added to \texttt{problems.m}. Add the following after line 56 to create a third problem:
\vspace{-12pt}
\begin{lstlisting}[numbers=none]
elseif (probtype == 3) % RUGBY BALL PROBLEM
if ( mod(nelx,5) > 0 || mod(nely,5) > 0 )
error('ERROR: Number of elements per side must be divisable by 5.');
end
\end{lstlisting}
The element number check is needed, since the solid square will be defined in terms of fifths of the domain height and width. Then define the boundary conditions and Reynolds number as follows:
\vspace{-12pt}
\begin{lstlisting}[numbers=none]
nodesOuter  = [1:nody-1 nody:nody:nodtot nody+1:nody:nodtot (nodx-1)*nody+1:nodtot-1];
nodesPressure = 1;
fixedDofsOuterX = 2*nodesOuter-1; fixedDofsOuterY = 2*nodesOuter;
fixedDofsU = [fixedDofsOuterX fixedDofsOuterY];
fixedDofsP = 2*nodtot+nodesPressure;
fixedDofs  = [fixedDofsU fixedDofsP];
% DIRICHLET VECTORS
DIRU=zeros(nodtot*2,1); DIRP=zeros(nodtot,1);
DIRU(fixedDofsOuterX) = Uin; DIR = [DIRU; DIRP];     
% INLET REYNOLDS NUMBER
Renum = Uin*Ly*rho/mu;
\end{lstlisting}
The fluid region will be defined as a single element wide layer around the outer edge:
\vspace{-12pt}
\begin{lstlisting}[numbers=none]
% PASSIVE ELEMENTS
fluid = [1:nely nely:nely:neltot (nely+1):nely:neltot (nelx-1)*nely+1:neltot-1];
\end{lstlisting}
The solid square is defined as the central one fifth of the domain height and width:
\vspace{-12pt}
\begin{lstlisting}[numbers=none]
x = 2*nelx/5:3*nelx/5; y = 2*nely/5:3*nely/5;
[X,Y] = meshgrid(x,y);
solid = sub2ind([nely nelx],Y(:),X(:))';
\end{lstlisting}
Finally, all elements of the computational domain will be classified as either \textit{passive} or \textit{active}:
\vspace{-12pt}
\begin{lstlisting}[numbers=none]
passive = [solid fluid]';
active = setdiff(1:neltot,passive)';
nactive = length(active);
\end{lstlisting}
where \texttt{nactive} is the number of active elements to be used in the main script.
In order to make the updated code backwards compatible with the first two problems defined in \texttt{problem.m}, the following must be added at the bottom of each definition:
\vspace{-12pt}
\begin{lstlisting}[numbers=none]
% PASSIVE ELEMENTS
solid = []; fluid = [];
passive = [solid fluid]'; ...
active = setdiff(1:neltot,passive)';
nactive = length(active);
\end{lstlisting}

Now, modifications needs to be made to the main script \texttt{topFlow.m} in order to ensure that passive elements are given their prescribed value and that the optimisation is only applied to the active elements. After line 57, where the initial design field is defined, add the corrections for the prescribed values:
\vspace{-12pt}
\begin{lstlisting}[numbers=none]
xPhys(solid) = 0.0; xPhys(fluid) = 1.0;
\end{lstlisting}
Next, the greyscale indicator should be calculated for the design domain only, so change line 77 to:
\vspace{-12pt}
\begin{lstlisting}[numbers=none]
Md = 100*full(4*sum(xPhys(active).* (1-xPhys(active)))/nactive); 
\end{lstlisting}
The volume constraint should also be modified to only account for the design domain. The evaluation on line 117 should be updated to:
\vspace{-12pt}
\begin{lstlisting}[numbers=none]
V = mean(xPhys(active));
\end{lstlisting}
and the sensitivities updated accordingly on line 144:
\vspace{-12pt}
\begin{lstlisting}[numbers=none]
dV = ones(nely,nelx)/nactive; dV(passive) = 0.0;
\end{lstlisting}
The sensitivities of the objective for the passive elements should also be set to 0, so add the following after the sensitivity calculation on line 142:
\vspace{-12pt}
\begin{lstlisting}[numbers=none]
sens(passive) = 0.0;
\end{lstlisting}
In order to only update the active elements, the pre- and post-steps of the OC solver should be updated. Lines 146-147 should be updated to:
\vspace{-12pt}
\begin{lstlisting}[numbers=none]
xnew = xPhys(active); ...
xlow = xPhys(active)-mvlim; ...
xupp = xPhys(active)+mvlim;
ocfac = xPhys(active).*max(1e-10, (-sens(active)./dV(active))).^(1/3);
\end{lstlisting}
and line 154 should be updated to:
\vspace{-12pt}
\begin{lstlisting}[numbers=none]
xPhys(active) = xnew;
\end{lstlisting}

In order to solve the rugby ball problem, it is not necessary to use a continuation strategy for the interpolation parameter. As noted by \citet{Borrvall2003}, the problem is very well-conditioned and it is plenty to use a single parameter value from the start. Therefore, the first definition of line 22 should be updated to \texttt{qavec = 10} and lines 20 and 21 can be deleted. Beware that changing $\alpha_\textrm{max}$ will now change the initial Brinkman penalty.

\begin{figure}
    \centering
    \begin{subfigure}{\columnwidth}
    \hspace{0.225\columnwidth}
    \includegraphics[height=0.55\columnwidth]{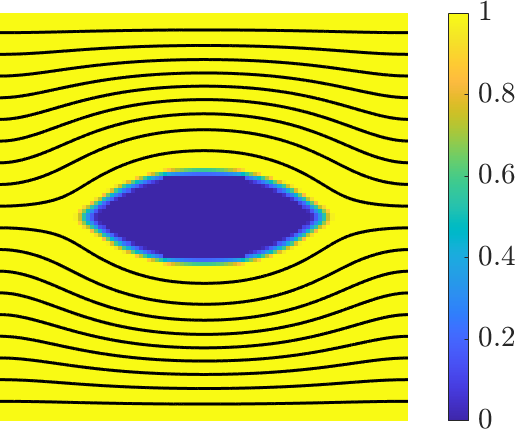}
    \caption{$V_{f} = 0.94$}
    \label{fig:rugbyball_result-a}
    \end{subfigure}
    \\
    \begin{subfigure}{\columnwidth}
    \hspace{0.225\columnwidth}
    \includegraphics[height=0.55\columnwidth]{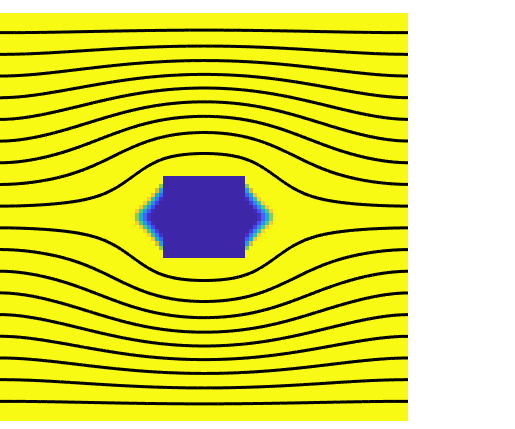}
    \caption{$V_{f} = 0.99$}
    \label{fig:rugbyball_result-b}
    \end{subfigure}
    \caption{Optimised designs for the modified rugby ball problem with different allowable fluid volume fractions. Final values: (a) $\phi = 13.503$ after $19$ iterations; (b) $\phi = 9.839$ after $17$ iterations.}
    \label{fig:rugbyball_result}
\end{figure}
Figure \ref{fig:rugbyball_result} shows the optimised design for \texttt{volfrac = 0.94} and \texttt{volfrac = 0.99}. In order to compare directly with the results of \citet{Borrvall2003}, the volume fraction has to account for the enforced solid square, which means $0.04$ has been added. Figure \ref{fig:rugbyball_result-a} is visually identical to the result obtained by \citet{Borrvall2003}, with a final dissipated energy of $\phi = 13.503$ compared to $14.44$. Figure \ref{fig:rugbyball_result-b} shows the optimised design for a higher fluid volume fraction, where the enforced solid square is clearly seen. In contrast to before, the design is now completely different since the solid square dominates the design and triangles have been added in the up- and downwind directions to minimise dissipation. This also causes a significantly higher dissipated energy of $\phi = 9.839$ compared to the freely-optimised design by \citet{Borrvall2003} with $8.35$.

\subsection{Other objective functionals} \label{sec:otherfuncs}

In order to treat other objective functionals than the dissipated energy, a general optimisation solver is needed. Herein the popular Method of Moving Asymptotes (MMA) will be implemented.

\subsubsection{Method of Moving Asymptotes (MMA)} \label{sec:MMA}

The Method of Moving Asymptotes (MMA) by \citet{Svanberg1987} is a very popular optimisation solver within the topolog optimisation community because it handles general optimisation problems with many design variables extremely well. The MATLAB version of MMA is freely available for download under the GNU GPLv3 license \cite{MMAwebsite} and the files \texttt{mmasub.m} and \texttt{subsolv.m} are needed.

In order to use MMA, there are a series of variables and parameters that must be defined under the initialisation block. After line 66, add the following:
\vspace{-12pt}
\begin{lstlisting}[numbers=none]
% Initialise MMA parameters
numconstr = 1;
xold1 = xPhys(:); xold2 = xPhys(:); xnew = xPhys(:);
low = 0; upp = 0;
a0 = 1; ai = 0*ones(numconstr,1);
c = 1000*ones(numconstr,1); ...
d = ones(numconstr,1);
\end{lstlisting}
where \texttt{numconstr} is the number of constraints and \texttt{a0}, \texttt{ai}, \texttt{c} and \texttt{d} all are parameters for setting the type of problem to be solved by MMA. Please see the document by \citet{SvanbergNote} for details on these parameters. In this document, it is also discussed that MMA works best when the objective and constraint functionals are scaled properly. A block computing the scaling should be added after line 117:
\vspace{-12pt}
\begin{lstlisting}[numbers=none]
%% MMA SCALING
if (loop == 0); obj0 = obj/10; end
f0 = obj/obj0; fi = V/volfrac - 1;
\end{lstlisting}
This scales the objective to start at 10 for the first iteration and the volume constraint to be at most 1 in magnitude.
The constraint is scaled to be at most around one, by normalising the volume:
\begin{equation}
    \texttt{fi} = \frac{V}{V_{f}} - 1
\end{equation}
which should be less than 0.
The sensitivities should of course be scaled accordingly, which will come later. In order to monitor the scaled values during the optimisation, add the following output in the result printing block of the code after line 128:
\vspace{-12pt}
\begin{lstlisting}[numbers=none]
fprintf('      MMA: f0 = %4.3e - f1 = %4.3e \n',f0,fi);
\end{lstlisting}
The entire OC solver block (lines 145-154) can now be replaced with:
\vspace{-12pt}
\begin{lstlisting}[numbers=none]
%% MMA UPDATE
df0 = sens(:)/obj0; dfi = dV(:)'/volfrac;
xlow = max(0,xPhys(:)-mvlim); ...
  xupp = min(1,xPhys(:)+mvlim);
[xnew,~,~,~,~,~,~,~,~,low,upp] = mmasub(numconstr,neltot,loop,xPhys(:), xlow,xupp,xold1,xold2,f0,df0,fi,dfi, ...
    low,upp, a0,ai,c,d);
xold2 = xold1; xold1 = xPhys(:); xPhys(:) = xnew;
\end{lstlisting}

\begin{figure}
    \centering
    \includegraphics[height=0.55\columnwidth]{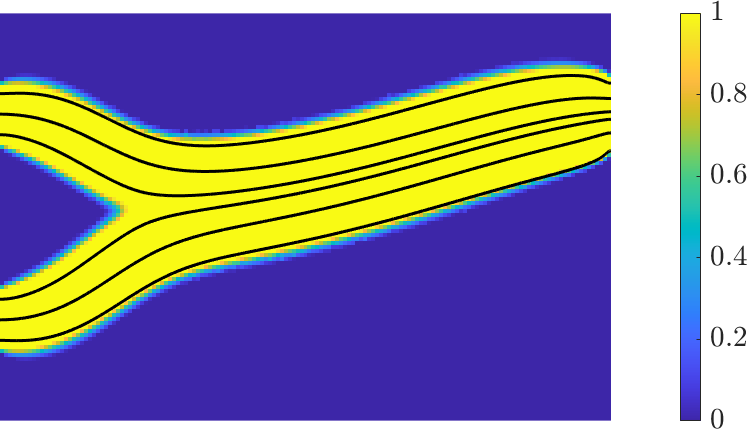}
    \caption{Optimised design for the double pipe problem using MMA and a random initial design field. Final values: $\phi = 0.0876$ after $200$ iterations (maximum).}
    \label{fig:doublepipe_MMA}
\end{figure}
As was shown by \cite{Papadopoulos2021} for Stokes flow, even better minima exists for the double pipe problem when using open boundary conditions at the outlet as in Figure \ref{fig:example1_bcs}. The better designs still uses merged channels, but only a single of the outlets are used. By decreasing the stopping tolerance for the relative change in the objective to \texttt{chlim = 1e-4}, MMA is actually able to find the better performing design shown in Figure \ref{fig:doublepipe_MMA}. This is likely because MMA uses all gradients, both positive and negative, whereas OC only uses positive gradients (add fluid). Removing one of the outlet channels requires to remove fluid caused by negative gradients.

\subsubsection{Flow reversal} \label{sec:flowreversal}

\begin{figure*}
    \centering
    \includegraphics[width=0.6\textwidth]{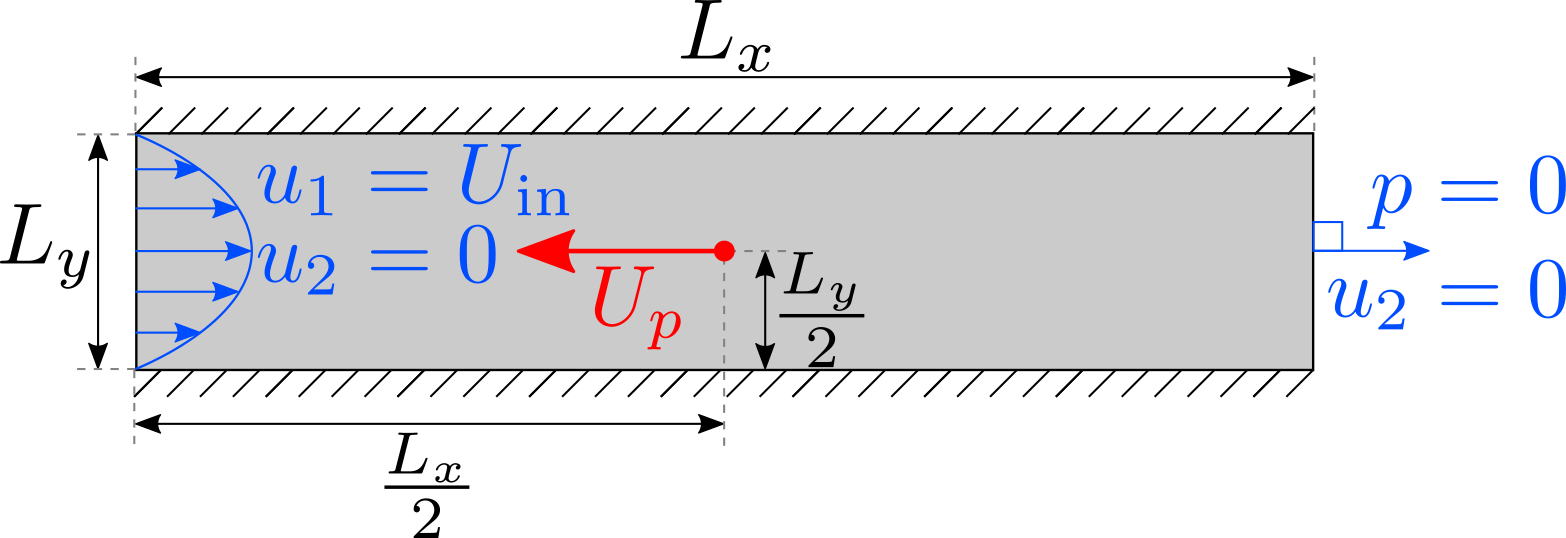}
    \caption{Problem setup for the flow reversal problem.}
    \label{fig:flowreversal_bc}
\end{figure*}
\citet{GersborgHansen2005} introduced the flow reversal problem, which is analogous to the compliant mechanism problem in topology optimisation of solid mechanics. Instead of giving the exact code modifications needed, instructions will be given and the implementation will be left as an exercise for the reader.

The goal is to maximise the velocity of the flow in the reverse direction of the incoming flow, denoted $U_{p}$ as seen in Figure \ref{fig:flowreversal_bc}. A parabolic flow profile enters the domain at the left-hand side and at the outlet a zero-pressure and straight-out flow condition is imposed.
The optimisation problem is formulated as a minimisation problem:
\begin{equation}\label{eq:optprob_reverse}
\begin{split}
 \underset{\bm{\gamma}}{\text{minimise:}} &\quad f_{u}\left(\textbf{s}(\bm{\gamma})\right) = -U_{p} \\
\text{subject to:} & \quad V(\bm{\gamma}) \leq V_{f}\\
                  & \quad p_{in}\left(\textbf{s}(\bm{\gamma})\right) \leq \beta p_{ref}\\
\text{with:}& \quad \textbf{r}(\textbf{s}(\bm{\gamma}),\bm{\gamma})= \mathbf{0}\\
& \quad 0 \leq \gamma_{i} \leq 1, \; i=1,\ldots,n_{el}
\end{split}
\end{equation}
where the minus infront of $U_{p}$ is to turn the maximisation problem into a minimisation problem.
The velocity to be maximised is defined as the negative of the velocity component in the $x_2$-direction at the point $(x_1,x_2) = \left( \frac{L_{x}}{2},\frac{L_{y}}{2} \right)$:
\begin{equation}
    U_{p} = - u_2\left( \frac{L_{x}}{2},\frac{L_{y}}{2} \right)
\end{equation}
For the discretised problem, the objective functional then becomes:
\begin{equation} \label{eq:fu_def}
    f_{u} = {\textbf{l}_{u}}^{T}\textbf{s}
\end{equation}
where $\textbf{l}_{u}$ is a vector of zeros except at the velocity DOF of interest, where a 1 is placed. The vector product then extracts the velocity value at the DOF of interest. According to Equation \ref{eq:adjointproblem}, the adjoint right-hand side is easily found to be simply this vector:
\begin{equation}
    \frac{\partial f_{u}}{\partial \mathbf{s}}^{T} = \textbf{l}_{u}
\end{equation}

This problem is more complex than minimum dissipation, since it includes a constraint on the inlet pressure, $p_{in}$, and thus requires solving an additional adjoint problem.  
The inlet pressure is defined as:
\begin{equation}
    p_{in} = \frac{1}{| \Gamma_{in} |}\int_{\Gamma_{in}} p \,dS
\end{equation}
which is equivalent to the pressure drop since zero pressure is imposed at the outlet. When discretised, this can be formulated as:
\begin{equation} \label{eq:pin_def}
    p_{in} = {\textbf{l}_{p}}^{T}\textbf{s}
\end{equation}
where $\textbf{l}_{p}$ results from the integration of the shape functions for pressure over the inlet boundary. This can be approximated as a vector of zeros with $\frac{1}{n_{in}}$ on the pressure DOFs along the inlet boundary, where $n_{in}$ is the number of nodes along the inlet boundary.
According to Equation \ref{eq:adjointproblem}, the adjoint right-hand side is easily found to be simply the vector:
\begin{equation}
    \frac{\partial p_{in}}{\partial \mathbf{s}}^{T} = \textbf{l}_{p}
\end{equation}

To evaluate the objective and inlet pressure, the user must first define the vectors $\textbf{l}_{u}$ and  $\textbf{l}_{p}$ under pre-processing and then the values can be evaluated in the optimisation loop using the vector products given by Equations \ref{eq:fu_def} and \ref{eq:pin_def}. To evaluate the sensitivities, two adjoint problems must be solved. In order to do so efficiently, the right-hand side of the adjoint problem can be defined as an array with two columns \texttt{RHS = [lu lp]}, where \texttt{lu} and \texttt{lp} are $\textbf{l}_{u}$ and  $\textbf{l}_{p}$, respectively.
When solving the adjoint problem, \texttt{L = J'\textbackslash RHS}, MATLAB solves both adjoint problems using the same factorisation of the Jacobian matrix. This yields two columns of \texttt{L} corresponding to the adjoint solution for the objective and inlet pressure, respectively. The sensitivities can then be computed separately for the two functionals. Please note that for both functionals, the partial derivative with respect to the design variables, $\frac{\partial f}{\partial \bm{\gamma}}$, are zero.

The problem now has two constraints, one on the fluid volume and one on the inlet pressure, so \texttt{numconstr} should be set to 2. Furthermore, the arrays of constraints and sensitivities, \texttt{fi} and \texttt{dfi}, should now have two rows each corresponding to the two constraints. The inlet pressure constraint should be scaled similarly to the volume constraint, \texttt{pin/pmax - 1}, where \texttt{pin} is $p_{in}$ and \texttt{pmax} is $p_{max}$, which should be defined under the definition of input parameters. 
Better behaviour of MMA has been observed by setting the scaling of the objective functional to just the initial objective value, rather than a tenth of this value.

It is possible for the optimiser to get stuck at a solution with zero velocity at the DOF of interest. This occurs mainly when the initial design field is constant and symmetric. In order to push the optimiser away from this local minima, a random initial design field is recommended for this problem. It is recommended to generate it using the \texttt{rand} function in MATLAB and save the generated design field once. 
In order to ensure the random design field is close to the prescribed design field value, in order to ensure the initial Brinkman penalty is hit, the design field can be defined as:
\vspace{-12pt}
\begin{lstlisting}[numbers=none]
xPhys = xinit + 0.1*(-0.5 + rand(nely,nelx)); save('xRand.mat');
\end{lstlisting}
The same field should then be loaded when changing design parameters, ensuring a fair comparison between different parameter settings:
\vspace{-12pt}
\begin{lstlisting}[numbers=none]
load('xRand.mat','xPhys');
\end{lstlisting}

The problem parameters are defined in Table \ref{tab:flowreversal}.
\begin{table}[]
    \centering
    \begin{tabular}{l|l|l}
        Parameter & Value & Code \\ \hline
        $\rho$ & $1$ & \texttt{rho = 1.0} \\
        $L_{x}$ & $5$ & \texttt{Lx = 5.0} \\
        $L_{y}$ & $1$ & \texttt{Ly = 1.0} \\
        $\mu$ & $\sfrac{1}{(Re_\textrm{in})}$ & \texttt{mu = 1/1} \\
        $\alpha_\textrm{min}$ & 0 & \texttt{alphamin = 0.0} \\
        $\alpha_\textrm{max}$ & $\sfrac{\mu}{10^{-5}}$ & \texttt{alphamax = mu/(1e-5)}\\
        $V_{f}$ & $0.6$ & \texttt{volfrac = 0.6} \\
        $n^{e}_{x}$ & $250$ & \texttt{nelx = 250} \\
        $n^{e}_{y}$ & $50$ & \texttt{nely = 50}
    \end{tabular}
    \caption{Parameter values for the flow reversal problem in Section \ref{sec:flowreversal}.}
    \label{tab:flowreversal}
\end{table}
The maximum Brinkman penalty is set one order of magnitude higher than \citet{GersborgHansen2005} used, since their value seems to still allow significant flow through solid feature, especially the thin features that appear. The maximum pressure drop is set to different factors, $\beta$, relative to the pressure drop of an empty channel, $p_{ref}$.

\begin{figure*}
    \centering
    \begin{subfigure}{\textwidth}
    \centering
    \includegraphics[width=0.8\textwidth]{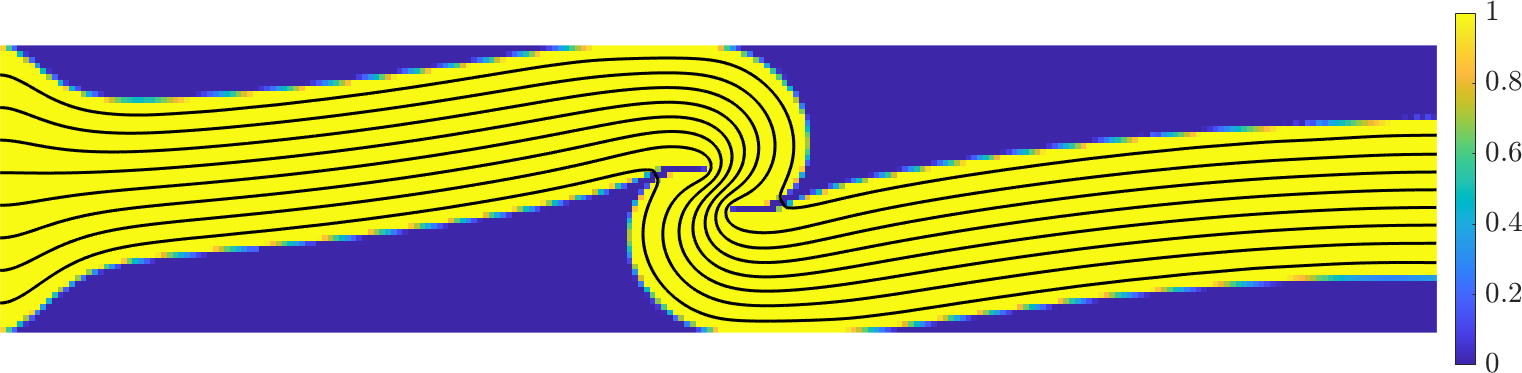}
    \caption{Design field - $\beta = 15$}
    \label{fig:flowreversal_re1-a}
    \end{subfigure}
    \\
    \begin{subfigure}{\textwidth}
    \centering
    \includegraphics[width=0.8\textwidth]{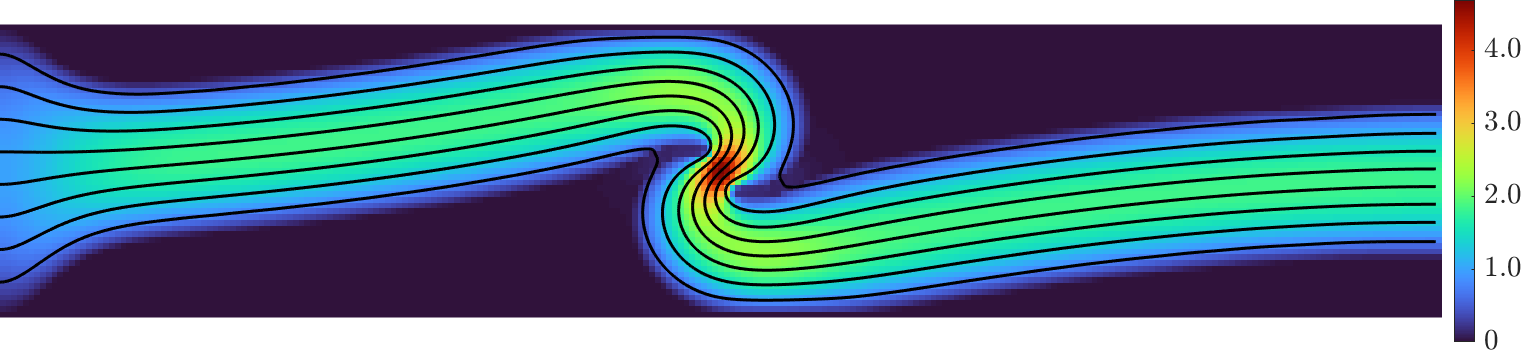}
    \caption{Velocity magnitude - $\beta = 15$}
    \label{fig:flowreversal_re1-b}
    \end{subfigure}
    \\
    \begin{subfigure}{\textwidth}
    \centering
    \includegraphics[width=0.8\textwidth]{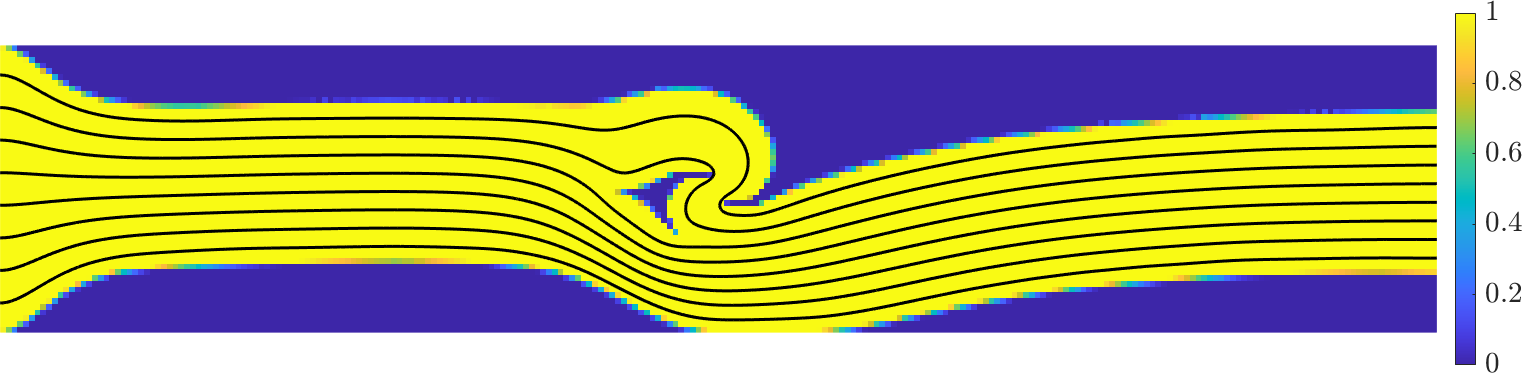}
    \caption{Design field - $\beta = 8$}
    \label{fig:flowreversal_re1-c}
    \end{subfigure}
    \\
    \begin{subfigure}{\textwidth}
    \centering
    \includegraphics[width=0.8\textwidth]{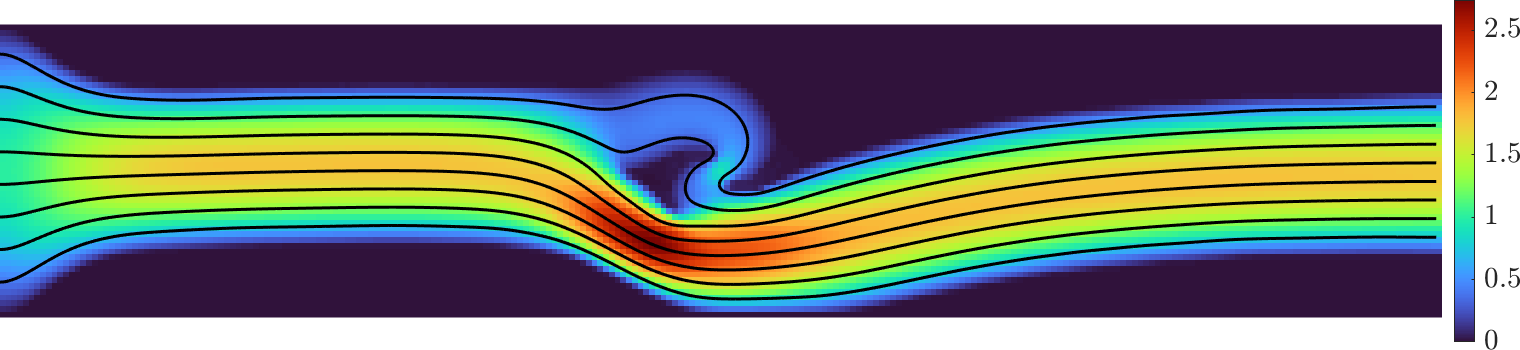}
    \caption{Velocity magnitude - $\beta = 8$}
    \label{fig:flowreversal_re1-d}
    \end{subfigure}
    \caption{Optimised designs and velocity fields for the flow reversal problem with $Re_\textrm{in} = 1$ and variable maximum pressure drop relative to $p_{ref} = 39.870$. Final values: (a,b) $U_{p} = 3.362$ after $80$ iterations; (c,d) $U_{p} = 0.693$ after $112$ iterations.}
    \label{fig:flowreversal_re1}
\end{figure*}
Figure \ref{fig:flowreversal_re1} shows two optimised designs for $Re_\textrm{in} = 1$ that qualitatively agree with those presented by \citet{GersborgHansen2005}. Both designs have a characteristic S-shaped flow feature at the centre, the purpose of which is to reverse the flow direction as requested. It can be seen that for the larger pressure drop, a single channel connects the inlet to the outlet and because all the fluid has to pass through, the velocity becomes large at the centre. For the smaller pressure drop, the topology changes and a bypass channel is included to reduce the pressure drop. The 46\% reduction in pressure drop leads to a significant reduction of 79\% in the objective value.

\begin{figure*}
    \centering
    \begin{subfigure}{\textwidth}
    \centering
    \includegraphics[width=0.8\textwidth]{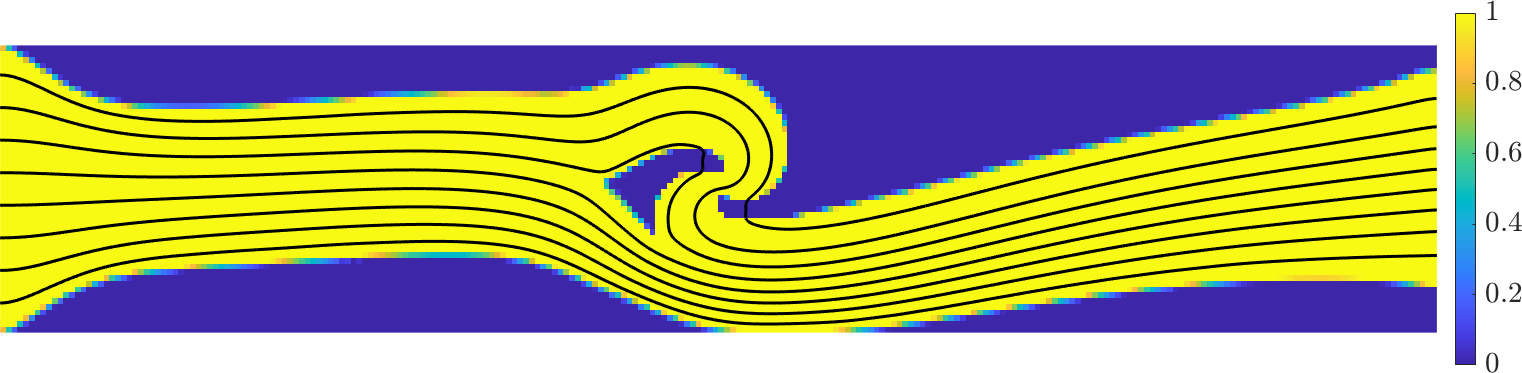}
    \caption{$\beta=15$}
    \label{fig:flowreversal_re100-a}
    \end{subfigure}
    \\
    \begin{subfigure}{\textwidth}
    \centering
    \includegraphics[width=0.8\textwidth]{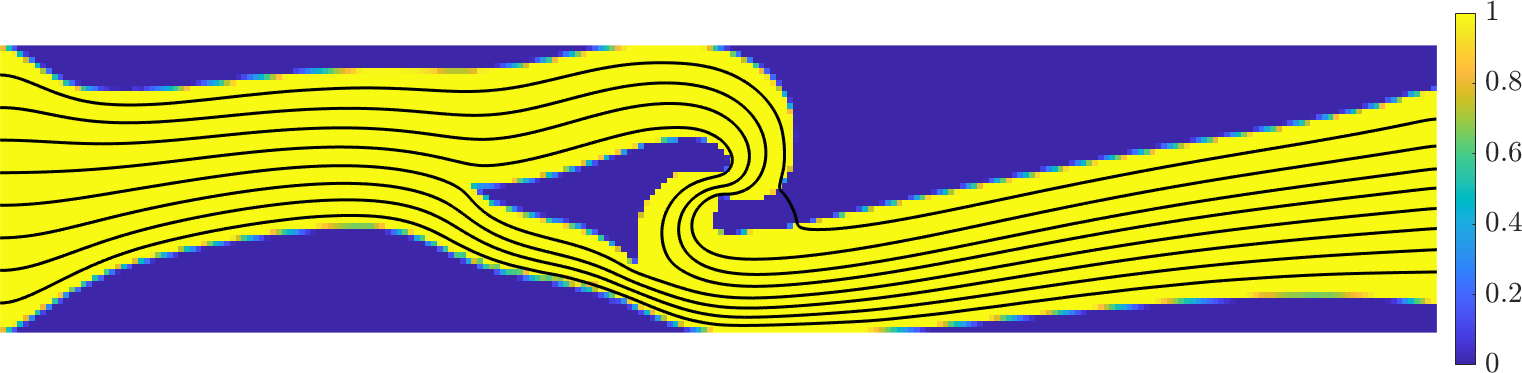}
    \caption{$\beta=30$}
    \label{fig:flowreversal_re100-b}
    \end{subfigure}
    \caption{Optimised designs and velocity fields for the flow reversal problem with $Re_\textrm{in} = 100$ and variable maximum pressure drop relative to $p_{ref} = 0.39819$. Final values: (a) $U_{p} = 2.264$ after $102$ iterations; (b) $U_{p} = 4.075$ after $88$ iterations.}
    \label{fig:flowreversal_re100}
\end{figure*}
Figure \ref{fig:flowreversal_re100} shows the optimised designs for $Re_\textrm{in} = 100$ for two different maximum pressure drops. It can be seen that for this higher Reynolds number, the bypass channel is present for both designs. This is because inertia is dominant in the flow and having a sharp S-turn for the entire flow volume as in Figure \ref{fig:flowreversal_re1-a}, causes a 12 times higher pressure drop compared to the design with the smallest bypass channel in Figure \ref{fig:flowreversal_re100-b}.

\subsubsection{Drag and lift} \label{sec:kondoh}

This example is inspired by the work of \citet{Kondoh2012}, who presented drag minimisation and lift maximisation using topology optimisation based on a variety of objective function formulations. The simplest definition of calculating the drag and lift force are based on body forces resulting from the Brinkman penalty:
\begin{subequations}
\begin{align}
    D &= \int_{\Omega} \alpha u_1 \, dV \\
    L &= \int_{\Omega} \alpha u_2 \, dV 
\end{align}
\end{subequations}
where $D$ is the drag force and $L$ is the lift force.
Being a volumetric integration over the computational domain, the minimum drag objective can easily be implemented by changing the objective functional in the element derivation code \texttt{analyticalElement.m} to:
\vspace{-12pt}
\begin{lstlisting}[numbers=none]
phi = alpha*ux(1);
\end{lstlisting}
for drag and equivalent for lift.
However, because the drag and lift only relate to either the $x_{1}$ or $x_{2}$ velocity DOFs, respectively, the automatically generated output from the Symbolic Toolbox for the partial derivatives $\frac{\partial D}{\partial \mathbf{s}}$ and $\frac{\partial L}{\partial \mathbf{s}}$ will contain single scalar zeros, instead of vectors of zeros for a vectorised function. The output should be modified from:
\vspace{-12pt}
\begin{lstlisting}[numbers=none]
out1 = [t2;0;t2;0;t20;t2;0];
\end{lstlisting}
to:
\vspace{-12pt}
\begin{lstlisting}[numbers=none]
out1 = [t2;zeros(1,length(t2));t2; (1,length(t2));t2;zeros(1,length(t2)); t2;zeros(1,length(t2))];
\end{lstlisting}

\begin{figure*}
    \centering
    \includegraphics[width=0.55\textwidth]{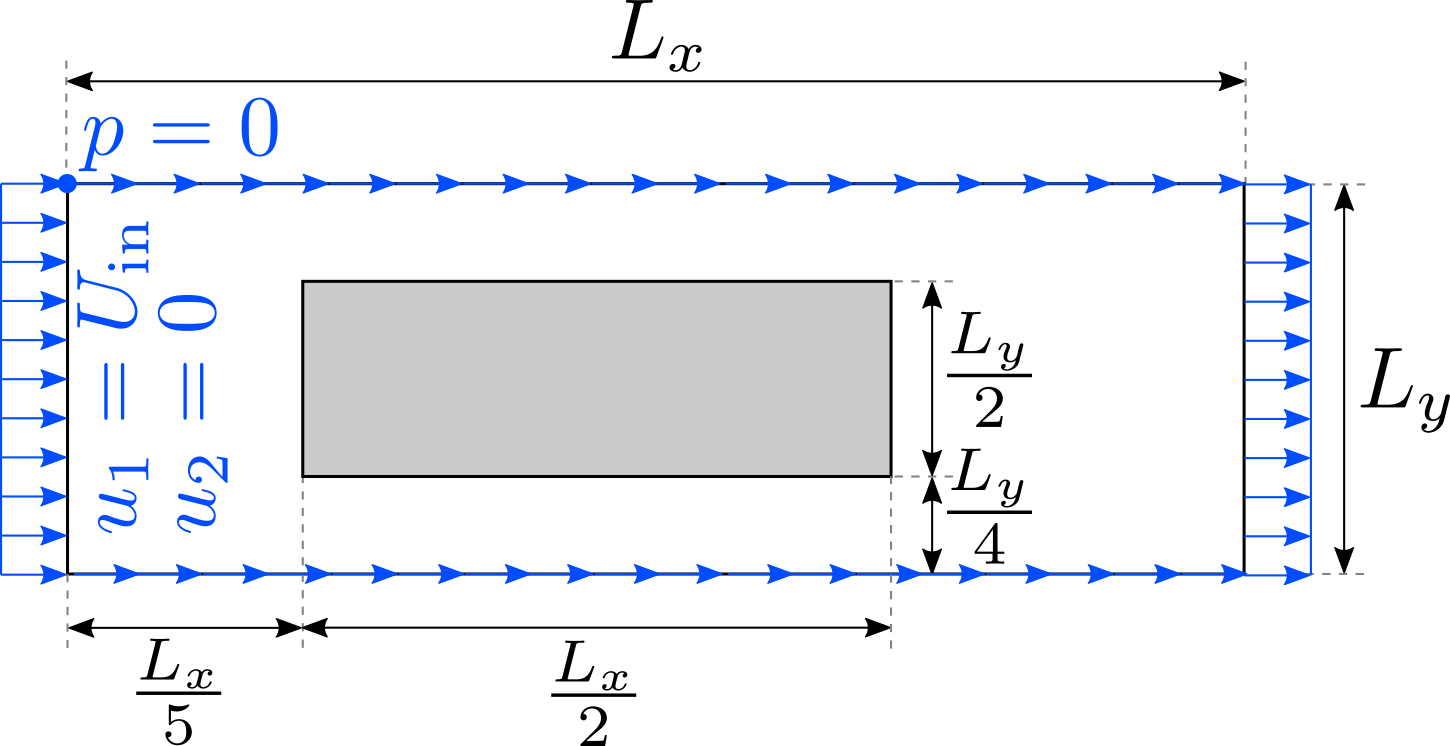}
    \caption{Problem setup for the drag and lift problem.}
    \label{fig:kondoh_bc}
\end{figure*}
Figure \ref{fig:kondoh_bc} shows the problem setup for the drag and lift problem. Similar to the rugby ball problem, a constant velocity is imposed in the $x_1$-direction along the entire outer boundary with a single reference pressure point in the upper-left corner. The design domain is a subset of the computational flow inspired by the work of \citet{Kondoh2012}.
\begin{table}[]
    \centering
    \begin{tabular}{l|l|l}
        Parameter & Value & Code \\ \hline
        $\rho$ & $1$ & \texttt{rho = 1.0} \\
        $L_{x}$ & $3$ & \texttt{Lx = 3.0} \\
        $L_{y}$ & $1$ & \texttt{Ly = 1.0} \\
        $L_{c}$ & $\sqrt{V_{f} | \Omega_{d} |}$ & \texttt{Lc = sqrt(volfrac*Lx/2*Ly/2)} \\
        $\mu$ & $\sfrac{U_\textrm{in}L_{c}\rho}{Re_\textrm{in}}$ & \texttt{mu = Uin*Lc*rho/1} \\
        $\alpha_\textrm{min}$ & 0 & \texttt{alphamin = 0.0} \\
        $\alpha_\textrm{max}$ & $\sfrac{\mu}{(10^{-5}L_{c})}$ & \texttt{alphamax = mu/(1e-5*Lc)}\\
        $V_{f}$ & $0.85$ & \texttt{volfrac = 0.85} \\
        $n^{e}_{x}$ & $300$ & \texttt{nelx = 300} \\
        $n^{e}_{y}$ & $100$ & \texttt{nely = 100} \\
        $q_\alpha$ & $10$ & \texttt{qavec = 10} \\
        Max. iter. & $100$ & \texttt{conit = 100}
    \end{tabular}
    \caption{Parameter values for the drag and lift problem in Section \ref{sec:kondoh}.}
    \label{tab:kondoh}
\end{table}
However, details of the physical dimensions are missing from the original work, so the values used herein are given in Table \ref{tab:kondoh}. \citet{Kondoh2012} defines the Reynolds number in term of a characteristic length given by the square root of the area of the solid profile:
\begin{equation}
    L_{c} = \sqrt{V_{f}| \Omega_{d} |}
\end{equation}
where $| \Omega_{d} |$ is the size of the design domain and $V_{f}$ the prescribed volume fraction. The maximum Brinkman penalty is defined as $\alpha_\textrm{max} = \frac{\mu}{(10^{-5}L_{c})}$ in order to ensure a value of $10^5$ for a Reynolds number of 1.

The optimisation problem for minimum drag is formulated as:
\begin{equation}\label{eq:optprob_drag}
\begin{split}
 \underset{\bm{\gamma}}{\text{minimise:}} &\quad f_{D}\left(\textbf{s}(\bm{\gamma}),\bm{\gamma}\right) = D \\
\text{subject to:} & \quad V(\bm{\gamma}) \leq V_{f}\\
\text{with:}& \quad \textbf{r}(\textbf{s}(\bm{\gamma}),\bm{\gamma})= \mathbf{0}\\
& \quad 0 \leq \gamma_{i} \leq 1, \; i=1,\ldots,n_{el}
\end{split}
\end{equation}
which only requires a single adjoint problem for the drag force, similar to the minimum dissipation problem. For maximum lift, a maximum constraint on the drag force is necessary to avoid very thin and sharply rotated designs:
\begin{equation}\label{eq:optprob_lift}
\begin{split}
 \underset{\bm{\gamma}}{\text{minimise:}} &\quad f_{L}\left(\textbf{s}(\bm{\gamma}),\bm{\gamma}\right) = -L \\
\text{subject to:} & \quad V(\bm{\gamma}) \leq V_{f}\\
                  & \quad D\left(\textbf{s}(\bm{\gamma}),\bm{\gamma}\right) \leq \beta D_\textrm{ref} \\
\text{with:}& \quad \textbf{r}(\textbf{s}(\bm{\gamma}),\bm{\gamma})= \mathbf{0}\\
& \quad 0 \leq \gamma_{i} \leq 1, \; i=1,\ldots,n_{el}
\end{split}
\end{equation}
which means that a second adjoint problem is necessary, similar to the flow reversal problem. The maximum drag is set to different factors, $\beta$, relative to that of the drag minimised design, $D_\textrm{ref}$. \citet{Kondoh2012} also impose a constraint on the centre-of-gravity of the profile\footnote{The constraint is not formulated explicitly anywhere in the paper.}, which is not included herein for simplicity. Furthermore, similar to the rugby ball problem, a single interpolation factor of $q_\alpha = 10$ is sufficient for this external flow problem.

The MMA version of the code is required and this can easily be updated to handle passive elements by following the same steps as in Section \ref{sec:passive} and basically replacing \texttt{xPhys}, \texttt{sens} and \texttt{dV} with only the active elements (\texttt{active}) in the definition of arrays for MMA and changing \texttt{neltot} to \texttt{nactive} in the call to MMA. It is important to scale the drag constraint similarly to the volume constraint, $\sfrac{D}{(\beta D_\textrm{ref})} - 1 \leq 0$, when inputting it to MMA. Furthermore, the objective functional should be scaled using the absolute of the initial value, especially for the maximisation of lift since otherwise the sign changes incorrectly.

\begin{figure*}
    \centering
    \begin{subfigure}{\textwidth}
    \centering
    \hspace{0.07\columnwidth}
    \includegraphics[width=0.6\textwidth]{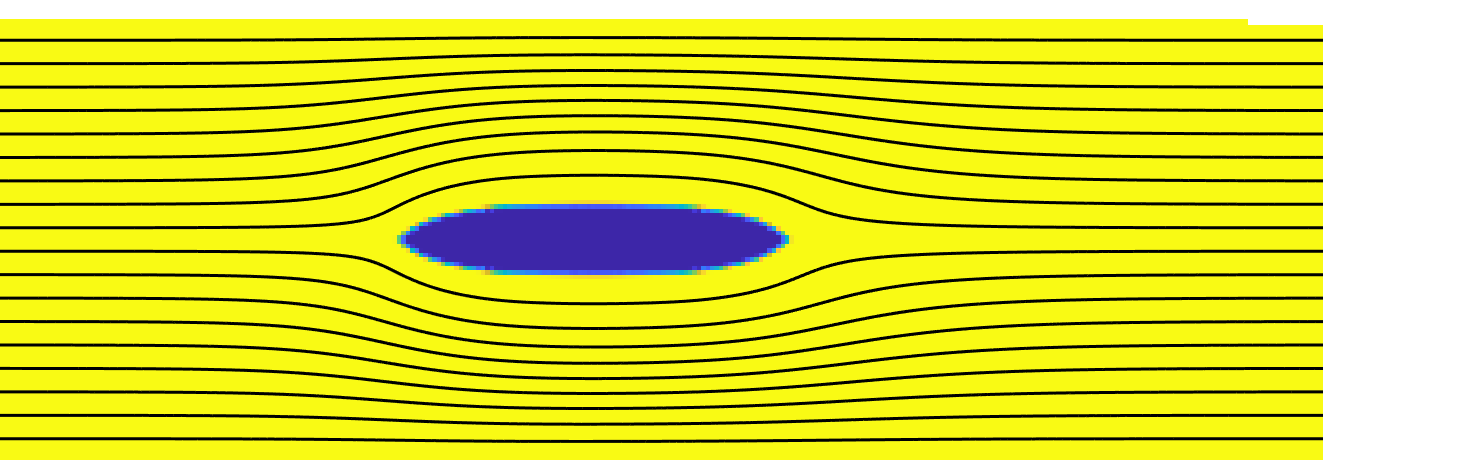}
    \caption{$Re = 10$}
    \label{fig:minDrag-a}
    \end{subfigure}
    \\
    \begin{subfigure}{\textwidth}
    \centering
    \hspace{0.07\columnwidth}
    \includegraphics[width=0.6\textwidth]{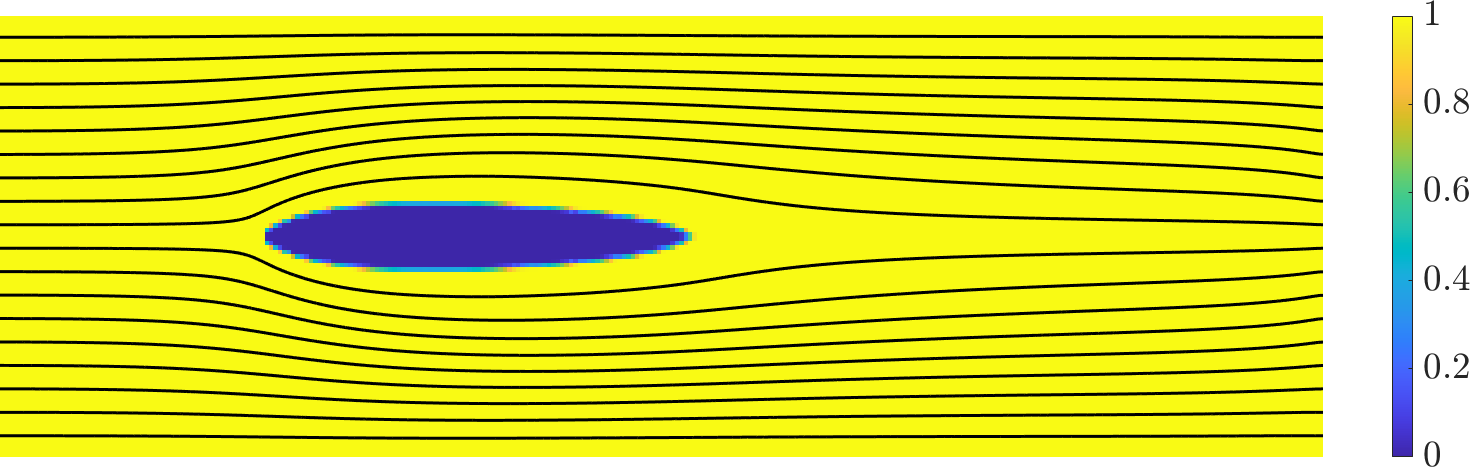}
    \caption{$Re = 100$}
    \label{fig:minDrag-b}
    \end{subfigure}
    \\
    \begin{subfigure}{\textwidth}
    \centering
    \hspace{0.07\columnwidth}
    \includegraphics[width=0.6\textwidth]{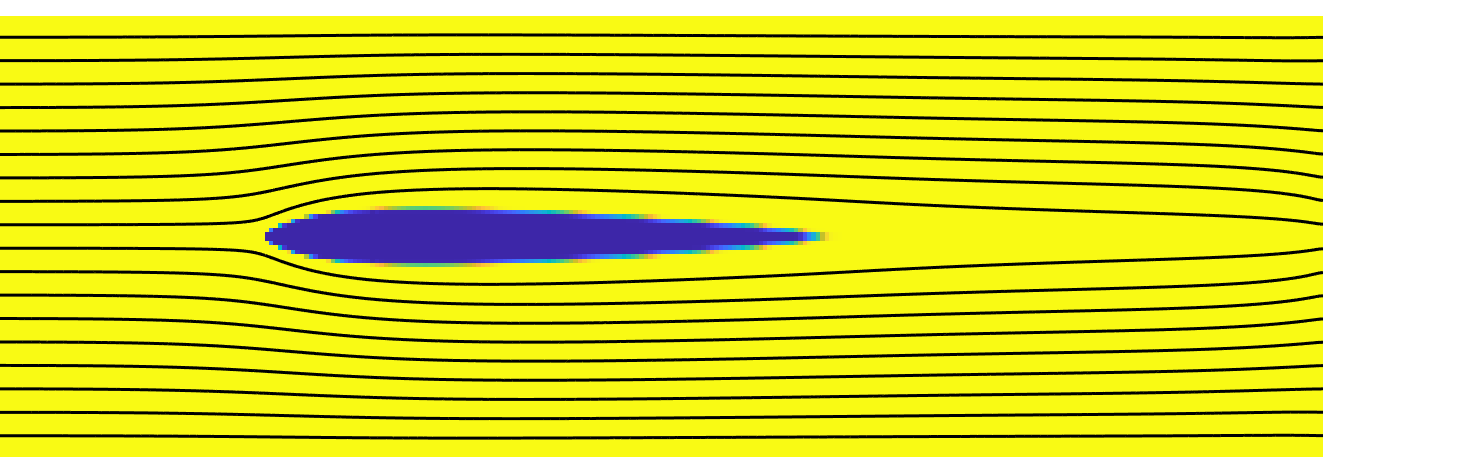}
    \caption{$Re = 1000$}
    \label{fig:minDrag-c}
    \end{subfigure}
    \caption{Optimised designs for minimum drag for increasing Reynolds number. Final values: (a) $D = 2.4207$ after $38$ iterations; (b) $D = 0.3399$ after $37$ iterations; (c) $D = 0.0800$ after $44$ iterations.}
    \label{fig:minDrag}
\end{figure*}
Figure \ref{fig:minDrag} shows optimised designs for increasing Reynolds number. The optimised designs are qualitatively similar to those presented by \citet{Kondoh2012}. It can be seen that for $Re=10$, a more or less symmetric profile is observed. For $Re=100$ and $Re=1000$, the profiles become elongated and tapered towards the trailing edge as is well-known from aerodynamics.

\begin{figure*}
    \centering
    \begin{subfigure}{\textwidth}
    \centering
    \hspace{0.1\columnwidth}
    \includegraphics[width=0.6\textwidth]{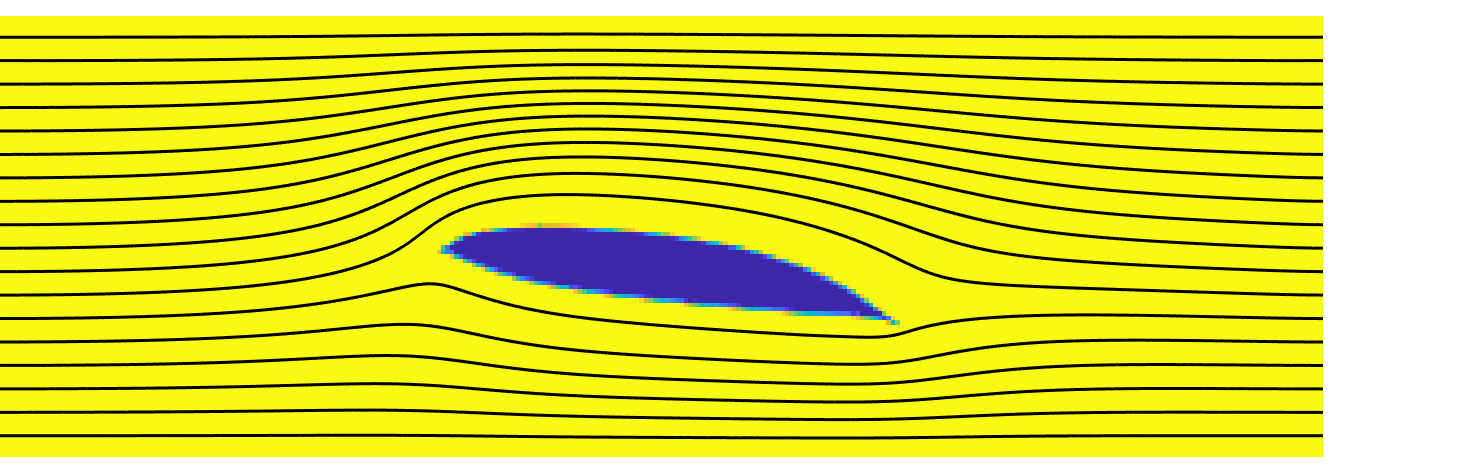}
    \caption{$\beta = 1.1$}
    \label{fig:maxLift-a}
    \end{subfigure}
    \\
    \begin{subfigure}{\textwidth}
    \centering
    \hspace{0.1\columnwidth}
    \includegraphics[width=0.6\textwidth]{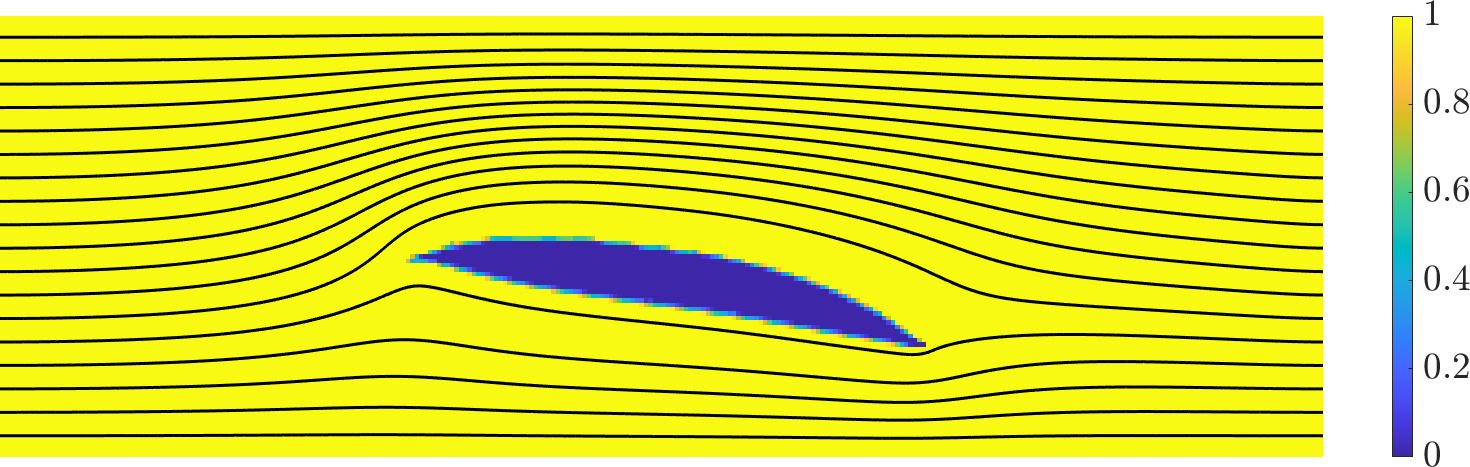}
    \caption{$\beta = 1.2$}
    \label{fig:maxLift-b}
    \end{subfigure}
    \\
    \begin{subfigure}{\textwidth}
    \centering
    \hspace{0.1\columnwidth}
    \includegraphics[width=0.6\textwidth]{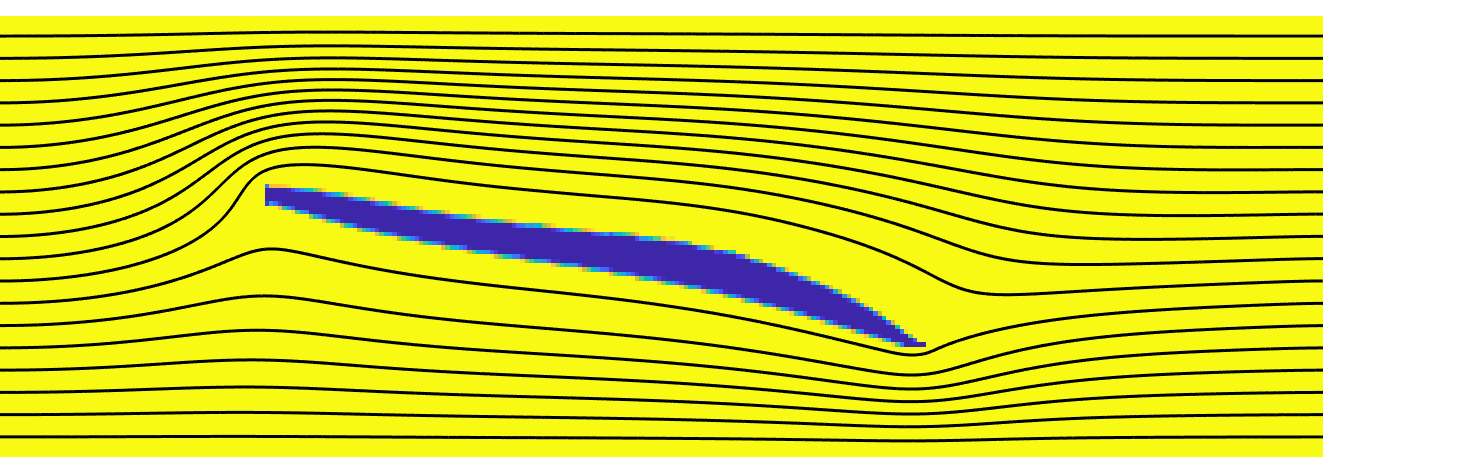}
    \caption{$\beta = 2.0$}
    \label{fig:maxLift-c}
    \end{subfigure}
    \caption{Optimised designs for maximum lift at $Re=10$ under increasing drag constraint relative to $D_\textrm{ref} = 2.4207$ on the drag. Final values: (a) $L = 2.0148$ after $100$ iterations (maximum); (b) $L = 3.1882$ after $100$ iterations (maximum); (c) $L = 7.2880$ after $60$ iterations.}
    \label{fig:maxLift}
\end{figure*}
Figure \ref{fig:maxLift} shows optimised designs for maximum lift at $Re=10$. The drag optimised design from Figure \ref{fig:minDrag-a} is used as the initial design field and the drag is constrained to be less than 10\%, 20\% and 100\% higher than the drag optimised case. It can be seen that the optimised design becomes further elongated as larger drag is allowed. The optimised designs are qualitatively similar to those presented by \citet{Kondoh2012}, except the design for $\beta=2.0$ which has a blunt tip due to reaching the edge of the design domain.

\subsection{Other extensions}

Due to the similarity in structure of the presented code to the well-known 88-line code for static mechanics, many of the extensions made available to that code are applicable here. For some problems, it might be relevant to include length-scale control. For that purpose, filtering, projection and robust formulations can be directly copied from extensions to the 88-line code. 

The purpose of the presented code is to stand on its own as an introductory tool for newcomers to the field. However, it is also the idea that it will serve as the basis for extension to flow-driven multiphysics. A sequel paper is planned treating fluid-structure-interaction and conjugate heat transfer.

\section{Computational performance} \label{sec:compperf}

This subsection covers the computational time and memory use of the code. The results are given using both a workstation and a laptop to compare the upper and lower bounds of performance. The workstation was running Ubuntu 18.04.6LTS with an Intel Core i9-9980XE CPU with 18 cores at $3.0\textrm{GHz}$ and $128\textrm{GB}$ of memory. The laptop was running Windows 10 with an Intel Core i5-8350U CPU with 4 cores at $1.7\textrm{GHz}$ and $8\textrm{GB}$ of memory. Both machines were running MATLAB R2021b and had all unnecessary processes shut down. MATLAB was completely shut down and restarted to clear all memory cache between runs.

\subsection{Computational time}

\begin{table*}
    \centering
    \begin{tabular}{c|c|c|c|c|c|c|c|c|c}
         &  &  &  & \multicolumn{3}{c|}{Workstation} & \multicolumn{3}{c}{Laptop} \\
        Problem & $Re$ & Elements & Iter. & Total [min] & Iteration [s] & Memory [GB] & Total [min] & Iteration [s] & \% slower \\ \hline
        Double pipe & $20$ & $15606$ & $102$ & $4.10$ & $2.41$ & $1.6$ & $6.08$ & $3.58$ & $48$ \\
        Pipe bend & $500$ & $10000$ & $40$ & $1.34$ & $2.01$ & $1.6$ & $1.2$ & $2.92$ & $44$ \\
        Rugby ball & $1$ & $10000$ & $19$ & $0.49$ & $1.55$ & $1.6$ & $1.3$ & $2.26$ & $47$ \\
        Flow reversal & $100$ & $12500$ & $90$ & $3.96$ & $2.33$ & $1.4$ & $6.00$ & $3.50$ & $52$ \\
        Maximum lift & $10$ & $30000$ & $100$ & $7.07$ & $4.24$ & $2.3$ & $10.60$ & $6.36$ & $50$ 
    \end{tabular}
    \caption{Computational time for selected results from the paper using both a workstation and a laptop with specifications given in the text. Total time is given in minutes and the average time per design iteration is given in second. The values correspond to the following results: double pipe = Figure \ref{fig:doublepipe_reinc-a}; pipe bend = Figure \ref{fig:example2_reinc}; rugby ball = Figure \ref{fig:rugbyball_result-a}; flow reversal = Figure \ref{fig:flowreversal_re100-a}; maximum lift = Figure \ref{fig:maxLift-b}.}
    \label{tab:comptime}
\end{table*}
Due to the non-linear governing equations and the vastly more complex finite element formulation, the present code is significantly slower than the available codes for static mechanics. The computational time of a selected number of results from the paper are listed in Table \ref{tab:comptime}. It can be seen that overall all the examples can be generated in 10 minutes or less on both of the types of machine. Generally, the laptop takes around 50\% longer than the workstation, which is mostly attributed to the significantly lower clock-speed of the CPU, but a high number of cores are likely also helpful for the vectorised assembly procedure.

The computational time can easily be decreased by doing one or both of the following:
\begin{enumerate}
    \item Loosening the tolerance of the non-linear solver
    \item Re-using the last Jacobian from the non-linear solver for the adjoint solver
\end{enumerate}
The tolerance of the non-linear solver, \texttt{nltol}, determines the number of iterations needed to provide a solution with the required accuracy. Loosening the tolerance (increasing the number) gives a less accurate solution to the non-linear system of equations in Equation \ref{eq:residual}. This means that errors will be introduced into the adjoint sensitivity analysis due to a non-zero residual. However, as detailed by \citet{Amir2010,Amir2014} for linear solvers, solver tolerances can be loosened significantly without impacting accuracy significantly. 

The transposed Jacobian is necessary for solving the adjoint problem and by default it is built using the final converged (to the desired accuracy) state solution. This ensures an accurate (to  solution accuracy) evaluation of the adjoint variables and, thus, the sensitivities. However, if the non-linear tolerance is tight (a low number), the state field will not have changed significantly from the second-to-last non-linear iteration to the last one. Therefore, the Jacobian already built for the last non-linear iteration can be used for solving the adjoint problem and this saves the time of assemblying the Jacobian a single time per design iteration.

\begin{table*}
    \centering
    \begin{tabular}{c|c|c|c|c|c|c|c|c|c|c|c}
         & \multicolumn{2}{c|}{Standard} & \multicolumn{3}{c|}{Re-use Jacobian} & \multicolumn{3}{c|}{Loose tolerance}  & \multicolumn{3}{c}{Both modifications} \\
        Problem & Obj. & Iter. & Obj. & Iter. & Time [\%] & Obj. & Iter. & Time [\%] & Obj. & Iter. & Time [\%] \\ \hline
        Double pipe & $2.126$ & $102$ & $2.129$ & $100$ & $89.7$ & $2.129$ & $100$ & $65.4$ & $2.130$ & $99$ & $55.9$ \\
        Flow reversal & $2.264$ & $102$ & $2.257$ & $105$ & $88.7$ & $2.232$ & $89$ & $67.4$ & $2.260$ & $83$ & $63.7$ \\
        Maximum lift & $3.188$ & $100m$ & $3.232$ & $100m$ & $91.2$ & $3.243$ & $100m$ & $68.6$ & $3.330$ & $100m$ & $58.3$ \\
    \end{tabular}
    \caption{Final objective value, number of design iterations, and average computational time per design iteration for selected results from the paper using a laptop with re-use of Jacobian from non-linear solver for the adjoint solver and/or loose tolerance of non-linear solver. All objective values are evaluated with the default tolerance of \texttt{nltol = 1e-6}. For the double pipe, the objective values should be multiplied by $10^{-1}$. All runs for maximum lift maxed out at 100 iterations (100m).}
    \label{tab:comptime_reduced}
\end{table*}
Table \ref{tab:comptime_reduced} shows the computational time for three examples using the two computational time reduction methods on the laptop. It can be seen that re-using the Jacobian can reduce the computational time per design iteration, but only by 8-11\%. Loosening the tolerance to \texttt{nltol = 1e-2}, compared to the default value of \texttt{nltol = 1e-6}, reduces the computational cost more significantly more by 31-35\%. Generally, the looser tolerance reduces the number of non-linear iterations by 1 or 2 per design iteration, depending on the non-linearity of the flow problem. Lastly, combining the two modifications together yields a total reduction of 36-44\%. The reduction in computational time does come at a cost of accuracy, which is reflected in the different number of design iterations and different final values of the  objective functionals. Despite the differences in objective values, visually-identical optimised designs are obtained for all cases. However, this may not be a general conclusion and it will depend on the physical problem and optimisation problem to be solved.

\subsection{Memory use}

The memory use of the code is of interest, because the local finite element matrices are formed all-at-once using the vectorised functionality of the auto-generated functions from the Symbolic Toolbox. Ideally, this means that the computer should be able to store all local finite element matrices in memory all at once - in addition to the factorisation of the Jacobian matrix for the solution of the systems of equations.

Table \ref{tab:comptime} also shows the amount of peak RAM memory used for the 5 examples, when run on the workstation (similar values were observed on the laptop). This is in addition to the approximately 1.2GB that MATLAB used in idle after startup. The laptop was able to handle a mesh of $90$ thousand elements ($300\times300$) without exceeding the 8GB of RAM and avoiding to resort to swap storage. A design iteration for the pipe bend problem took around 30 seconds for 5 Newton iterations. The workstation could easily handle a mesh of 1 million elements ($1000\times1000$), peaking at 50GB of RAM memory use and solving one design iteration in around 400 seconds (6.7 minutes).

By default, the \texttt{topFlow.m} script starts with the \texttt{clear} command. This ensures that any cached functions are not cleared and should speed up running of the code, after the first initial run. However, memory use can build up and if time is not an issue, the command can be updated to \texttt{clear all}. This clears the memory everytime the script is run. Beware that in this case, the initial design iterations will be slower, because MATLAB realises many of the functions are being called repeatedly.

\section{Concluding remarks} \label{sec:conclusion}

This article presents a detailed introduction to density-based topology optimisation of fluid flow problems. The goal is to be the first point of contact for new students and researchers, allowing them to quickly get started in the research area and to skip many of the initial steps, often consuming unnecessarily long time from the scientific advancement of the field. 

A step-by-step guide is provided to the components necessary to understand and implement topology optimisation for fluid flow. The continuous design representation and Brinkman penalty approach are examined in detail using parametric simulations of a reference geometry, as well as through optimisation examples. Under the way, several recommendations are given to the choice of minimum, maximum and initial Brinkman penalty value.

The guide is aided by a MATLAB code based on the well-known ``88-line'' code for static mechanics, but with significant modifications to treat Navier-Stokes flow. The code uses vectorised functions for building all element-level matrices and vectors, which have been generated using the MATLAB Symbolic Toolbox. All partial derivatives necessary for both the Newton solver and adjoint sensitivity analysis are automatically computed using symbolic differentiation.

The extendability of the code is demonstrated through various additional modifications and explanations. The code is shown to be efficient, but the computational time is significantly higher than that for static mechanics, due to the significantly more complex finite element formulation and the non-linear nature of the governing equations.

\begin{acknowledgements}
The author would like to acknowledge Yupeng Sun, for providing feedback on a close-to-finished manuscript, and Christian Lundgaard, for providing feedback on a very early draft. The author also wishes to thank the DTU TopOpt group for discussions over the many years it has taken to form the experienced embodied in this article.
\end{acknowledgements}

\section*{Conflict of interest}
The author has no conflict of interest.

\section*{Reproduction of results}
The code to reproduce all results is either directly provided or the required modifications are explained thoroughly. Upon proof-of-attempt, the code for the examples not provided can be obtained from the author. Lastly, manual derivations of the analytical sensitivities can be provided upon request.


\appendix

\section{Alternative scaling of Brinkman penalty factor} \label{app:kondoh}

The order of magnitude of the convective term is constant due to the non-dimensional formulation of Equation \ref{eq:navierstokes}. Thus, the Brinkman penalty factor should no longer become smaller with increasing Reynolds number. \citet{Kondoh2012} suggested the following scaling:
\begin{equation} \label{eq:alphakappa_kondoh}
    \alpha_{\rm{max}} = \left( 1 + \frac{1}{Re} \right) \frac{1}{Da}
\end{equation}
which aims to ensure that the Brinkman penalty factor is large enough even for increasing Reynolds numbers. From a scaling point of view, it can be argued that $\frac{1}{Da}$ makes sense for purely convective flow, which is recovered from the above for $Re \longrightarrow \infty$.

\begin{figure}
    \centering
    \includegraphics[width=0.95\linewidth]{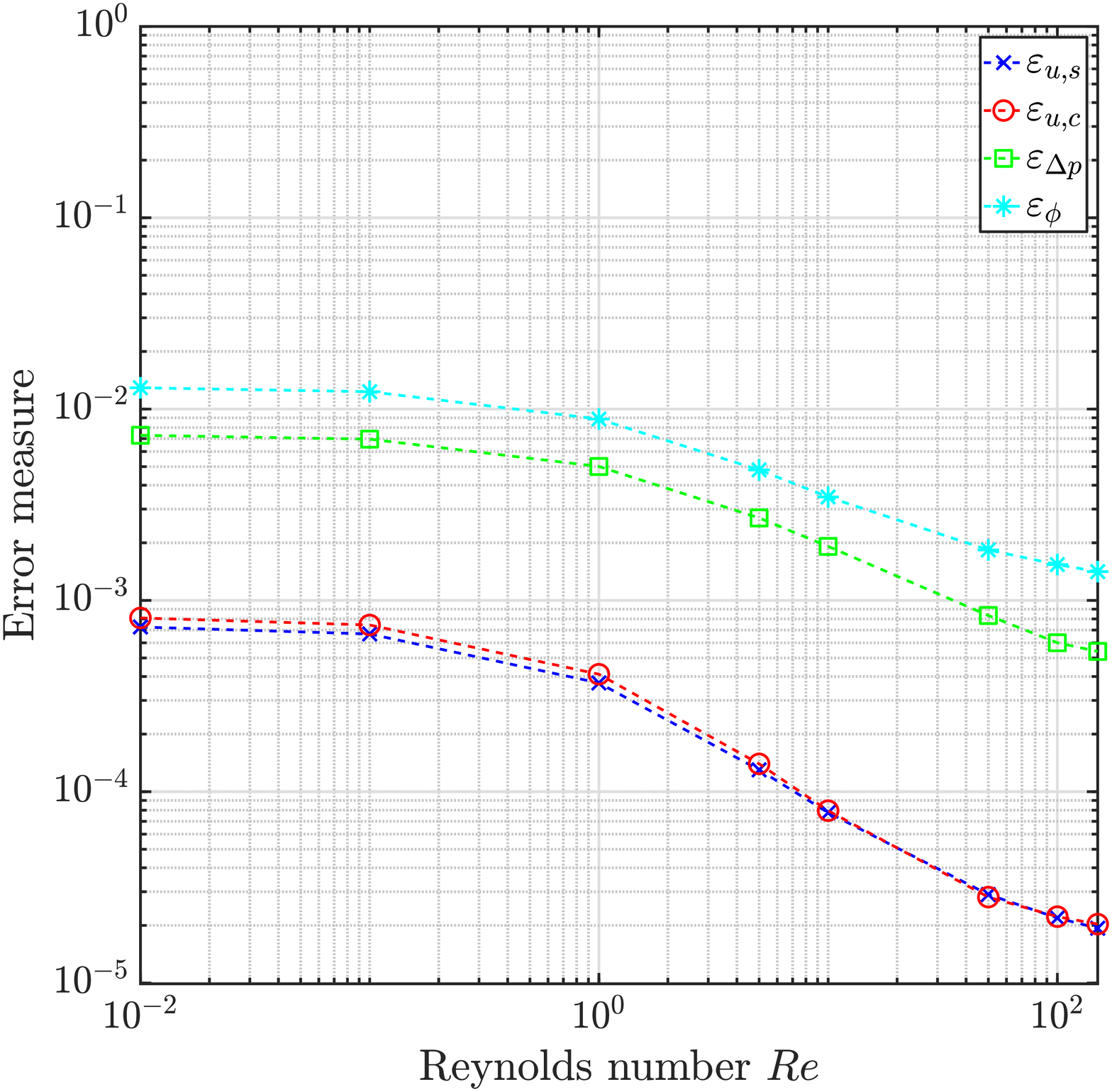}
    \caption{Error measures, when penalising using Equation \ref{eq:alphakappa_kondoh}, as a function of Reynolds number for a Darcy number of $Da=10^{-6}$ and $Re\in \{ 0.01,0.1,1,5,10,50,100,150 \}$.}
    \label{fig:headless_errorVrenum_kondoh}
\end{figure}
Figure \ref{fig:headless_errorVrenum_kondoh} shows the error measures as a function of Reynolds number when using Equation \ref{eq:alphakappa_kondoh}. In practise, it is seen that the error decreases for increasing Reynolds number over 1, meaning the flow in the solid is penalised increasingly hard. It appears that the error is beginning to stagnate. However, the flow becomes non-steady for $Re>150$ and, thus, the limit cannot be investigated presently. This is an area worth more investigation, as the community transitions to treating larger Reynolds number flows.

\section{Finite element formulation} \label{app:FEM}

The weak form of the momentum conservation equations is derived by multiplying the strong form by the test function $w_i$ for the velocity field, integrating over the volume, applying integration-by-parts and introducing a zero normal stress natural boundary condition:
\begin{multline}
    \int_\Omega \rho w_{i} u_{j}\dfrac{\partial u_i}{\partial x_j} \,dV + \int_\Omega \mu \dfrac{\partial w_{i}}{\partial x_j}\left( \dfrac{\partial u_i}{\partial x_j} + \dfrac{\partial u_j}{\partial x_i} \right) \,dV \\ - \int_\Omega  \dfrac{\partial w_{i}}{\partial x_i} p \,dV + \int_\Omega \alpha w_{i} u_{i} \,dV \\
    + \sum_{e=1}^{n_e} \int_{\Omega_{e}} \tau w_{k}\dfrac{\partial u_i}{\partial x_k} \left( \rho u_{j}\dfrac{\partial u_i}{\partial x_j} + \dfrac{\partial p}{\partial x_i} + \alpha u_{i} \right) \,dV= 0
\end{multline}
where the last integral is the additional SUPG stabilisation terms with $\tau$ as the stabilisation parameter.
Likewise, the weak form of the mass conservation equation is derived by multiplying the strong form with the test function $q$ for the pressure field:
\begin{multline}
    \int_\Omega q \dfrac{\partial u_i}{\partial x_i} \,dV \\ + \sum_{e=1}^{n_e} \int_{\Omega_{e}} \dfrac{\tau}{\rho} \dfrac{\partial q}{\partial x_i} \left( u_{j}\dfrac{\partial u_i}{\partial x_j} + \dfrac{\partial p}{\partial x_i} + \alpha u_{i} \right) \,dV = 0
\end{multline}
where the last integral is the additional PSPG stabilisation terms with $\tau$ as the stabilisation parameter. 
The diffusive term has been left out of the SUPG and PSPG stabilisation, since it is negligible for bi-linear interpolation functions due to the second-order derivative.

The stabilisation parameter is computed using an approximate minimum function:
\begin{equation}
    \tau = \left( {\tau_{1}}^{-2} + {\tau_{3}}^{-2} + {\tau_{4}}^{-2} \right)^{-1/2}
\end{equation}
based on three limit cases:
\begin{subequations}
\begin{align}
    \tau_{1} & = \frac{h}{2\sqrt{u_{i}u_{i}}}  \\
    \tau_{3} & = \frac{\rho h^{2}}{12\mu} \\
    \tau_{4} & = \frac{\rho}{\alpha}
\end{align}
\end{subequations}
where $\tau_{1}$ is the convective limit, $\tau_{3}$ is the diffusive limit, and $\tau_{4}$ is the reactive limit\footnote{$\tau_{2}$ is for the transient case, which is traditionally numbered as number 2, and left out since steady-state flow is treated herein.}. The reactive limit $\tau_{4}$ is very important to ensure stability in the solid domain and at the interface, especially for large Brinkman penalty parameters. The stabilisation parameters are assumed to be constant within each element and $\tau_{1}$ is, thus, computed based on the velocity components evaluated in the element centroid.

\section{Adjoint sensitivity analysis} \label{app:adjoint}
Adjoint sensitivity analysis need not be difficult or especially derived for every type of problem. The derivation is very simply by posing the system of equations (be they uncoupled, weakly coupled or strongly coupled) as a common residual:
\begin{equation}
\mathbf{r} = \mathbf{A} \, \mathbf{s} - \mathbf{b} = \mathbf{0}
\end{equation}
where $\mathbf{A}$ is the system coefficient matrix, $\mathbf{s}$ is the vector of all state variables and $\mathbf{b}$ is the forcing vector.
To find the derivatives of a given functional $f$, the Lagrangian $\mathcal{L}$ is defined as: 
\begin{equation}
\mathcal{L}= f - \bm{\lambda}^T\mathbf{r}
\end{equation}
where $\bm{\lambda}$ is the vector of adjoint variables. The total derivative with respect to a design variable $\gamma_{e}$ is then taken of the Lagrangian:
\begin{equation}
\frac{d\mathcal{L}}{d\gamma_{e}} = \frac{d f}{d\gamma_{e}} - \bm{\lambda}^T \frac{d\mathbf{r}}{d\gamma_{e}}
\label{eq:totalDerLag}
\end{equation}
where the total derivative is given by:
\begin{equation}
\frac{df}{d\gamma_{e}} = \frac{\partial f}{\partial\gamma_{e}} + \frac{\partial f}{\partial\mathbf{s}} \frac{\partial\mathbf{s}}{\partial\gamma_{e}}
\end{equation}
due to the implicit dependence of $f$ on the state field. Expanding the total derivative of the Lagrangian gives:
\begin{equation}
\frac{d\mathcal{L}}{d\gamma_{e}} = \frac{\partial f}{\partial\gamma_{e}} + \frac{\partial f}{\partial\mathbf{s}}\frac{\partial\mathbf{s}}{\partial\gamma_{e}} - \bm{\lambda}^T \left( \frac{\partial\mathbf{r}}{\partial\gamma_{e}} + \frac{\partial\mathbf{r}}{\partial\mathbf{s}} \frac{\partial\mathbf{s}}{\partial\gamma_{e}} \right)
\end{equation}
which can be rewritten to:
\begin{equation}
\frac{d\mathcal{L}}{d\gamma_{e}} = \frac{\partial f}{\partial\gamma_{e}} - \bm{\lambda}^T \frac{\partial\mathbf{r}}{\partial\gamma_{e}} + \left(  \frac{\partial f}{\partial\mathbf{s}} - \bm{\lambda}^T \frac{\partial\mathbf{r}}{\partial\mathbf{s}} \right) \frac{\partial\mathbf{s}}{\partial\gamma_{e}}
\end{equation}
by collecting the terms multiplied by the derivative of the state field. The adjoint problem is then defined as what is inside the brackets:
\begin{equation}
\frac{\partial \mathbf{r}}{\partial\mathbf{s}}^{T} \bm{\lambda} = \frac{\partial f}{\partial \mathbf{s}}^{T} 
\end{equation}
When $\bm{\lambda}$ is the solution to the adjoint problem, the terms inside the brackets become zero and it is avoided to compute the design sensitivities of the state field:
\begin{equation}
\frac{d\mathcal{L}}{d\gamma_{e}} = \frac{\partial f}{\partial\gamma_{e}} - \bm{\lambda}^T \frac{\partial\mathbf{r}}{\partial\gamma_{e}}
\end{equation}
Since the state solution will be updated after a design change to make the residual equal to zero, the total derivative of the residual with respect to the design variable is equal to zero. Thus, the total derivative of the Lagrangian will be equal to that of the functional and Equation \ref{eq:totalDerLag} gives:
\begin{equation}
    \frac{d\mathcal{L}}{d\gamma_{e}} = \frac{d f}{d\gamma_{e}}
\end{equation}
Thus, the final sensitivities of the given functional become:
\begin{equation}
\frac{df}{d\gamma_{e}} = \frac{\partial f}{\partial\gamma_{e}} - \mathbf{\lambda}^T \frac{\partial\mathbf{r}}{\partial\gamma_{e}}
\end{equation}
The above result is valid for ALL systems of equations, be they linear/non-linear or un/weakly/strongly coupled. This is the methodology laid forth in various textbooks, e.g. the “Structural Sensitivity Analysis and Optimization” series by \citet{Choi2005,Choi2005a}, and papers from the 1990s \citet{Michaleris1994}.

\onecolumn{
\section{MATLAB code: \texttt{topFlow.m}} \label{app:topFlow}
\vspace{-12pt}
\begin{lstlisting}
%%%%%%%%%%%%%%%%%%%%%%%%%%%%%%%%%%%%%%%%%%%%%%%%%%%%%%%%%%%%
%    NAVIER-STOKES TOPOLOGY OPTIMISATION CODE, MAY 2022    %
% COPYRIGHT (c) 2022, J ALEXANDERSEN. BSD 3-CLAUSE LICENSE %
%%%%%%%%%%%%%%%%%%%%%%%%%%%%%%%%%%%%%%%%%%%%%%%%%%%%%%%%%%%%
clear; close all; clc;
%% DEFINITION OF INPUT PARAMETERS
% PROBLEM TO SOLVE (1 = DOUBLE PIPE; 2 = PIPE BEND)
probtype = 1;
% DOMAIN SIZE
Lx = 1.0; Ly = 1.0;
% DOMAIN DISCRETISATION
nely = 30; nelx = nely*Lx/Ly;
% ALLOWABLE FLUID VOLUME FRACTION
volfrac = 1/3; xinit = volfrac;
% PHYSICAL PARAMETERS
Uin = 1e0; rho = 1e0; mu = 1e0;
% BRINKMAN PENALISATION
alphamax = 2.5*mu/(0.01^2); alphamin = 2.5*mu/(100^2);
% CONTINUATION STRATEGY
ainit = 2.5*mu/(0.1^2);
qinit = (-xinit*(alphamax-alphamin) - ainit + alphamax)/(xinit*(ainit-alphamin));
qavec = qinit./[1 2 10 20]; qanum = length(qavec); conit = 50;
% OPTIMISATION PARAMETERS
maxiter = qanum*conit; mvlim = 0.2; plotdes = 0;
chlim = 1e-3; chnum = 5;
% NEWTON SOLVER PARAMETERS
nltol = 1e-6; nlmax = 25; plotres = 0;
% EXPORT FILE
filename='output'; exportdxf = 0;
%% PREPARE FINITE ELEMENT ANALYSIS
dx = Lx/nelx; dy = Ly/nely;
nodx = nelx+1; nody = nely+1; nodtot = nodx*nody;
neltot = nelx*nely; doftot = 3*nodtot;
% NODAL CONNECTIVITY
nodenrs = reshape(1:nodtot,nody,nodx);
edofVecU = reshape(2*nodenrs(1:end-1,1:end-1)+1,neltot,1);
edofMatU = repmat(edofVecU,1,8)+repmat([0 1 2*nely+[2 3 0 1] -2 -1],neltot,1);
edofVecP = reshape(nodenrs(1:end-1,1:end-1),neltot,1);
edofMatP = repmat(edofVecP,1,4)+repmat([1 nely+[2 1] 0],neltot,1);
edofMat = [edofMatU 2*nodtot+edofMatP];
iJ = reshape(kron(edofMat,ones(12,1))',144*neltot,1);
jJ = reshape(kron(edofMat,ones(1,12))',144*neltot,1);
iR = reshape(edofMat',12*neltot,1); jR = ones(12*neltot,1); 
jE = repmat(1:neltot,12,1);
%% DEFINE BOUNDARY CONDITIONS
% DEFINE THE PROBLEMS IN SEPARATE MATLAB FILE
run('problems.m');
% NULLSPACE MATRICES FOR IMPOSING BOUNDARY CONDITIONS
EN=speye(doftot); ND=EN; ND(fixedDofs,fixedDofs)=0.0; EN=EN-ND;
% VECTORS FOR FREE DOFS
alldofs = 1:doftot; freedofs = setdiff(alldofs,fixedDofs);
%% INITIALISATION
% SOLUTION VECTOR
S = zeros(doftot,1); dS = S; L = S; 
S(fixedDofs) = DIR(fixedDofs);
% DESIGN FIELD
xPhys = xinit*ones(nely,nelx); 
% COUNTERS
loop = 0; loopcont = 0; nlittot = 0; chcnt = 0;
% CHANGE
change = Inf; objOld = Inf;
% CONTINUATION
qastep = 1; qa = qavec(1);
% VECTORISED CONSTANTS
dxv = dx*ones(1,neltot); dyv = dy*ones(1,neltot);
muv = mu*ones(1,neltot); rhov = rho*ones(1,neltot);
%% OUTPUT PROBLEM INFORMATION
fprintf('=========================================================\n');
fprintf('      Problem number: %2i - Reynolds number: %3.2e\n',probtype,Renum);
fprintf('=========================================================\n');
fprintf('      Design it.:   0\n');
%% START ITERATION
destime = tic; ittime = tic;
while (loop <= maxiter)
    if (plotdes); figure(1); imagesc(xPhys); colorbar; caxis([0 1]); axis equal; axis off; drawnow; end
    %% GREYSCALE INDICATOR
    Md = 100*full(4*sum(xPhys(:).*(1-xPhys(:)))/neltot); 
    %% MATERIAL INTERPOLATION
    alpha = alphamin + (alphamax-alphamin)*(1-xPhys(:))./(1+qa*xPhys(:));
    dalpha = (qa*(alphamax - alphamin)*(xPhys(:) - 1))./(xPhys(:)*qa + 1).^2 - (alphamax - alphamin)./(xPhys(:)*qa + 1);
    %% NON-LINEAR NEWTON SOLVER
    normR = 1; nlit = 0; fail = -1; nltime = tic;
    while (fail ~= 1)
        nlit = nlit+1; nlittot = nlittot+1;
        % BUILD RESIDUAL AND JACOBIAN
        sR = RES(dxv,dyv,muv,rhov,alpha(:)',S(edofMat'));
        R = sparse(iR,jR,sR(:)); R(fixedDofs) = 0;
        if (nlit == 1); r0 = norm(R); end
        r1 = norm(R); normR = r1/r0;
        if (plotres); figure(6); semilogy(nlittot,normR,'x'); axis square; grid on; hold on; end
        if (normR < nltol); break; end
        sJ = JAC(dxv,dyv,muv,rhov,alpha(:)',S(edofMat'));
        J = sparse(iJ,jJ,sJ(:)); J = (ND'*J*ND+EN);  
        % CALCULATE NEWTON STEP
        dS = -J\R;
        % L2-NORM LINE SEARCH
        Sp = S + 0.5*dS;
        sR = RES(dxv,dyv,muv,rhov,alpha(:)',Sp(edofMat'));
        R = sparse(iR,jR,sR(:)); R(fixedDofs) = 0; r2 = norm(R);
        Sp = S + 1.0*dS;
        sR = RES(dxv,dyv,muv,rhov,alpha(:)',Sp(edofMat'));
        R = sparse(iR,jR,sR(:)); R(fixedDofs) = 0; r3 = norm(R);
        % SOLUTION UPDATE WITH "OPTIMAL" DAMPING
        lambda = max(0.01,min(1.0,(3*r1 + r3 - 4*r2)/(4*r1 + 4*r3 - 8*r2)));
        S = S + lambda*dS;
        % IF FAIL, RETRY FROM ZERO SOLUTION
        if (nlit == nlmax && fail < 0); nlit = 0; S(freedofs) = 0.0; normR=1; fail = fail+1; end
        if (nlit == nlmax && fail < 1); fail = fail+1; end
    end
    nltime=toc(nltime);
    fprintf('      Newton it.: %2i - Res. norm: %3.2e - Sol. time: %6.3f sec\n',nlit,normR,nltime);
    if (fail == 1); error('ERROR: Solver did not converge after retry from zero!\n      Stopping optimisation.\n'); end
    %% OBJECTIVE EVALUATION
    obj = sum( PHI(dxv,dyv,muv,alpha(:)',S(edofMat')) ); 
    change = abs(objOld-obj)/objOld; objOld = obj;
    %% VOLUME CONSTRAINT
    V = mean(xPhys(:));
    %% PRINT RESULTS
    ittime = toc(ittime);
    fprintf('      Obj.: %3.2e - Constr.: %3.2e - Md: %3.2f\n',obj,V,Md);
    fprintf('      Change: %4.3e - It. time: %6.3f sec\n',change,ittime);
    fprintf('      Contin. step: %2i - qa: %4.3e\n',qastep,qa);
    ittime = tic;
    %% EVALUATE CURRENT ITERATE - CONTINUE UNLESS CONSIDERED CONVERGED
    if (change < chlim); chcnt = chcnt + 1; else; chcnt = 0; end
    if (qastep == qanum && ( (chcnt == chnum) || (loopcont == conit) ) ); break; end
    %% PRINT HEADER FOR ITERATION
    loop = loop + 1; loopcont = loopcont + 1;
    fprintf('---------------------------------------------------------\n');
    fprintf('      Design it.:%4i\n',loop);
    %% ADJOINT SOLVER
    sR = [dPHIds(dxv,dyv,muv,alpha(:)',S(edofMat')); zeros(4,neltot)];
    RHS = sparse(iR,jR,sR(:)); RHS(fixedDofs) = 0;
    sJ = JAC(dxv,dyv,muv,rhov,alpha(:)',S(edofMat'));
    J = sparse(iJ,jJ,sJ(:)); J = (ND'*J*ND+EN);
    L = J'\RHS;
    %% COMPUTE SENSITIVITIES
    % OBJECTIVE
    sR = dRESdg(dxv,dyv,muv,rhov,alpha(:)',dalpha(:)',S(edofMat'));
    dRdg = sparse(iR(:),jE(:),sR(:));
    dphidg = dPHIdg(dxv,dyv,muv,alpha(:)',dalpha(:)',S(edofMat'));
    sens = reshape(dphidg - L'*dRdg,nely,nelx);
    % VOLUME CONSTRAINT
    dV = ones(nely,nelx)/neltot;
    %% OPTIMALITY CRITERIA UPDATE OF DESIGN VARIABLES AND PHYSICAL DENSITIES
    xnew = xPhys; xlow = xPhys(:)-mvlim; xupp = xPhys(:)+mvlim;
    ocfac = xPhys(:).*max(1e-10,(-sens(:)./dV(:))).^(1/3);
    l1 = 0; l2 = ( 1/(neltot*volfrac)*sum( ocfac ) )^3;
    while (l2-l1)/(l1+l2) > 1e-3
        lmid = 0.5*(l2+l1);
        xnew(:) = max(0,max(xlow,min(1,min(xupp,ocfac/(lmid^(1/3))))));
        if mean(xnew(:)) > volfrac; l1 = lmid; else; l2 = lmid; end
    end
    xPhys = xnew;
    %% CONTINUATION UPDATE
    if (qastep < qanum && (loopcont == conit || chcnt == chnum) )
        loopcont = 0; chcnt = 0;
        qastep = qastep + 1; qa = qavec(qastep);
    end
end
%% PRINT FINAL INFORMATION
destime = toc(destime);
fprintf('=========================================================\n');
fprintf('      Number of design iterations: %4i\n',loop);
fprintf('      Final objective: %4.3e\n',obj);
fprintf('      Total time taken: %6.2f min\n',destime/60);
fprintf('=========================================================\n');
%% PLOT RESULTS
run('postproc.m');
if (exportdxf); run('export.m'); end
%% %%%%%%%%%%%%%%%%%%%%%%%%%%%%%%%%%%%%%%%%%%%%%%%%%%%%%%%%%%%%%%%%%%%%%%%%
% This code was written by: Joe Alexandersen                              %
%                           Department of Mechanical and                  %
%                                         Electrical Engineering          %
%                           University of Southern Denmark                %
%                           DK-5230 Odense M, Denmark.                    %
% Please send your comments and questions to: joal@sdu.dk                 %
%                                                                         %
% The code is intended for educational purposes and theoretical details   %
% are discussed in the paper: "A detailed introduction to density-based   %
% topology optimisation of fluid flow problems including implementation   %
% in MATLAB", J. Alexandersen, SMO 2022, doi:                             %                          
%                                                                         %
% A preprint version of the paper can be downloaded from the author's     %
% website: joealexandersen.com                                            %
% The code is available from GitHub: github.com/sdu-multiphysics/topflow  %
%                                                                         %
% The basic structure of the code is based on the 88-line code for        %
% elastic compliance from: "Efficient topology optimization in MATLAB     %
% using 88 lines of code", E. Andreassen, A. Clausen, M. Schevenels,      %
% B. S. Lazarov and O. Sigmund, SMO 2010, doi:10.1007/s00158-010-0594-7   %
%                                                                         %
% Disclaimer:                                                             %
% The author does not guarantee that the code is free from errors.        %
% Furthermore, the author shall not be liable in any event caused by the  %
% use of the program.                                                     %      
%%%%%%%%%%%%%%%%%%%%%%%%%%%%%%%%%%%%%%%%%%%%%%%%%%%%%%%%%%%%%%%%%%%%%%%%%%%

\end{lstlisting}

\section{MATLAB code: \texttt{problems.m}} \label{app:problems}
\vspace{-12pt}
\begin{lstlisting}[firstline=1,lastline=57]
%%%%%%%%%%%%%%%%%%%%%%%%%%%%%%%%%%%%%%%%%%%%%%%%%%%%%%%%%%%%
%    NAVIER-STOKES TOPOLOGY OPTIMISATION CODE, MAY 2022    %
% COPYRIGHT (c) 2022, J ALEXANDERSEN. BSD 3-CLAUSE LICENSE %
%%%%%%%%%%%%%%%%%%%%%%%%%%%%%%%%%%%%%%%%%%%%%%%%%%%%%%%%%%%%
%% DEFINE THE PROBLEMS TO BE SOLVED  %%%%
if (probtype == 1) % DOUBLE PIPE PROBLEM
    if ( mod(nely,6) > 0 > 0 )
        error('ERROR: Number of elements in y-dir. must be divisable by 6.');
    elseif ( nely < 30 )
        error('ERROR: Number of elements in y-dir. must be above 30.');
    end
    inletLength = 1/6*nely; inlet1 = 1/6*nely+1; inlet2 = 4/6*nely+1;
    outletLength = 1/6*nely; outlet1 = 1/6*nely+1; outlet2 = 4/6*nely+1; 
    nodesInlet  = [inlet1:(inlet1+inletLength)' inlet2:(inlet2+inletLength)'];
    nodesOutlet = (nodx-1)*nody+[outlet1:(outlet1+outletLength)' outlet2:(outlet2+outletLength)'];
    nodesTopBot = [1:nely+1:nodtot nody:nody:nodtot];
    nodesLefRig = [2:nely (nelx)*nody+2:nodtot-1];
    fixedDofsTBx = 2*nodesTopBot-1; fixedDofsTBy = 2*nodesTopBot;
    fixedDofsLRx = 2*setdiff(nodesLefRig,[nodesInlet nodesOutlet])-1;
    fixedDofsLRy = 2*nodesLefRig;
    fixedDofsInX  = 2*nodesInlet-1; fixedDofsInY  = 2*nodesInlet;
    fixedDofsOutY = 2*nodesOutlet; fixedDofsOutP = 2*nodtot+nodesOutlet;
    fixedDofsU = [fixedDofsTBx fixedDofsTBy fixedDofsLRx fixedDofsLRy ...
                  fixedDofsInX fixedDofsInY fixedDofsOutY];
    fixedDofsP = [fixedDofsOutP];
    fixedDofs  = [fixedDofsU fixedDofsP];
    % DIRICHLET VECTORS
    u = @(y) -4*y.^2+4*y; Uinlet = Uin*u([0:inletLength]'/inletLength);
    DIRU=zeros(nodtot*2,1); DIRU(fixedDofsInX) = [Uinlet' Uinlet'];
    DIRP=zeros(nodtot,1); DIR = [DIRU; DIRP];     
    % INLET REYNOLDS NUMBER
    Renum = Uin*(inletLength*Ly/nely)*rho/mu;
elseif (probtype == 2) % PIPE BEND PROBLEM
    if ( mod(nelx,10) > 0 || mod(nely,10) > 0 )
        error('ERROR: Number of elements per side must be divisable by 10.');
    end
    inletLength = 2/10*nely; inlet1 = 1/10*nely+1; 
    outletLength = 2/10*nelx; outlet1 = 7/10*nelx+1; 
    nodesInlet  = [inlet1:(inlet1+inletLength)];
    nodesOutlet = nody*[outlet1:(outlet1+outletLength)];
    nodesTopBot = [1:nely+1:nodtot nody:nody:nodtot];
    nodesLefRig = [2:nely (nodx-1)*nody+2:nodtot-1];
    fixedDofsTBx = 2*nodesTopBot-1; fixedDofsTBy = 2*setdiff(nodesTopBot,nodesOutlet);
    fixedDofsLRx = 2*setdiff(nodesLefRig,nodesInlet)-1; fixedDofsLRy = 2*nodesLefRig;
    fixedDofsInX  = 2*nodesInlet-1; fixedDofsInY  = 2*nodesInlet;
    fixedDofsOutX = 2*nodesOutlet-1; fixedDofsOutP = 2*nodtot+nodesOutlet;
    fixedDofsU = [fixedDofsTBx fixedDofsTBy fixedDofsLRx fixedDofsLRy ...
                  fixedDofsInX fixedDofsInY fixedDofsOutX];
    fixedDofsP = [fixedDofsOutP];
    fixedDofs  = [fixedDofsU fixedDofsP];
    % DIRICHLET VECTORS
    DIRU=zeros(nodtot*2,1); DIRP=zeros(nodtot,1);
    u = @(y) -4*y.^2+4*y; Uinlet = Uin*u([0:inletLength]'/inletLength);
    DIRU(fixedDofsInX) = Uinlet'; DIR = [DIRU; DIRP];     
    % INLET REYNOLDS NUMBER
    Renum = Uin*(inletLength*Ly/nely)*rho/mu;
end
\end{lstlisting}

\section{MATLAB code: \texttt{postproc.m}} \label{app:postproc}
\vspace{-12pt}
\begin{lstlisting}[firstline=1,lastline=30]
%%%%%%%%%%%%%%%%%%%%%%%%%%%%%%%%%%%%%%%%%%%%%%%%%%%%%%%%%%%%
%    NAVIER-STOKES TOPOLOGY OPTIMISATION CODE, MAY 2022    %
% COPYRIGHT (c) 2022, J ALEXANDERSEN. BSD 3-CLAUSE LICENSE %
%%%%%%%%%%%%%%%%%%%%%%%%%%%%%%%%%%%%%%%%%%%%%%%%%%%%%%%%%%%%
%% POST-PROCESSING
% SETTING UP NODAL COORDINATES FOR STREAMLINES
[X,Y] = meshgrid(0.5:nodx,0.5:nody); sx = 0.5*ones(1,21); sy = linspace(0,nody,21);
U = S(1:(nely+1)*(nelx+1)*2); umag=reshape(sqrt(U(1:2:end).^2+U(2:2:end).^2),nely+1,nelx+1);
% DESIGN FIELD
figure(1); imagesc(xPhys); colorbar; caxis([0 1]); axis equal; axis off;
h = streamline(X,Y,reshape(U(1:2:end),nody,nodx),-reshape(U(2:2:end),nody,nodx),sx,sy); set(h,'Color','black');
% BRINKMAN PENALTY FACTOR
figure(2); imagesc(reshape(log10(alpha),nely,nelx)); colorbar; caxis([0 log10(alphamax)]); axis equal; axis off; %colormap turbo;
% VELOCITY MAGNITUDE FIELD
figure(3); imagesc(umag); colorbar; axis equal; axis on; hold on; %colormap turbo;
h = streamline(X,Y,reshape(U(1:2:end),nody,nodx),-reshape(U(2:2:end),nody,nodx),sx,sy); set(h,'Color','black');
% PRESSURE FIELD
P = S(2*nodtot+1:3*nodtot);
figure(4); imagesc(reshape(P,nody,nodx)); colorbar; axis equal; axis off; %colormap turbo;
h = streamline(X,Y,reshape(U(1:2:end),nody,nodx),-reshape(U(2:2:end),nody,nodx),sx,sy); set(h,'Color','black');
% VELOCITY ALONG A LINE
if (probtype == 2)
    uline=flipud(diag(fliplr(umag))); xline=flipud(diag(fliplr(xPhys)));
else
    uline=umag(:,floor((end-1)/2)); xline=xPhys(:,floor(end/2));
end
figure(5);
subplot(3,1,1); plot(uline,'-x'); grid on; 
subplot(3,1,2); plot(log10(uline),'-x'); grid on;
subplot(3,1,3); plot(xline,'-x'); grid on; drawnow
\end{lstlisting}

\section{MATLAB code: \texttt{export.m}} \label{app:export}
\vspace{-12pt}
\begin{lstlisting}[firstline=1,lastline=40]
%%%%%%%%%%%%%%%%%%%%%%%%%%%%%%%%%%%%%%%%%%%%%%%%%%%%%%%%%%%%
%    NAVIER-STOKES TOPOLOGY OPTIMISATION CODE, MAY 2022    %
% COPYRIGHT (c) 2022, J ALEXANDERSEN. BSD 3-CLAUSE LICENSE %
%%%%%%%%%%%%%%%%%%%%%%%%%%%%%%%%%%%%%%%%%%%%%%%%%%%%%%%%%%%%
%% DXF
xMod = padarray(padarray(xPhys,[1 1],'replicate'),[1 1],1);
figure('visible','off');
C = contour(flipud(xMod),[0.5 0.5]);
C = unique(C','rows','stable')';
C = C-2.5; C(1,:) = C(1,:)*dx; C(2,:) = C(2,:)*dy;
h = sqrt(dx^2+dy^2);
ind = 1:size(C,2);
ind1 = intersect(find(C(1,:)<Lx+2*dx),find(C(2,:)<Ly+2*dy));
ind2 = intersect(find(C(1,:)>0-dx),find(C(2,:)>0-dy));
ind = intersect(ind1,ind2);
X = []; Y = []; c = 0;
indused = [];
for i = 1:(length(ind))
    x0 = C(1,ind(i)); y0 = C(2,ind(i));
    redind = ind([1:max(1,i-5) i+1:min(length(ind),i+2)]);
    redind = setdiff(redind,indused(max(1,c-10):end));
    R = sqrt( (C(1,redind)-x0).^2 + (C(2,redind)-y0).^2 );
    [minR,idx] = min(R);
    if (minR < 2*h && minR > 0)
        c = c + 1;
        indused(c) = i;
        X(c,:) = [x0 C(1,redind(idx))];
        Y(c,:) = [y0 C(2,redind(idx))];
    end
end
fid=fopen([filename '.dxf'],'w');
fprintf(fid,'0\nSECTION\n2\nENTITIES\n0\n');
for i=1:size(X,1)
      fprintf(fid,'LINE\n8\n0\n');
      fprintf(fid,'10\n%.4f\n20\n%.4f\n30\n%.4f\n',X(i,1),Y(i,1),0);
      fprintf(fid,'11\n%.4f\n21\n%.4f\n31\n%.4f\n',X(i,2),Y(i,2),0); 
      fprintf(fid,'0\n');
end
fprintf(fid,'ENDSEC\n0\nEOF\n');
fclose(fid);
\end{lstlisting}

\section{MATLAB code: \texttt{analyticalElement.m}} \label{app:analytical}
\vspace{-12pt}
\begin{lstlisting}[firstline=1,lastline=148]
%%%%%%%%%%%%%%%%%%%%%%%%%%%%%%%%%%%%%%%%%%%%%%%%%%%%%%%%%%%%
%    NAVIER-STOKES TOPOLOGY OPTIMISATION CODE, MAY 2022    %
% COPYRIGHT (c) 2022, J ALEXANDERSEN. BSD 3-CLAUSE LICENSE %
%%%%%%%%%%%%%%%%%%%%%%%%%%%%%%%%%%%%%%%%%%%%%%%%%%%%%%%%%%%%
clear; close all; clc;
%% Switches
buildJR = 1; exportJR = 1;
buildPhi = 1; exportPhi = 1;
%% Initialisation
syms rho mu alpha dalpha xi eta dx dy;
syms u1 u2 u3 u4 u5 u6 u7 u8 p1 p2 p3 p4;
time = tic;
%% Shape functions and matrices
xv = [-1 1 1 -1]'; yv = [-1 -1 1 1]';
Np = 1/4*(1+xi*xv').*(1+eta*yv');
Nu = sym(zeros(2,8));
Nu(1,1:2:end-1) = Np;
Nu(2,2:2:end) = Np;
%% Nodal coordinates, interpolation and coordinate transforms
xc = dx/2*xv; yc = dy/2*yv;
x = Np*xc; y = Np*yc;
J = [diff(x,xi) diff(y,xi); diff(x,eta) diff(y,eta)];
iJ = inv(J); detJ = det(J);
%% Derivatives of shape functions
dNpdx = iJ(:,1)*diff(Np,xi) + iJ(:,2)*diff(Np,eta);
dNudx = sym(zeros(2,8,2));
for i = 1:2
   dNudx(i,:,:) = transpose(iJ(:,1)*diff(Nu(i,:),xi) + iJ(:,2)*diff(Nu(i,:),eta));
end
%% Nodal DOFs
u = [u1; u2; u3; u4; u5; u6; u7; u8];
p = [p1; p2; p3; p4]; s = [u; p];
ux = Nu*u; px = Np*p;
dudx = [dNudx(:,:,1)*u dNudx(:,:,2)*u];
dpdx = dNpdx*p;
%% Stabilisation parameters
h = sqrt(dx^2 + dy^2);
u0 = subs(ux,[xi,eta],[0,0]);
ue = sqrt(transpose(u0)*u0);
tau1 = h/(2*ue);
tau3 = rho*h^2/(12*mu);
tau4 = rho/alpha;
tau  = (tau1^(-2) + tau3^(-2) + tau4^(-2))^(-1/2);
%% Loop over the tensor weak form to form residual
if (buildJR)
    Ru = sym(zeros(8,1)); Rp = sym(zeros(4,1));
    % Momentum equations
    for g = 1:8
        for i = 1:2
            Ru(g) = Ru(g) + alpha*Nu(i,g)*ux(i); % Brinkman term
            for j = 1:2
                Ru(g) = Ru(g) + mu*dNudx(i,g,j)*( dudx(i,j) + dudx(j,i) ); % Viscous term
                Ru(g) = Ru(g) + rho*Nu(i,g)*ux(j)*dudx(i,j); % Convection term
                Ru(g) = Ru(g) + tau*ux(j)*dNudx(i,g,j)*alpha*ux(i); % SUPG Brinkman term
                for k = 1:2
                    Ru(g) = Ru(g) + tau*ux(j)*dNudx(i,g,j)*rho*ux(k)*dudx(i,k); % SUPG convection term
                end
                Ru(g) = Ru(g) + tau*ux(j)*dNudx(i,g,j)*dpdx(i); % SUPG pressure term
            end
            Ru(g) = Ru(g) - dNudx(i,g,i)*px; % Pressure term
        end
    end
    % Incompressibility equations
    for g = 1:4
        for i = 1:2
            Rp(g) = Rp(g) + Np(1,g)*dudx(i,i); % Divergence term
            Rp(g) = Rp(g) + tau/rho*dNpdx(i,g)*alpha*ux(i); % PSPG Brinkman term
            for j = 1:2
                Rp(g) = Rp(g) + tau*dNpdx(i,g)*ux(j)*dudx(i,j); % PSPG convection term
            end
            Rp(g) = Rp(g) + tau/rho*dNpdx(i,g)*dpdx(i); % PSPG pressure term
        end
    end
    fprintf('Simplifying Ru... \n');
    Ru = simplify(detJ*Ru);
    fprintf('Simplifying Rp... \n');
    Rp = simplify(detJ*Rp);
    %% Integrate analytically
    fprintf('Integrating Ru... \n');
    Ru = int(int(Ru,xi,[-1 1]),eta,[-1 1]);
    fprintf('Integrating Rp... \n');
    Rp = int(int(Rp,xi,[-1 1]),eta,[-1 1]);
    fprintf('Simplifying Ru... \n');
    Ru = simplify(Ru);
    fprintf('Simplifying Rp... \n');
    Rp = simplify(Rp);
    %% Differentiate residual to form Jacobian
    Re = [Ru; Rp];
    Je = sym(zeros(12,12));
    for b = 1:12
        fprintf('Computing dR/ds%2i... \n',b);
        Je(:,b) = diff(Re,s(b));
    end
end
%% Export residual and Jacobian
if (exportJR)
    fprintf('Exporting residual... \n');
    f = matlabFunction(Re,'File','RES','Comments','Version: 0.99','Vars',{dx,dy,mu,rho,alpha,transpose([u1 u2 u3 u4 u5 u6 u7 u8 p1 p2 p3 p4])});
    fprintf('Exporting Jacobian... \n');
    f = matlabFunction(Je(:),'File','JAC','Comments','Version: 0.99','Vars',{dx,dy,mu,rho,alpha,transpose([u1 u2 u3 u4 u5 u6 u7 u8 p1 p2 p3 p4])});
end
%%
time = toc(time);
fprintf('Finished analysis part in %1.3e seconds. \n',time)
%% OPTIMISATION PART
time = tic;
if (buildPhi)
    %% Compute objective functional
    fprintf('Computing phi... \n');
    phi = 1/2*alpha*transpose(ux)*ux;
    for i = 1:2
        for j = 1:2
            phi = phi + 1/2*mu*dudx(i,j)*( dudx(i,j) + dudx(j,i) );
        end
    end
    %% Intregrate and simplify objective functional
    fprintf('Integrating phi... \n');
    phi = int(int(detJ*phi,xi,[-1 1]),eta,[-1 1]);
    phi = simplify(phi);
    %% Compute the partial derivative wrt. design field
    fprintf('Computing dphi/dgamma... \n');
    dphidg = simplify( diff(phi,alpha)*dalpha );
    %% Compute the partial derivative wrt. state field
    dphids = sym(zeros(12,1));
    for a = 1:12
        fprintf('Computing dphi/ds%2i... \n',a);
        dphids(a) = simplify(diff(phi,s(a)));
    end
end
%% Compute partial derivative of residual wrt. design field
if (buildJR)
    fprintf('Computing dr/dgamma... \n');
    drdg = simplify( diff(Re,alpha)*dalpha );
end
%% Export optimisation functions
if (exportPhi)
    fprintf('Exporting phi... \n');
    f = matlabFunction(phi,'File','PHI','Comments','Version: 0.9','Vars',{dx,dy,mu,alpha,transpose([u1 u2 u3 u4 u5 u6 u7 u8 p1 p2 p3 p4])});
    fprintf('Exporting dphi/dg... \n');
    f = matlabFunction(dphidg,'File','dPHIdg','Comments','Version: 0.9','Vars',{dx,dy,mu,alpha,dalpha,transpose([u1 u2 u3 u4 u5 u6 u7 u8 p1 p2 p3 p4])});
    fprintf('Exporting dphi/ds... \n');
    f = matlabFunction(dphids(1:8),'File','dPHIds','Comments','Version: 0.9','Vars',{dx,dy,mu,alpha,transpose([u1 u2 u3 u4 u5 u6 u7 u8 p1 p2 p3 p4])});
end
if (exportJR)
    fprintf('Exporting dr/ds... \n');
    f = matlabFunction(drdg,'File','dRESdg','Comments','Version: 0.9','Vars',{dx,dy,mu,rho,alpha,dalpha,transpose([u1 u2 u3 u4 u5 u6 u7 u8 p1 p2 p3 p4])});
end
%%
\end{lstlisting}
}


\end{document}